\documentclass[%
nofootinbib,
amsmath,amssymb,
aps,
]{revtex4-2}

\usepackage{graphicx}
\usepackage{dcolumn}
\usepackage{bm}

\usepackage{array}
\usepackage{booktabs}
\usepackage{tabu}
\usepackage{dcolumn}
\usepackage{amsmath}
\usepackage{amsfonts}
\usepackage{amssymb}
\usepackage{graphicx}
\usepackage{subfigure}
\usepackage{graphicx}
\usepackage{dcolumn}
\usepackage{bm}
\usepackage{xcolor}

\graphicspath{{figs/}} 

\usepackage{orcidlink} 

\begin{document}
	
%

\title{Spinning Primordial Black Holes and Scalar Induced Gravitational Waves from Single Field Inflation}

\author{Abolhassan Mohammadi\orcidlink{0000-0003-1228-9107}}
\email{abolhassanm@zjut.edu.cn}
\affiliation{Institute for Theoretical Physics \& Cosmology, Zhejiang University of Technology, Hangzhou, 310023, China}

\author{Yogesh\orcidlink{0000-0002-7638-3082}}
\email{yogesh@zjut.edu.cn}
\affiliation{Institute for Theoretical Physics \& Cosmology, Zhejiang University of Technology, Hangzhou, 310023, China}

\author{Qiang Wu\orcidlink{0000-0002-5483-4903}}
\email{wuq@zjut.edu.cn}
\affiliation{Institute for Theoretical Physics \& Cosmology, Zhejiang University of Technology, Hangzhou, 310023, China}

\author{Tao Zhu\orcidlink{0000-0003-2286-9009}}
\email{zhut05@zjut.edu.cn}
\affiliation{Institute for Theoretical Physics \& Cosmology, Zhejiang University of Technology, Hangzhou, 310023, China}
 
\begin{abstract}
We investigate the formation of primordial black holes (PBHs), their spin and abundance, in a single-field inflationary model based on a mutated hilltop potential inserted with a small step-like feature. This step induces a brief phase of ultra-slow-roll inflation, producing the large enhancement of the scalar power spectrum required for an appreciable amount of PBH abundance. Instead of the commonly used analytical power spectra, we compute the primordial power spectrum accurately by numerically solving the Mukhanov-Sasaki equation. Using the obtained power spectrum, we apply peak theory with $\nabla^2 \zeta$ treated as a Gaussian random field and parametrize the curvature profile by its amplitude $\mu$ and characteristic width $K$. Confining the study to Type-I PBH, the threshold value is calculated using two robust methods: the average of the compaction function and the q-function method. Using the result, the dimensionless spin parameter of the resulting PBHs is calculated at linear order and found to be $\sqrt{\langle a_\star^2 \rangle} \sim 10^{-3}$; however, it can be higher for smaller masses. We present detailed predictions for two representative parameter sets, calculate the present-day PBH mass function $f_{\rm PBH}(M)$ and the associated scalar-induced gravitational waves (SIGW). The first produces PBHs of mass $M \simeq 10^{-13}\,M_\odot$ that can account for $100\%$ of dark matter, while the second yields $M \simeq 10^{-2}\,M_\odot$ PBHs contributing approximately $2.4\%$ of the dark matter density. The predicted signals of SIGWs lie within the sensitivity bands of future experiments such as LISA, DECIGO, BBO, and SKA. In particular, the second parameter set produces a SIGWs compatible with the recent NANOGrav evidence for a low-frequency gravitational-wave signal.

\end{abstract}
\maketitle


\section{Introduction \label{intro}} 

Since the detection of the first gravitational waves (GWs) by LIGO/Virgo~\cite{LIGOScientific:2016dsl, LIGOScientific:2016aoc, LIGOScientific:2016sjg, LIGOScientific:2016wyt, LIGOScientific:2017bnn, LIGOScientific:2017vox, LIGOScientific:2017ycc}, interest in Primordial Black Holes(PBHs) has increased tremendously~\cite{Zeldovich:1967lct, Hawking:1971ei, Carr:1974nx, Carr:1975qj}. There is a possibility that the merger of (P)BH observed by GW surveys might have a primordial origin~\cite{Fernandez:2019kyb}. 
Another intriguing feature that PBHs exhibit is that they can be a natural non-particle candidate for dark matter~\cite{Carr:2020gox, Carr:2021bzv}. Unlike the astrophysical black holes, PBHs do not follow the stringent Chandrasekhar limit; they span a wide mass range from the Planck Mass to thousands of solar masses. PBHs in the mass range $10^{-16}\,M_\odot \lesssim M_{\rm PBH} \lesssim 5 \times 10^{-12}\,M_\odot$ could in principle account for all of the dark matter in the present Universe~\cite{Carr:2021bzv}. Several observational constraints have been imposed on the abundance of PBH based on their mass. Lighter PBHs of $M_{\rm PBH} \leq 10^{-16} M_\odot$ are constrained by the observations of galactic and extra-galactic $\gamma-$surveys~\cite{Laha:2019ssq, 2011MNRAS.411.1727C, Siegert:2016ijv, Dasgupta:2019cae, Laha:2020ivk}. On the other hand, heavy mass PBHs $M_{\rm PBH} \geq 10^{-11} M_\odot \ $ are constrained by the lensing and binary mergers~\cite{Carr:2021bzv}. 

In the early universe, PBHs are formed through the collapse of oversized densities after inflation. Several mechanisms for the formation of PBHs have been explored in the past, including the formation of PBHs through density fluctuations generated during the inflationary epoch~\cite{PhysRevD.50.7173, Yokoyama:1998pt, Garcia-Bellido:2017mdw, Ballesteros:2017fsr, Hertzberg:2017dkh, Kinney:2005vj, Germani:2017bcs, Pattison:2017mbe, Ezquiaga:2018gbw, Biagetti:2018pjj, Kohri:2007qn, Garcia-Bellido:1996mdl,Pi:2022zxs,Zhou:2020kkf, Cacciapaglia:2025xqd,Wang:2025lti,Kim:2025dyi,Cai:2018tuh,Kawasaki:1997ju, Lyth:2001nq, Kawasaki:2012wr, Kohri:2012yw, Yokoyama:1995ex, PhysRevLett.132.221003, Bhattacharya:2023ysp, Sharma:2024whg, Gangopadhyay:2021kmf,Correa:2022ngq, Braglia:2020eai, Bhattacharya:2022fze, Papanikolaou:2024fzf, Teimoori:2021thk, Solbi:2021rse, Heydari:2023rmq, Kawai:2021edk, Braglia:2020eai, Ragavendra:2020sop, Ragavendra:2023ret,Choudhury:2023vuj,Riotto:2023hoz,Choudhury:2023hvf}. Quantum fluctuations are generated at each scale during inflation, and as inflation proceeds, corresponding modes cross the horizon, becoming super-horizon fluctuations. These fluctuations then freeze and evolve into classical density fluctuations. After the end of inflation, when the comoving Hubble horizon starts to increase, the modes that were at the superhorizon start re-entering. If a mode has enough overdensity to overcome the pressure gradient of the background, it can collapse and form a PBH. The mass of the formed PBHs is characterized by the enhancement scale ($k_p$) of the primordial power spectrum. Formation of PBHs crucially depends on the collapse of the density perturbation. The criteria used in the literature, for a perturbation collapse based on Jeans length instability in a radiation-dominated universe, is that density contrasts exceeding a critical value $\delta_c \sim 1/3$ can collapse to form PBHs~\cite{Carr:1974nx}. 
In \cite{Harada:2013epa}, the authors present an analytical expression for calculating the collapse threshold, which depends solely on the background equation of state parameter ($w$)(see also~\cite{Escriva:2020tak} ). This calculation has been widely used to estimate the abundance of PBHs. The threshold signifies the minimum value of overdensities required to overcome the pressure and collapse to form a PBH. The abundance of PBHs based on $\delta_c$ has been previously discussed using the Press-Schechter (PS) formalism \cite{1974ApJ...187..425P}. 

However, in \cite{Shibata:1999zs} a new robust method based on the compaction function approach has been proposed to calculate the threshold for the PBH formation. The compaction function measures the mass excess in a given region, which provides a more accurate criterion for determining whether a region is likely to collapse into a PBH~\cite{Escriva:2022duf, Harada:2013epa, Harada:2023ffo}. In general, the compaction function is defined as $\mathcal{C} = 2 G\delta M (R)/R$, where $\delta M (R)$ is the mass excess within a region of areal radius $R$. This prescription is based on the Poisson equation, which establishes a relation between the gravitational potential and the density perturbation. A combination of the PS method along with the compaction function (PSC) has also been studied in the past. Similar to PS$\delta$, this method lacks the information on the profile of the density fields. These limitations can be circumvented in the Peaks Theory (PT) approach, where a customary profile of the density field is dictated by the power spectrum and higher-order statistical moments. In the PT approach, the number density of overdense regions is given via the ensemble average on the comoving volume of the maxima of a Gaussian field. Number density of PBHs can be calculated from PT by incorporating either the comoving curvature perturbations~\cite{Yoo:2018kvb, Atal:2019cdz, Yoo:2019pma, Atal:2019erb, Germani:2019zez, Young:2022phe} or its Laplacian~\cite{Yoo:2020dkz, Kitajima:2021fpq}. Another advantage of adopting the compaction function approach is that it naturally encodes the information of the non-Gaussianity arising from the nonlinear relation between comoving curvature perturbation ($\mathcal{R}$) and density contrast ($\delta$). 

A widely accepted assumption is the spherical symmetry of the collapsing region. Once the perturbation enters the horizon, it can generate some torque, which can result in the asymmetry of the collapsing region. Thus, leading to a non-zero angular momentum of the region once the region decouples for the background it preserves the angular momentum which leads to the spin of the formed PBH. Several papers have explored the spin of PBHs in the past~\cite{DeLuca:2019buf, Harada:2020pzb, Harada:2024jxl, Banerjee:2024nkv, Chongchitnan:2021ehn, Eroshenko:2021sez, Mirbabayi:2019uph, Chiba:2017rvs,Banerjee:2023qya}.  The initial spin of the PBHs depends on the background equation of state $w$, PBHs forming in matter-dominated or a soft equation of state epoch are expected to have a larger spin~\cite{Ye:2025wif,deJong:2023gsx, Saito:2023fpt, Saito:2024hlj}. Assuming that the no-hair conjecture is true in astrophysics, an isolated (P)BH can be considered as a Kerr BH, which is characterized by the mass ($M_{\rm PBH}$) and spin angular momentum $S$. Alternatively, a non-dimensional spin angular momentum can be defined: $\textbf{a}_{*} = S c/G M^2$ to analyze the spin. A non dimensionless Kerr parameter can be defined as $a_* = \sqrt{\textbf{a}_{*}~^.~\textbf{a}_{*}}$. 

If we observe a PBH that has a mass higher than the Chandrasekhar limit, it would be extremely difficult to distinguish it from a BH. However, spin can potentially differentiate between the two populations of (P)BHs \footnote{It is noteworthy to mention that accretion and merger history can severely alter the spin of the (P)BH, so while comparing the spin with the observed spin of the BH in GW, one should be cautious.}.
The PT presents a robust and accurate formalism for calculating the PBH abundance. According to the PT, PBHs can only originate from the rare peaks. Furthermore, these rare peaks have very limited freedom to be non-symmetric in nature.  In \cite {DeLuca:2019buf}, the perturbative analysis based on peak theory has shown that PBHs forming in the radiation-dominated era can have $a_{*} \sim 10^{-2}$. Whereas in~\cite{Harada:2020pzb}, the author obtained a smaller value $a_{*} \sim 10^{-3}$.    

The GW survey measures the chirp mass and effective spin $\chi_{\rm eff}$ of the BH binary. Thus, it becomes crucial to consider the spin of the PBHs to confront the GW observations. With a few exceptions~\cite{2020PhRvD.102d3015A, LIGOScientific:2025rsn}, most of the BH binary merger observations are consistent with the $\chi_{\rm eff} \sim 0$~\cite{2019ApJ...882L..24A}. 
It is a well-established fact that the formation of PBH requires an enhancement in the primordial power spectrum. In a single-field inflationary model, enhancing the power spectrum requires some exotic features to be present in the inflation potential, such as a phase of ultra-slow roll, a tiny bump/dip, or a step. In this article, we adopt a similar modification to the inflationary potential. A careful selection of the parameters of the underlying model would allow us to determine the mass and abundance of the PBHs. Today, PBHs serve as a probe to small-scale physics, which would otherwise be impossible to probe through the CMB observations. Another compelling phenomenon associated with the PBH formation is the scalar-induced gravitational waves (SIGWs). In perturbation theory, at tree level, scalar and tensor perturbations are decoupled. However, in the second order, scalar and tensor perturbations are coupled, and an enhancement in the scalar power spectrum causes an increment in the tensor perturbations, which results in the SIGWs \cite{Caprini:2018mtu, Christensen:2018iqi, Baumann:2007zm, Domenech:2024rks, Kohri:2018awv, Domenech:2021ztg, Domenech:2019quo, Domenech:2020kqm, Domenech:2021wkk} As SIGWs have originated from the inflationary sector and have a primordial nature, they appear to be stochastic. These SIGWs can be observed in both present and future detectors. Combining PBHs with SIGWs can further constrain the physics of the early universe. \\

The article is organized as follows: In section \ref{model}, we briefly discuss the background of inflation and the model under consideration. The peak profiles of curvature and density perturbation are discussed in the section~\ref{peak_profile}. In section~\ref{angular_momentum}, we discuss the angular momentum and obtain the spin for our model, followed by section~\ref{sec_abun} and section~\ref{sec_sigw}, where we discuss the PBH abundance and scalar-induced gravitational waves, respectively. Finally, with section~\ref{conclusion}, we concluded the manuscript. 

\section{The Model \label{model}} 

We begin by briefly introducing the model. The action of the model is given as 
\begin{equation}\label{action}
	S = \int d^4x \; \sqrt{-g} \; \left( \frac{M_p^2}{2} \; R + \nabla_\mu \phi \nabla^\mu \phi - V(\phi) \right), 
\end{equation}
where $M_p^2 = 1/8\pi G$ is the reduced Planck mass, $G$ is the Newton gravitational constant, $g$ is the determinant of the metric $g_{\mu\nu}$, $R$ is the Ricci scalar, $\phi$ is the scalar field, and $V(\phi)$ is the potential of the scalar field. Assuming a spatially flat FLRW metric, the main dynamical equations of the model are given as
\begin{equation}\label{friedmann}
	\begin{aligned}
		H^2 & = \frac{\rho_\phi}{3 M_p^2} = \frac{1}{3 M_p^2} \; \left( \frac{1}{2} \; \dot\phi^2 + V(\phi) \right),  \\
		\dot{H} & = \frac{-1}{2 M_p^2} \; \dot\phi^2, \\
		\ddot{\phi} & + 3 H \dot\phi + V'(\phi) = 0,
	\end{aligned}
\end{equation}
in which dot indicates derivative with respect to the cosmic time, and prime displays derivative with respect to the scalar field. During the inflationary time, the scalar field slowly rolls down toward the minimum of potential and provides a quasi-de Sitter expansion. The slow-roll movement is described by the slow-roll parameters, usually defined through a hierarchy, given as
\begin{equation}\label{srp}
	\epsilon_1 = \frac{-\dot{H}}{H^2}, \quad \epsilon_{n+1} = \frac{\dot{\epsilon}_n}{H \; \epsilon_n}.
\end{equation}
The parameters measure the deviation from the exact de Sitter expansion. Smallness of these parameters during the inflationary time provides a quasi-de Sitter expansion. The first slow-roll parameter, $\epsilon_1$, is required to be smaller than one to maintain a positive accelerated expansion. Inflation ends as the first slow-roll parameter reaches unity, at which point $\ddot{a} = 0$ and the accelerated expansion ceases. The amount of expansion is measured by the $N$ parameters, known as the number of e-folds, which is read as
\begin{equation}\label{efold}
	N = \ln\left( \frac{a_e}{a_i} \right) = \int_{t_i}^{t_e} H \; dt,
\end{equation}
where the subscripts ``i'' and ``e'' stand for the beginning and end of inflation, respectively. 

Besides resolving the problem of hot big bang model, inflation predicts the generation of quantum fluctuations, which are classified into three types: scalar, vector, and tensor perturbations. Up to the linear order, these perturbations evolve independently. By expanding the universe, these perturbation modes cross the horizon so that the smaller modes cross the horizon earlier and the larger modes cross at later times. The evolutions of the modes are described by the Mukhanov-Sasaki equation, in which, for the scalar perturbations, we have
\begin{equation}\label{MSequation}
	v''_k + \left( k^2 + \frac{z''}{z} \right) \; v_k = 0,
\end{equation}
where $v = z \; \zeta$, $z = a\dot\phi/H$, and $\zeta$ is the comoving curvature perturbations. As the modes cross the horizon, they freeze deep outside the horizon. The slow-roll inflation predicts an almost scale-invariant power spectrum given by 
\begin{equation}\label{Ps_sr}
	\mathcal{P}_s = A_\star \; \left( \frac{k}{k_\star} \right)^{n_s - 1},
\end{equation}
where $A_\star$ is the value of the power spectrum as the pivot scale $k_\star = 0.05 \; {\rm Mpc^{-1}}$ crosses the horizon. $n_s$ is the scalar spectral index expressed in terms of the slow-roll parameters as 
\begin{equation}
	n_s - 1 = \frac{d \ln(\mathcal{P}_s)}{d\ln(k)} = -2\epsilon_1 - \epsilon_2.
\end{equation}
However, a more accurate estimation of the power spectrum is given by solving the Mukhanov-Sasaki equation \eqref{MSequation} for the mode function $v_k$. The scalar power spectrum is then given by
\begin{equation}\label{ps_ms}
	\mathcal{P}_\zeta(k) = \frac{k^3}{2\pi^2} |\zeta_k|^2 = \frac{k^3}{2\pi^2} \; \frac{|v_k|^2}{z^2}\Big|_{k \ll aH}.
\end{equation}

To obtain an accurate power spectrum, we will numerically solve the Mukhanov-Sasaki equation. In this regard, one also needs to solve the background equations \eqref{friedmann}. We assumed that the pivot mode $k_\star = 0.05 \; {\rm Mpc^{-1}}$ crosses the horizon at $N = 0$, and inflation last for around $N = 60$ e-folds of expansion. 

\begin{figure}
	\centering
	\subfigure[\label{background_plot_hubble}]{\includegraphics[width=0.3\linewidth]{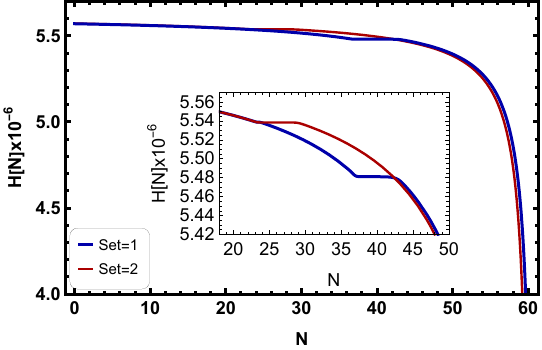}}
	\subfigure[\label{background_plot_phi}]{\includegraphics[width=0.3\linewidth]{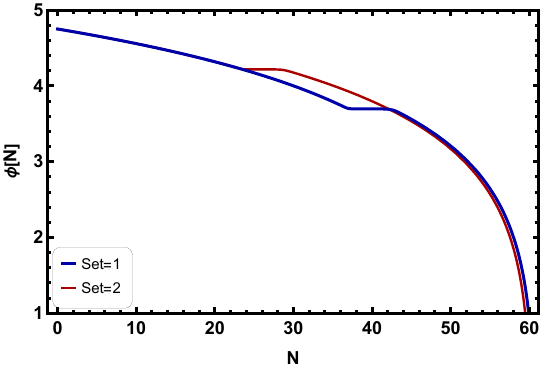}}
	\subfigure[\label{background_plot_epsilon}]{\includegraphics[width=0.3\linewidth]{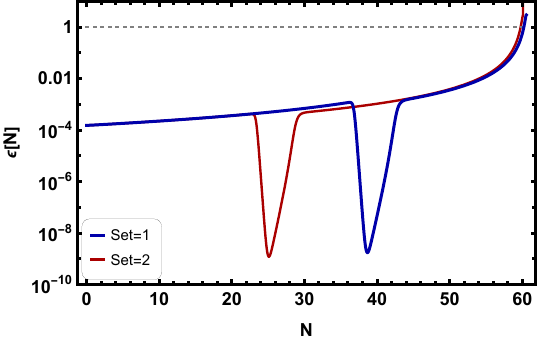}}
	\caption{The plot shows the behavior of the a)Hubble parameter, b) scalar field, and c) first slow-roll parameter versus the number of e-folds for two sets of the parameters: i) $(c, \phi_s, \delta) = (7.0043 \times 10^{-4}, 3.7, 1.26 \times 10^{-2})$, and ii)$(c, \phi_s, \delta) = (2.3658 \times 10^{-4}, 4.22, 7.2548 \times 10^{-3})$. The other parameter are taken as $\alpha = 1$, $V_0 = 4.75$, and $\phi_\star = 4.75$ where $\phi_\star$ is the field at the crossing time. The parameters are chosen to have around $ N=60$ e-folds of expansion for the inflationary phase. }
	\label{background_plot}
\end{figure}

To solve the background equation, one must first determine the potential of the scalar field. The potential is assumed to be given by~\cite{Pal:2009sd,Yogesh:2025hll,Hazra:2010ve, Thomas:2024ezg}
\begin{equation}\label{pot_step}
	V(\phi) = V_0 \; \big( 1 - \text{Sech}(\alpha \; \phi) \big) \; \left( 1 + c \; \tanh\Big[ \frac{\phi - \phi_s}{\delta} \Big] \right).
\end{equation}
The second term on the right-hand side of the equation generates a step on the original potential. 
Due to the step terms, there is an almost flat part for a small range of $\phi$, whose width is controlled by the $\delta$, height by $c$, and $\phi_s$ decides the position of the step. Substituting the potential into the background equations \eqref{friedmann} and rewriting the equations in terms of the number of e-folds instead of time, we can numerically obtain the background quantities by setting the initial conditions at the crossing time. The results are presented in Fig.~\ref{background_plot}, where the behavior of the Hubble parameter, the scalar field, and the first slow-roll parameter is displayed versus the number of e-folds. Due to the flat part of the potential, we expect the Hubble parameter to be almost constant for that specific part, as shown in Fig.~\ref{background_plot_hubble}. Then, the slow-roll parameter $\epsilon_1$ drops down to order $\mathcal{O}(10^{-9})$ because $\dot{H}$ becomes very small. The power spectrum for the case of slow-roll inflation is given as $\mathcal{P}_s = H^2 / 8 \pi M_p^2 \epsilon_1$. Therefore, if we have a sudden drop for the slow-roll parameter $\epsilon_1$, the magnitude of the power spectrum enhances. 

\begin{figure}
	\centering
	\includegraphics[width=0.5\linewidth]{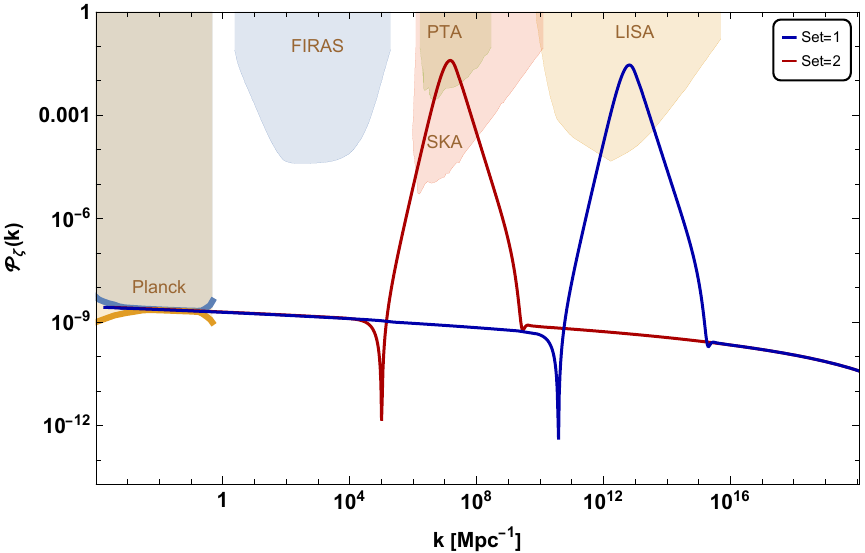}
	\caption{The plot displays the power spectrum as a function of wavenumber $k$ for 2 different sets of parameters: i) $(c, \phi_s, \delta) = (7.0043 \times 10^{-4}, 3.7, 1.26 \times 10^{-2})$, and ii)$(c, \phi_s, \delta) = (2.3658 \times 10^{-4}, 4.22, 7.2548 \times 10^{-3})$. The other parameter are taken as $\alpha = 1$, $\phi_\star = 4.75$,  and $V_0 = 9.48 \times 10^{-11}$. The parameters are chosen to yield approximately $ N=60$ e-folds of expansion for the inflationary phase and to reproduce the power spectrum $\mathcal{P}_s = 2.1 \times 10^{-9}$ at the crossing time of the pivot mode. The different constraints on the power spectrum are also plotted~Planck~\cite{Planck:2018jri}, FIRAS~\cite{Fixsen:1996nj}, SKA~\cite{Chluba:2019nxa,Inomata:2018epa}, PTA~\cite{Byrnes:2018txb} and LISA~\cite{Inomata:2018epa}. Plot legends are self-explanatory. We have obtained $r=0.0024$ and $n_s=0.975$, which are consistent with the recent ACT observations~\cite{ACT:2025fju,ACT:2025tim}.}
	\label{ps}
\end{figure}

Substituting the obtained background solution in the Mukhanov-Sasaki equation, we solve it for a wide range of modes by setting the initial condition at the crossing time. The resulting power spectrum is plotted in Fig.~\ref{ps}, where an enhancement in the power spectrum is visible. Considering the behavior of the power spectrum, it is evident that the model has the potential to predict the formation of PBHs. Additionally, from the shape of the power spectrum, one can realize that we are dealing with a broad power spectrum. 

\section{Profile in peak theory \label{peak_profile}} 

On the comoving slice, the spatial part of the perturbed metric could be given as
\begin{equation}
	ds_3^2 = a^2(t) \; e^{2 \zeta(r)} \; \left( dr^2 + r^2 \; d\Omega^2 \right), 
\end{equation}
in which $\zeta(r)$ is the comoving curvature, $r$ stands for the radial coordinate, and $d\Omega^2$ is an element of a unit two-dimensional sphere given by $d\Omega^2 = d\theta^2 + \sin^2(\theta) \; d\phi^2$. We confine our consideration to the case of a spherically symmetric peak, then, the comoving curvature $\zeta(r)$ is considered to depend only on $r$. This approximation is believed to be kept for high peaks. Randomly distributed curvature perturbations are produced during the inflationary phase of the universe. Considering only the Gaussian $\zeta$, the probability distribution function is read as
\begin{equation}
	\mathbb{P}_G(\zeta) = \frac{1}{\sqrt{2\pi} \; \sigma_\zeta} \; \exp\left( \frac{-\zeta^2}{2 \sigma_\zeta^2} \right),
\end{equation}
in which the variance of the comoving curvature $\zeta$ is determined by $\sigma_\zeta^2$ with a power spectrum as
\begin{equation}
	\sigma_{\zeta}^2(k_s)=\int \frac{\mathrm{d} k}{k} \mathcal{P}_{\zeta}(k) \widetilde{W}^2\left(k , k_s\right),\qquad
	\mathcal{P}_{\zeta}(k)=\frac{k^3}{2\pi^2}\left|\tilde{\zeta}\left(\boldsymbol{k}\right)\right|^2.
\end{equation}
where $\widetilde{W}\left(k , k_s\right)= {\rm exp}(-k/2k_s)$ is the window function to smoothen the power spectrum ~\cite{Pi:2024ert}. As the quantity that characterizes the over-dense region in a chosen area exceeds the critical threshold, PBHs form.

\begin{figure}[h]
	\centering
	\includegraphics[width=0.5\linewidth]{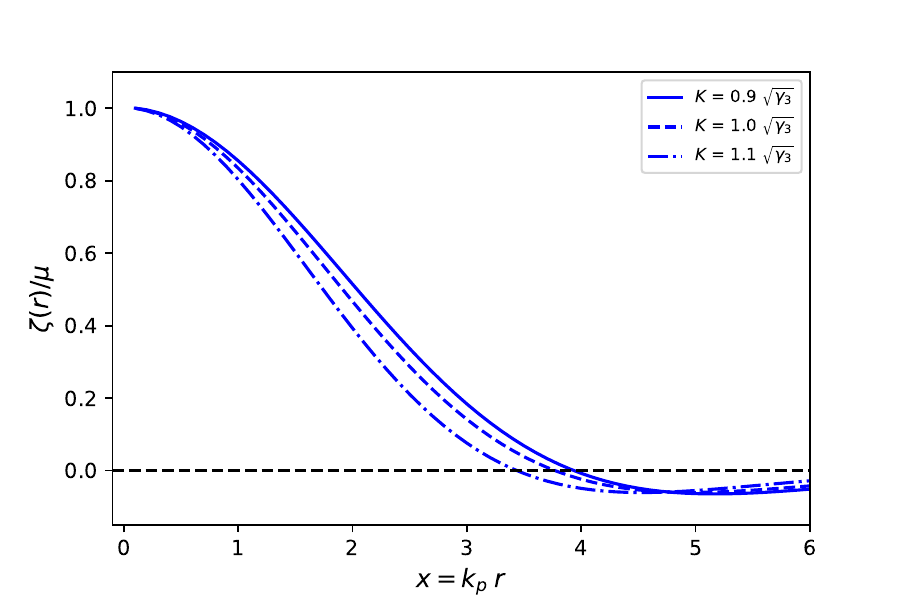}
	\caption{The plot displays the normalized profile versus $x = k_p r$ coordinate for different values of $K$ parameter. Here, $k_p$ is the peak mode where the power spectrum reaches the maximum value.}
	\label{profile}
\end{figure}

As first developed and shown in~\cite{Bardeen:1985tr, Peacock:1990zz}, peak theory can produce the shape of the local maxima and number density of a Gaussian random field~\cite{Yoo:2018kvb,Germani:2019zez,Young:2020xmk}. Different Gaussian random fields could be studied for consideration. For more detailed consideration, different Gaussian random fields could be chosen. Following ~\cite{Yoo:2020dkz,Kitajima:2021fpq,Pi:2024ert}, $\nabla^2 \zeta$ is taken as the Gaussian random field, as first discussed in~\cite{Yoo:2020dkz}. The main reason for this choice is that the formation of PBHs is a local process, and by this choice, a redefinition of the scale factor can absorb the long-wavelength component of $\zeta$. In addition, there is a physical motivation so that at linear order $\nabla^2\zeta$ is proportional to $ -\delta \rho / \rho$. The $-\nabla^2\zeta$ peak profile around the over-dense region can be expressed as~\cite{Pi:2024ert}
\begin{equation}\label{d2zeta_profile}
	-\widehat{\nabla^2 \zeta}(r)=\frac{\mu_2}{1-\gamma_3^2}\left[\psi_2(r)+\frac{R_3^2}{3} \nabla^2 \psi_2(r)-\frac{K_3^2}{\gamma_3} \frac{\sigma_2}{\sigma_4}\left(\gamma_3^2 \psi_2(r)+\frac{R_3^2}{3} \nabla^2  \psi_2(r)\right)\right],
\end{equation}
where the parameters $\mu$ is the height of the $-\nabla^2\zeta$ peak profile, and $K_3$ stands for its width. These two parameters are positive definite and given by (the reader could refer to ~\cite{Pi:2024ert} for more details)
\begin{equation}
	\mu_2\equiv\left.-\nabla^2 \zeta\right|_{r=0}, \quad K_3^2\equiv -\dfrac{1}{\mu_2} \left.\nabla^2\left(-\nabla^2 \zeta \right)\right|_{r=0} ~.
\end{equation}
There are some other statistical quantities that are used in peak theory, as the multiple moments\footnote{In case of non-monochromatic power spectrum, PBHs usually form at different scales, leading to a range of produced masses, which makes it necessary to employ a window function~\cite{Ando:2018qdb,Young:2019osy,Yoo:2020dkz} }
\begin{equation}\label{spectral}
	\sigma_n^2 = \int \frac{\mathrm{d} k}{k} k^{2 n} \mathcal{P}_\zeta(k) \; \widetilde{W}^2\left(k , k_s\right),
\end{equation}
two-point correlation
\begin{equation}\label{twopoint}
	\psi_n(r)=\frac{1}{\sigma_n^2} \int \frac{\mathrm{d} k}{k} k^{2 n} \frac{\sin (k r)}{k r} \mathcal{P}_\zeta(k) \; \widetilde{W}^2\left(k , k_s\right),
\end{equation}
and also, we have
\begin{equation}
	\gamma_n = \frac{\sigma_n^2}{\sigma_{n-1} \sigma_{n+1}}, \qquad
	R_n = \sqrt{3} \frac{\sigma_n}{\sigma_{n+1}},
\end{equation}
that are only valid for odd values of $n$ ~\cite{Pi:2024ert}. Using the iteration relation of $\psi_n$, the peak profile of $\zeta$ can be obtained from Eq.\eqref{d2zeta_profile} as
\begin{equation}\label{zeta_profile}
	\hat{\zeta}(r)=\frac{\mu}{1-\gamma_3^2}\left[\psi_1(r)+\frac{R_3^2}{3} \nabla^2 \psi_1(r)-\frac{K^2}{\gamma_3}\left(\gamma_3^2 \psi_1(r)+\frac{R_3^2}{3} \nabla^2 \psi_1(r)\right)\right]+\zeta_{\infty},
\end{equation}
where the rescaled parameter $\mu\equiv\mu_2 \frac{\sigma_1^2}{\sigma_2^2}$ and $K^2 \equiv K_3^2 \frac{\sigma_2}{\sigma_4}$ determine the height and width of $\zeta$-profile, respectively. Note that these two parameters are dimensionless. The last term on the right-hand side, $\zeta_{\infty}$, is an integration constant and carries the effect of long-wavelength perturbations. As one can safely absorb it into the redefinition of the scale factor, the term will be set to $\zeta_{\infty} = 0$ in subsequent analysis. Fig.~\ref{profile} displays the normalized profile versus coordinate $r$ for different values of the $K$ parameter. It can be seen that by increasing the $K$ parameter, the profile dope more slowly. 


\subsection{Compaction function and threshold}

\begin{figure}[!h]
	\centering
	\subfigure[]{\includegraphics[width=0.45\linewidth]{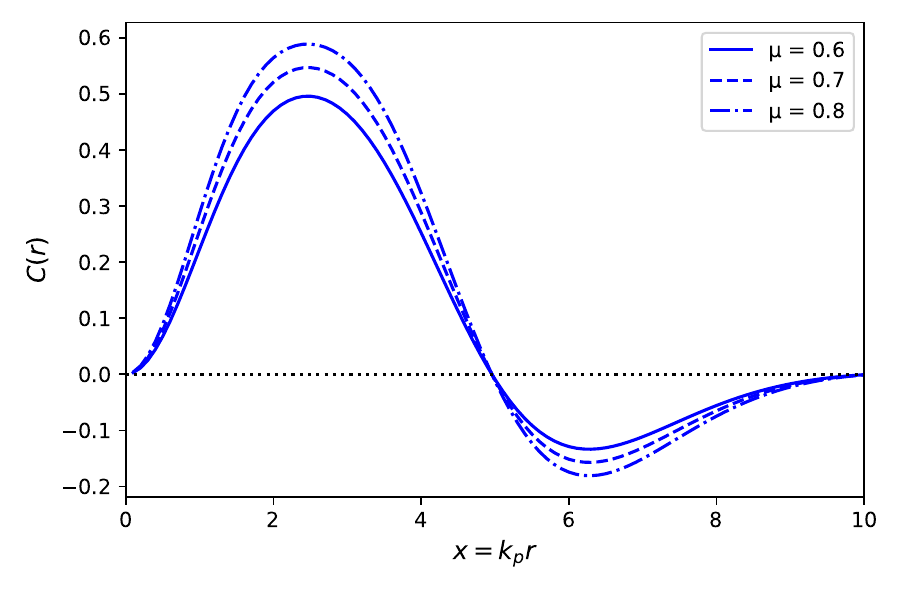}}
	\subfigure[\label{compaction_K}]{\includegraphics[width=0.45\linewidth]{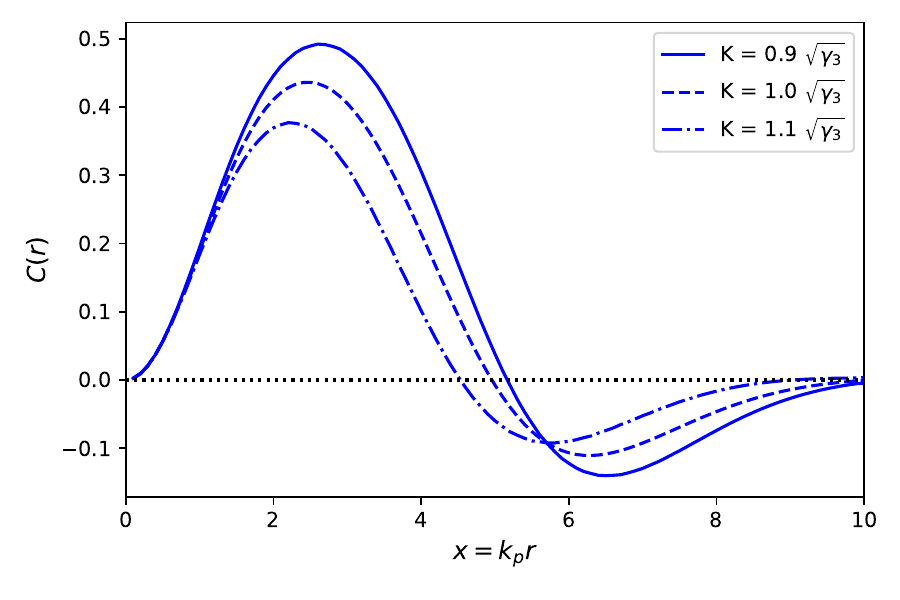}}
	\caption{Behavior of the compaction function versus $x = k_p r$ is illustrated: a) for different values of $\mu$ and $K = \sqrt{\gamma_3}$, and b) for different values of $K$ and $\mu = 0.5$ is shown. By increasing $\mu$, the amplitude of the compaction function increases; however, the location of the peak never changes. On the other hand, by increasing the $K$ parameter, the amplitude of the compaction function decreases and the location of the peak tends to lower values of $r$.}
	\label{compaction}
\end{figure}

The formation of PBHs is a nonlinear process. One of the main steps in the study of PBHs is to determine whether a peak collapses into a black hole. The compaction function quantity is a useful tool to estimate the status of collapse. The compaction function can be expressed in terms of the curvature perturbations, in which in the radiation-dominant phase it is given as
\begin{equation}\label{compaction}
	\mathcal{C}(r) = \frac{2}{3} \; \Big( 1 - \big(1 + r \; \zeta'(r) \big)^2 \Big).
\end{equation}
In general, the formation depends on the initial condition and the critical value of the compaction function, $\mathcal{C}_{th}$. Numerical studies show that, depending on the curvature profile, the critical value of the compaction function stands in the range $0.41$ to $0.67$. The behavior of the compaction function versus the $r$ coordinate for different values of $\mu$ and $K$ parameters is illustrated in Fig.\ref{compaction}. It is realized that the magnitude of the compaction function increases with the enhancement of the $\mu$ parameter; however, $r_m$ does not depend on $\mu$; $r_m$ is the value of $r$ where the compaction function reaches its maximum. On the other hand, Fig.\ref{compaction_K} shows compaction function $\mathcal{C}(r)$ for different values of $K$, where one can find that both the magnitude of the compaction function and $r_m$ depend on $K$. By increasing the $K$ parameter, the magnitude of the compaction function decreases, and its maximum occurs at lower values of $r$.

\begin{figure}[!h]
	\centering
	\includegraphics[width=0.5\linewidth]{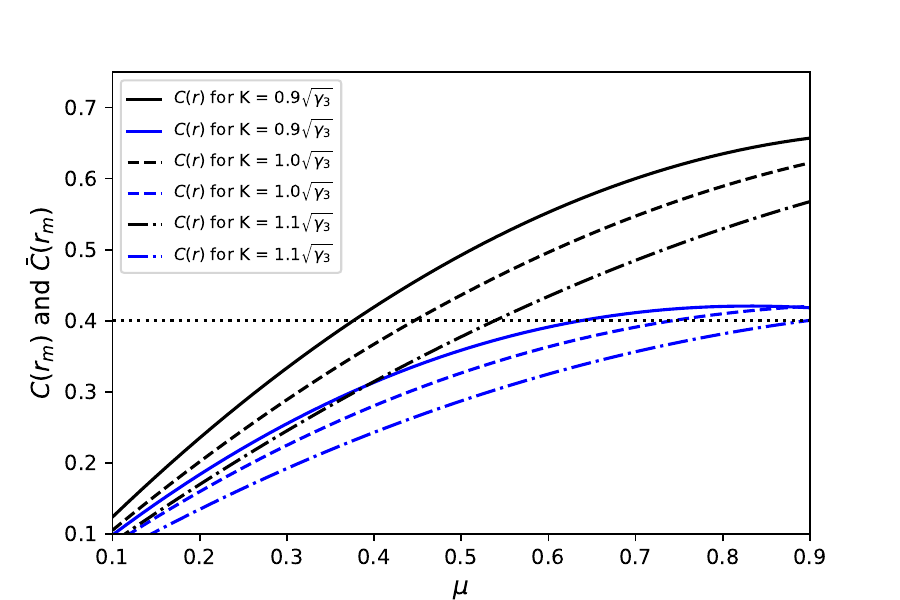}
	\caption{The plot shows the compaction function, black color lines, and the average of the compaction function, blue color lines, for the numerical power spectrum obtained in Sec.\ref{model}. The dotted horizontal line stands for the threshold value as $\bar{\mathcal{C}(r_m)} = 2/5$. }
	\label{Cr_Cr_avg}
\end{figure}

On the other hand, in \cite{Escriva:2019phb}, it was determined that the threshold on the average compaction function is universal and independent of the shape of the curvature perturbation. The average compaction function is defined as
\begin{equation}\label{C_average}
	\bar{\mathcal{C}}_m = \frac{4\pi \; \displaystyle\int_0^{R(r_m)} \mathcal{C}(r) \; R^2(r) \; dR}{\displaystyle\frac{4\pi}{3} \; R^3(r_m)},
\end{equation}
where $R(r) = a(t) \; r \; e^{\zeta(r)}$ is the areal radius which is related to the proper size of the over-density region, and $r_m$ is the point where the compaction function reaches the maximum value, which is obtained by solving the equation $\zeta'(r) + r \; \zeta''(r)  = 0$. As discussed in \cite{Escriva:2019phb}, as the average compaction function exceeds the threshold value, i.e, $\bar{\mathcal{C}}_m > \bar{\mathcal{C}}_{th} = 2/5$, the peak could collapse, and PBHs can form.  Fig.\ref{Cr_Cr_avg} displays the average compaction function versus the $\mu$ parameter for different values of $K$. It is realized that by increasing the $K$ parameter, the average compaction function reaches the threshold value at a larger value of $\mu$. On the other side, Fig.\ref{Cr_Cr_avg} also provides a comparison between the maximum of the compaction function and the average compaction function versus $\mu$, in which black color lines stand for the maximum of the compaction function, and the blue color curves indicate the average compaction function. The solid, dashed, and dot-dashed line styles stand for $K = 0.9$, $1.0$, and $1.1$, respectively. One can find that for a specific value of $K$, the maximum of the compaction function is always larger than the average compaction function. 

In addition to the above formalism for determining the threshold value, the $q$-function method is another approach. In this approach, the profile dependence of the threshold is considered more accurately, and it was found that the dependency can be determined by an analytical indicator, known as $q$-function, which contains the second derivative of the compaction function at its maximum \cite{Escriva:2019phb, Escriva:2022pnz}, given by
\begin{equation}\label{q_function}
	q = \frac{- r_m^2}{4} \; \frac{\mathcal{C}''(r_m)}{\mathcal{C}^2(r_m) - \frac{3}{2} \; \mathcal{C}^2(r_m)}.
\end{equation}
Then, the criterion $\bar{\mathcal{C}}_m = 2/5$ can be recast into  the threshold on the maximal compaction function $\delta_c$, read as
\begin{equation}\label{delta_c}
	\delta_c = \frac{4}{15} \; e^{-1/q} \; \frac{q^{1-\tilde{q}}}{\Gamma(\tilde{q}) - \Gamma(\tilde{q}, 1/q)},
\end{equation}
where $\tilde{q}$ is defined as $\tilde{q} = \frac{5}{2q}$, $\Gamma(x)$ is the Gamma function, and $\Gamma(x, z)$ is the upper incomplete Gamma function. 
\begin{figure}
	\centering
	\includegraphics[width=0.5\linewidth]{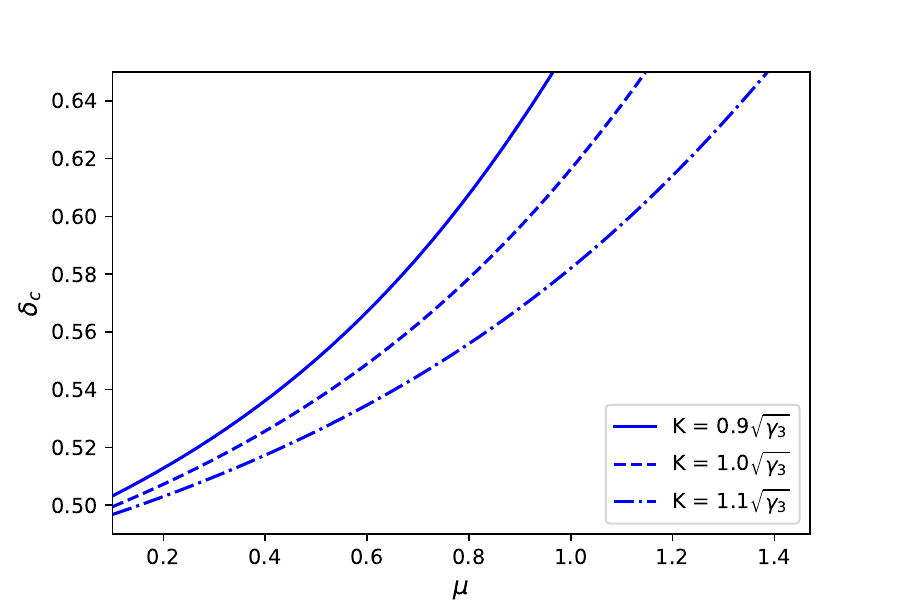}
	\caption{Dependency of $\delta_c$ on $\mu$ for different values of $K$ is illustrated. The plot shows that $\delta_c$ grows faster for smaller values of $K$.}
	\label{d_c}
\end{figure}
Fig.\ref{d_c} displays $\delta_c$ as a function of $\mu$ for different values of $K$, which shows that $\delta_c$ increases faster for smaller values of $K$. \\
The threshold then is defined as the moment when the criterion $\mathcal{C}(r_m) > \delta_c$ is satisfied. Note that the above result holds for different profiles of the spatial curvature $K$. Additionally, the numerical investigation revealed that the above threshold is consistent with the numerical results within a $2\%$ error~\cite{Escriva:2022pnz}.

It is worth noting that, depending on the behavior of the areal radius, there may be type I or type II PBHs. For type II PBHs, the condition $dR(r)/dr < 0$. This condition could be expressed in terms of the curvature perturbations, so that if the following condition
\begin{equation}\label{type_condition}
	\big( 1 + r \; \zeta'(r) \;\big) < 0
\end{equation}
satisfied, the curvature perturbations $\zeta(r)$ on super-horizon scale is called type II~\cite{Uehara:2025idq,Escriva:2025rja,Shimada:2024eec,Uehara:2024yyp}. 

\begin{figure}[!h]
	\centering
	\includegraphics[width=0.5\linewidth]{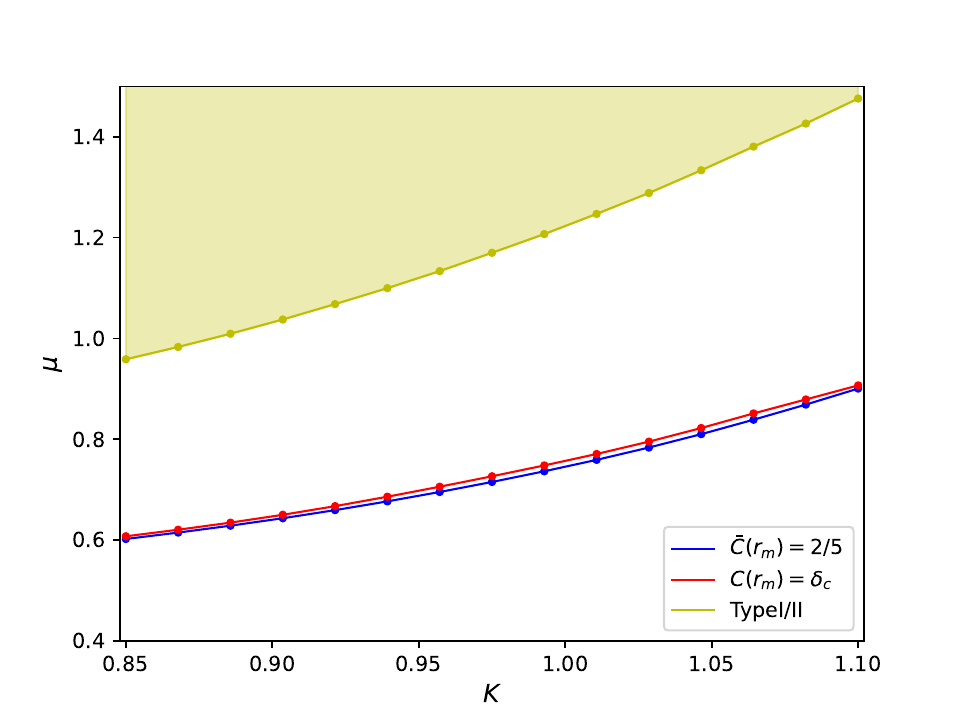}
	\caption{The figure displays the results of the model for the amplitude threshold for PBH formation across different values for $K$. The yellow line displays the boundary for type I and II fluctuation, so that the yellow-shaded area above this line stands for the set of $(\mu, K)$ points that lead to type II PBHs. The blue line represents the analytical estimation of the threshold using the average compaction function, and the red curves represent the analytical calculation of the threshold using the $q$-function method.}
	\label{threshold}
\end{figure}

The results of the above discussion can be found in Fig.\ref{threshold}. The yellow-shaded area shows the region where fluctuations lead to type II PBHs. The yellow line stands for the boundary of type I and II PBHs. The blue line displays the threshold value based on the average compaction function, as $\bar{\mathcal{C}}(r_m) = 2/5$. The red line represents the threshold values obtained using the $q$-function method. In both cases, the threshold value $\mu_{th}$ increases with $K$; however, the values are close to each other. The resulting threshold value based on the $q-$function approach is obtained as $\mu_{th} = 0.643, 0.753$, and $0.902$ for $K = 0.9\sqrt{\gamma_3}, \sqrt{\gamma_3}$, and $1.1\sqrt{\gamma_3}$, respectively. 

\subsection{Density perturbation}

\begin{figure}
	\centering
	\includegraphics[width=0.45\linewidth]{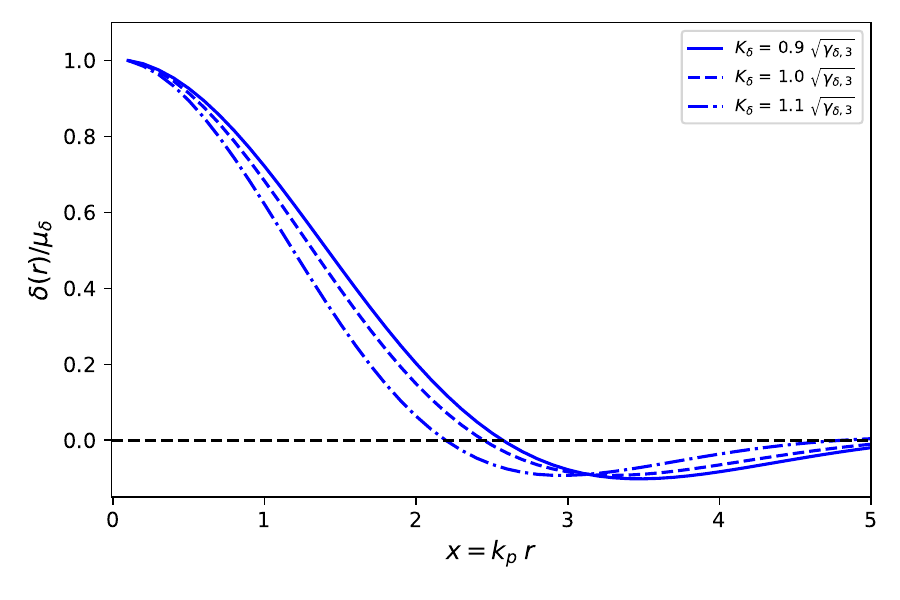}
	\caption{The plot displays the profile of the density perturbation versus $x = k_p r$ for different values of $K_\delta$ parameter.}
	\label{d_profile}
\end{figure}

Following a similar process, one can define the profile of the density profile. It has been shown in ~\cite{Harada:2015yda} that the perturbations, such as density perturbation $\delta(\eta)$, can be written in terms of the curvature perturbations $\zeta(\eta)$ in the long-wavelength regime. These perturbations are proportional to $\zeta(\eta)$ and its spatial derivative; therefore, one can also consider them as the Gaussian fields. \\ 
In particular, in the constant mean curvature (CMC) slicing, the density perturbation is proportional to the curvature perturbation by~\cite{Harada:2020pzb, Banerjee:2024nkv,Saito:2023fpt}
\begin{equation}\label{density_curvature}
	\delta_{cmc}(r, \eta_{init}) = \frac{2}{3  \; a^2 H_b^2} \; \nabla \zeta(r, 0),
\end{equation}
where $H_b$ is the background Hubble parameter. And
\begin{equation}
	\delta_{k, cmc}(\eta_{init}) = - \frac{2}{3a^2 H_b^2} k^2 \zeta_k(0), \qquad 
	\big| \delta_{k, cmc}\big|^2 = \frac{4}{9} \; \frac{2 \pi^2 k}{a^4 H_b^4} \; \mathcal{P}_\zeta(k),
\end{equation}
where $H_b$ is the background Hubble parameter. Then, the spectral moment of the density is defined as
\begin{equation}\label{sigma_d}
	\sigma_{\delta, n}= \int \frac{\mathrm{d} k}{k} k^{2 n} \mathcal{P}_\delta(k) = \frac{4}{9} \; \frac{1}{a^4 H_b^4}\int \frac{\mathrm{d} k}{k} k^{2 n + 4} \mathcal{P}_\zeta(k) \; \widetilde{W}^2\left(k , k_s\right).
\end{equation}
Similarly, one can define the two-point correlation for the density perturbations, as
\begin{equation}\label{twopoint_d}
	\psi_\delta(r) = \frac{4}{9} \; \frac{1}{a^4 \; H_b^4} \; \frac{1}{\sigma_{\delta, n}^2} \int \frac{\mathrm{d} k}{k} k^{2 n + 4} \frac{\sin (k r)}{k r} \mathcal{P}_\zeta(k) \; \widetilde{W}^2\left(k , k_s\right),
\end{equation}
and the profile of the density perturbations could be defined similarly to that in Eq.\eqref{zeta_profile}. Unlike the monochromatic power spectrum, where the Dirac delta function simplifies integration, we have a broad power spectrum here and must perform the integration to obtain a proper density perturbation profile. It is important in determining the overdense region. The overdense region is defined as $\Sigma_{O} = \big\{ \; x \; | \; r < r_0 \; \big\}$, where $r_0$ is the radius of the region determined as the point where the density profile vanishes. The profile of the density perturbations is plotted in Fig.\ref{d_profile} for different values of the $K_\delta$ parameter. The profile has the same behavior as the profile of the curvature perturbation; however, it is realized that the for the case of density perturbations, it vanishes for smaller values of $r$ so that the radius of the overdense region is given as $r_0 = 2.57, \; 2.44$ and $2.18$ for $K_\delta = 0.9\sqrt{\gamma_{\delta,3}}, \; \sqrt{\gamma_{\delta,3}}$, and $1.1 \sqrt{\gamma_{\delta,3}}$, respectively. 

\begin{figure}[h]
	\centering
	\includegraphics[width = 0.45\linewidth]{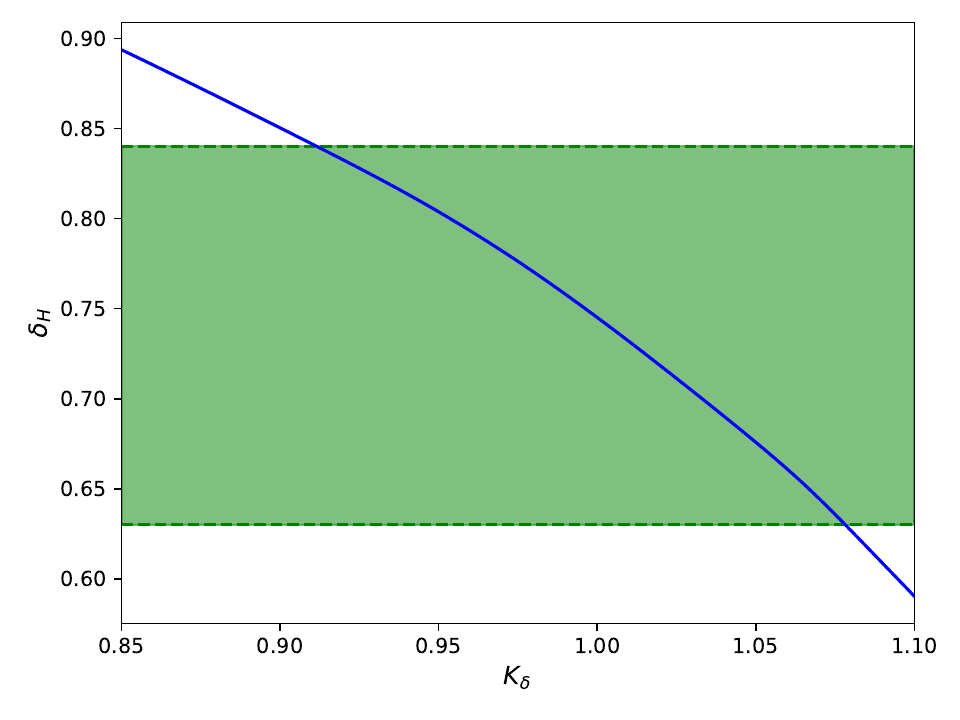}
	\caption{The plot shows the overage density perturbation, $\delta_H$, over the overdense region with the radius $r_0$ versus $K_\delta$. The colored shaded region clarifies the values between $0.63$ and $0.84$. To remain within this range of $\delta_H$, the parameter $K_\delta$ should stand in the range $0.9$ and $1.07$. }
	\label{delta_H_plot}
\end{figure}

By determining $\delta_{pk}$ and $r_0$, the average of the density perturbation can be calculated over the overdense region with radius $r_0$. The resulting average density perturbation, $\delta_H$, is plotted in Fig.\ref{delta_H_plot} versus $K_\delta$, where $K_\delta$ is the dimensionless parameter of the density perturbation profile. It is realized that, for the range of $K_\delta = [0.9, 1.07]$, the value of the average density perturbations stands in the range $\delta_H \in [0.63, 0.84]$~\cite{Harada:2020pzb, Banerjee:2024nkv} shown by the colored shaded region in the figure. 

\section{Angular momentum \label{angular_momentum}} 
In this section, we briefly review the definition of the angular momentum of the PBH and introduce the spin of the PBH. Our discussion is based on peak theory~\cite{Bardeen:1985tr, Peacock:1990zz}, where the curvature perturbation, density perturbation, and velocity could be considered as a Gaussian random field~\cite{Bardeen:1985tr, Peacock:1990zz}.

Solving the Einstein equations in the first order of perturbations leads to the following gauge invariant quantities~\cite{Harada:2020pzb}
\begin{align}
	\Delta(x) &= D\sqrt{3}\left(\dfrac{\sin z}{z} - \cos z\right),\nonumber\\
	V(x) &= D\left[\dfrac{3}{4}\left(\dfrac{2}{z^2}-1\right)\sin z-\dfrac{3}{2}\dfrac{\cos z}{z}\right],
\end{align}
in which $z = x/\sqrt{3} = k\eta / \sqrt{3}$ and $D$ is an arbitrary constant. The value of the constant $D$ in general depends on the shape of the perturbations. The density perturbation and velocity of the region in the CMC gauge are given as
\begin{align}
	\delta_{\mathrm{cmc}} &= D\dfrac{\sqrt{3}z^2}{z^2+2}\left(2\dfrac{\sin z}{z}-\cos z\right), \nonumber\\
	v_{\mathrm{cmc}} &= -\dfrac{3}{4}D\dfrac{(z^2-2)\sin z+2z\cos z}{z^2+2},
\end{align}
and in the Newtonian gauge are expressed as
\begin{align}
	\delta_{\mathrm{CN}} &= \sqrt{3}D\dfrac{2(z^2-1)\sin z+(2-z^2)z\cos z}{z^4},\nonumber\\
	v_{\mathrm{CN}} &= \dfrac{3}{4}D\dfrac{(2-z^2)\sin z - 2z\cos z}{z^2}.
\end{align}

Following ~\cite{DeLuca:2019buf, Harada:2020pzb}, if a spacetime has a Killing vector field $\phi^a_i$ that is tangent to a spacelike hypersurface and produces spatial rotation on it, the angular momentum can be defined as~\cite{Wald:1984rg}
\begin{equation}
	S_{i}(\Sigma)=-\int_{\Sigma} T^{ab} n_{a}(\phi_{i})_{b}d\Sigma,
\end{equation}
where $\Sigma$ is the region where PBH forms, $n^a$ is the unit vector normal to $\Sigma$, and $T_{ab}$ is the energy-momentum of the matter field, which is taken to be a perfect fluid. Determination of the $\Sigma$ region is non-trivial, however, based on ~\cite{Peacock:1990zz,DeLuca:2019buf} it is assumed that the region is defined as $\Sigma=\left\{{\bf x}|\delta({\bf x})>f\delta_{pk}\right\}$. 
After doing Taylor series expansion, truncation, and manipulation, the root mean square value of the angular momentum is obtained as
\begin{equation}
	\sqrt{\langle S_{i}S^{i}\rangle }=S_{{\rm ref}}\sqrt{\langle s_{e}^{i}s_{ei}\rangle}.
\end{equation}
Although $s_e$ depends on the shape and height of the peak, the reference angular momentum $S_{\rm ref}$ is a common feature of all the regions with the peak, given by 
\begin{equation}
	S_{\rm ref}(\eta) = \frac{4}{3} \; a^{4} \; \rho_{b} \; g(\eta) \; (1-f)^{5/2} \; R_{1}^{5},
\end{equation}
where $R_{\delta, 1} = \sqrt{3} \; \sigma_{\delta, 1} / \sigma_{\delta, 2}$, the term $g(\eta)$ is given as 
\begin{equation}
	g^{2}(\eta)=\frac{4}{9}\int\frac{dk}{k}k^{2}T_{v}^{2}(k,\eta)P_{\zeta}(k),
\end{equation}
and $T_v(k, \eta)$ is the transfer function for the velocity $v_k(\eta)$. It has been shown that the mean square root of the $s_e$ quantity can be obtained as~\cite{DeLuca:2019buf,Harada:2020pzb}
\begin{equation}
	\sqrt{\langle s_{e}^2\rangle} = 5.96 \; \frac{\sqrt{1 - \gamma^2_\delta}}{\gamma_\delta^6 \; \nu},
\end{equation}
where $\gamma_\delta = \sigma_{\delta,1}^2 / \sigma_{\delta,0} \sigma_{\delta,2}$ denotes the height of the peak, and $\nu = \delta_{pk} / \sigma_{\delta,0}$ characterize the width of the density perturbation. Using the above discussion, one can introduce the dimensionless Kerr parameter as 
\begin{equation}
	\sqrt{\langle a_\star^2 \rangle} = A_{\rm ref}(\eta_{ta}) \; \sqrt{\langle s_{e}^2\rangle},
\end{equation}
where $A_{\rm ref}$ is the dimensionless reference angular momentum defined as 
\begin{equation}
	A_{{\rm ref}}(\eta_{\rm ta})= \frac{S_{{\rm ref}}(\eta_{\rm ta})}{G \; M_{\rm ta}^{2}} = 
	\frac{4}{3 \; G \;M_{\rm ta}^{2}} \; \big( a^{4}\rho_{b} \; g_{\rm CN} \big)\big|_{\eta = \eta_{ta}} \; (1-f)^{5/2}R_{*}^{5},
\end{equation}
and $\eta_{ta}$, indicating the turnaround point. $M_{ta}$ is the mass of the region $\Sigma$ at the turnaround point given as $M_{ta} = \frac{4}{3} \pi r_f^3 \; \big(a^3 \rho_b \big)\big|_{\eta_{ta}}$. The turnaround moment is the moment when the region under consideration decouples from the background. It is the point where the perturbations are large enough to decouple from the background, gravitationally attracted, and acceleration becomes toward the center of the overdense region. There is some ambiguity in precisely specifying the turnaround point, and here, we follow ~\cite{Harada:2020pzb, Banerjee:2024nkv} and determine this moment as the time when the velocity of the region reaches its minimum value, i.e. $v'_{CN}(\eta = \eta_{ta}) = 0$. 

\begin{figure}
	\centering
	\subfigure[\label{vcn}]{\includegraphics[width=0.45\linewidth]{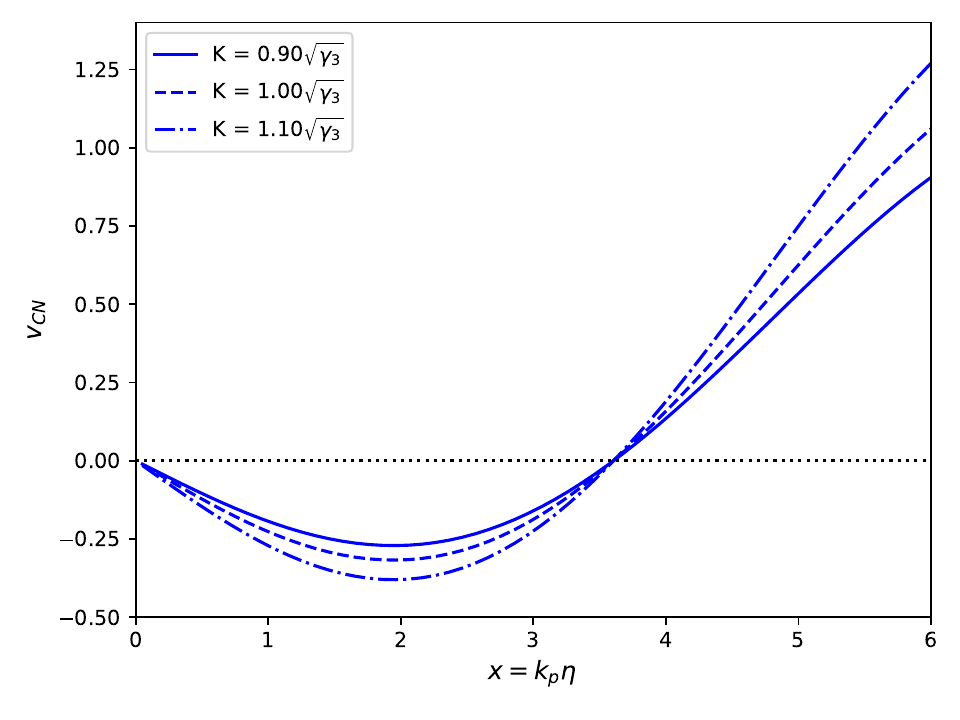}}
	\subfigure[\label{dvcn}]{\includegraphics[width=0.45\linewidth]{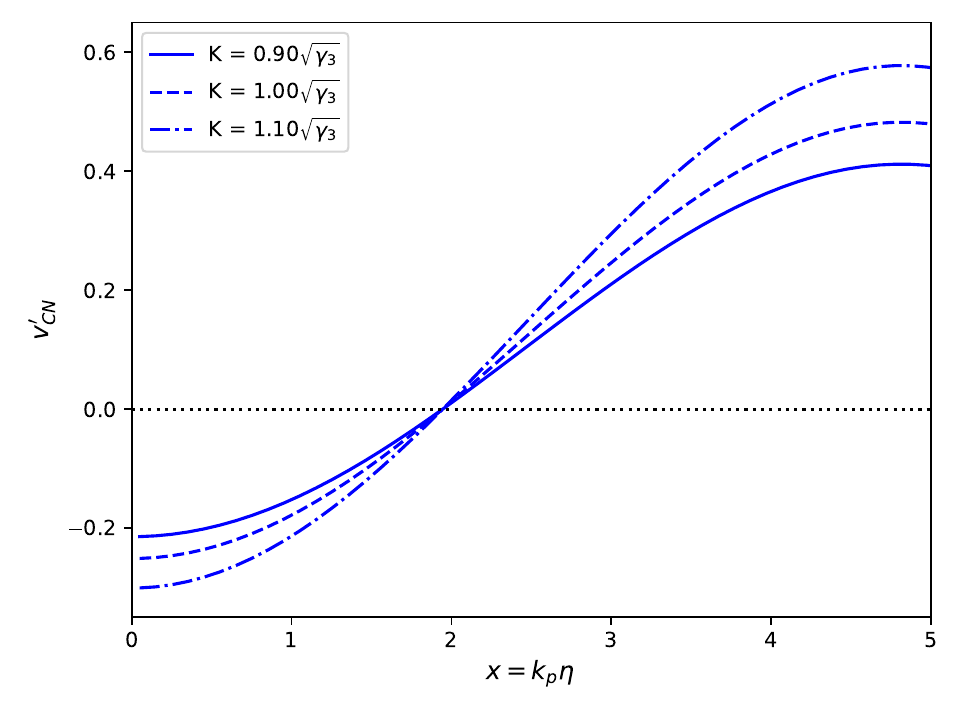}}
	\caption{the plot shows a) velocisty $v_{CN}$, and b) its derivative $v'_{CN}$ versus $x = k_p \eta$ for different values of $K$. Although by changing the $K$ parameter, the amplitude of the velocity is changing, the minimum of the velocity does not change, and it occurs at $x = 1.946$ for all values of $K$.  }
	\label{velocity_cn}
\end{figure}

\begin{figure}
	\centering
	\includegraphics[width=0.5\linewidth]{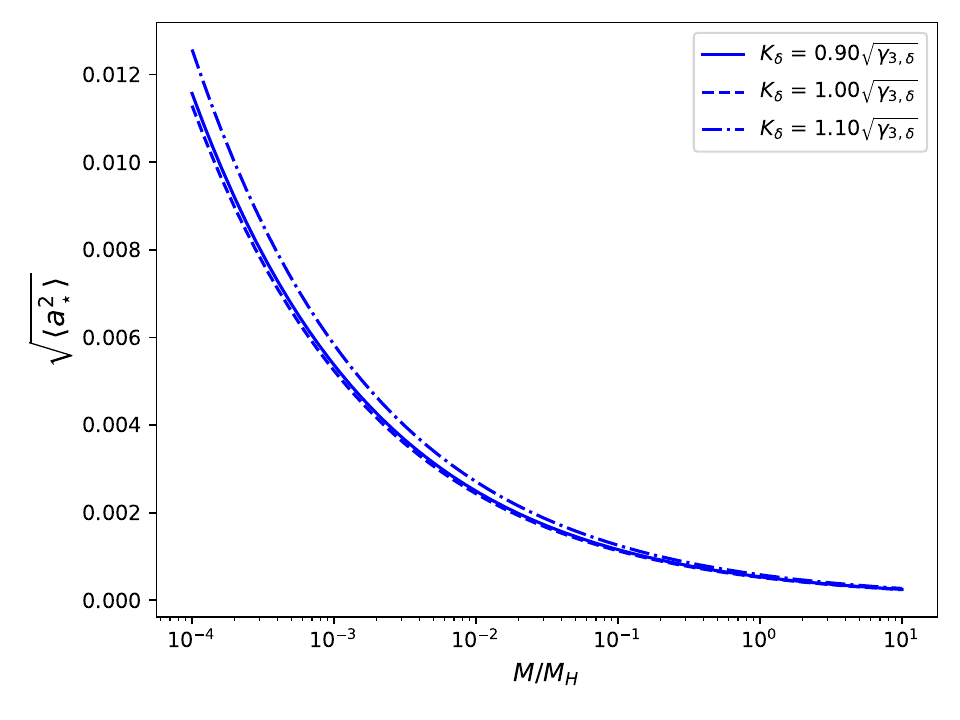}
	\caption{Dependency of the $\sqrt{\langle a_\star^2 \rangle}$ on mass $M/M_H$ is displayed for different values of $K_\delta$. The plot shows that the spin can change by varying $K_\delta$; however, there is no monotonic relation between them. The magnitude of the spin is of the order of $\mathcal{O}(10^{-3})$, and it increases by decreasing mass so that it can reach to the order of $\mathcal{O}(10^{-2})$ for $M \ll M_H$. }
	\label{a_star}
\end{figure}

The velocity and its derivative (in the conformal Newtonian gauge) are illustrated in Fig.\ref{velocity_cn} versus $x = k_p \eta$ for different values of the $K$ parameter. It is realized that by increasing $K$, the magnitude of the velocity increases; however, for all values of $K$, the minimum point is $x_{ta} = 1.946$, which is the turnaround point. This is clear from Fig.\ref{dvcn} where all the curves cross the zero line at the same point. Fig.\ref{a_star} displays $a_\star$ versus $M/M_H$ for different values of $K$. It is found that the order of $a_\star$ is $\mathcal{O}(10^{-3})$; however, as the mass decreases, the $a_\star$ parameter increases so that it can reach the order of $\mathcal{O}(10^{-2})$. Additionally, the magnitude of $a_\star$ depends on the value of the $K$ parameter; however, there is no monotonic relation between values of $a_\star$ and $K$. 

\section{PBH abundance \label{sec_abun}} 

A peak with the threshold height $\mu_{th}$ leads to the formation of PBHs with zero mass. As the peak height exceeds the threshold, a PBH forms in which its mass obeys a universal power-law formula as~\cite{Choptuik:1992jv,Niemeyer:1997mt, Inui:2024fgk}
\begin{equation}\label{mass_mu1}
	M_{PBH} = \mathcal{K} \; \big( \mu - \mu_{th} \big)^p \;M_H,
\end{equation}
where $\mathcal{K}$ is a dimensionless mass ratio of the order of unity, which we take to be $\mathcal{K} = 1$ for simplicity, and $p = 0.36$. $M_H$ stands for the Hubble mass at the Hubble re-entry of the maximal radius, $R(r_m) H = 1$. It can be related to the mass $M_{k_\star}$ so that the above equation can be rewritten as
\begin{equation}\label{mass_mu}
	M_{PBH} = \mathcal{K} \; \big( \mu - \mu_{th} \big)^p \; \big(k_\star \; r_m \big)^2 \; e^{2 \zeta(r_m)} \; M_{k_\star}.
\end{equation}
where $M_{k_\star}$ is the Hubble mass at the Hubble re-entry of the mode scale $k_\star$ (in which $k_\star$ is just a reference mode scale)
, given by
\begin{equation}\label{k_star}
	M_{k_\star} = 10^{20} \; \left( \frac{g_\star}{106.75} \right)^{-1/6} \; \left( \frac{k_\star}{1.56 \times 10^{13} {\rm Mpc^{-1}}} \right)^{-2} \; {\rm gr},
\end{equation}
where $g_\star$ is the effective number of relativistic degrees of freedom. 

\begin{figure}
	\centering
	\includegraphics[width=0.5\linewidth]{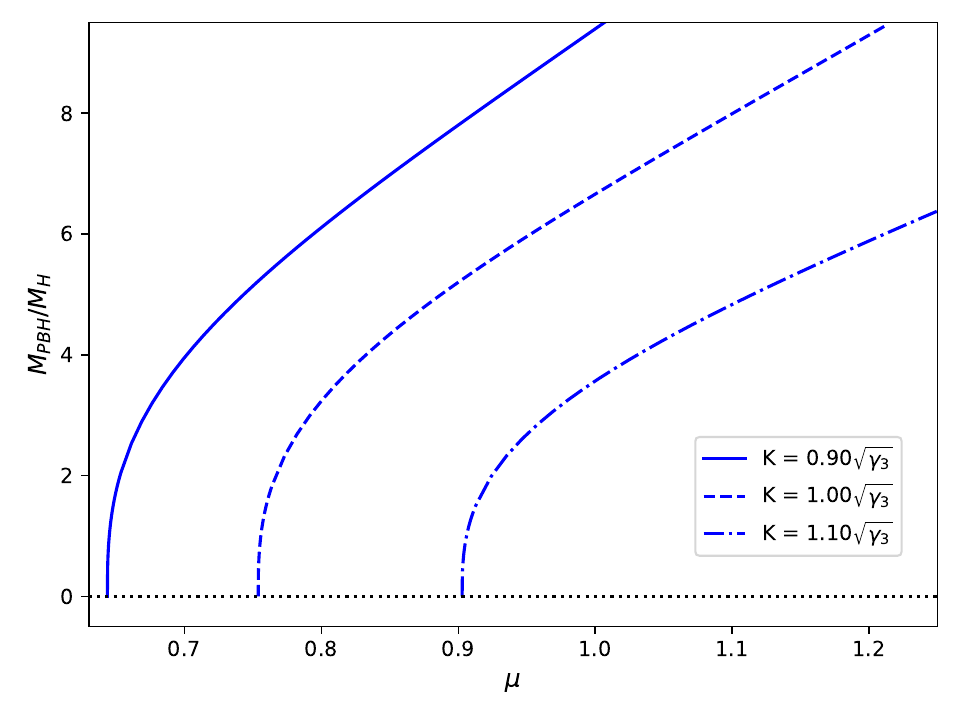}
	\caption{The figure shows the normalized PBH mass in terms of $\mu$ for different values of $K$.}
	\label{mass}
\end{figure}

Using the peak theory, one can obtain the number density of PBHs of mass $M$ in a comoving volume as \cite{Pi:2024ert,Inui:2024fgk}
\begin{equation}\label{N_pbh}
	\mathcal{N}_{PBH} = \int dK \int d\mu \; \delta_D\left( \frac{M}{M(\mu, K)} \right) \; \mathcal{N}_{\rm pk}(\mu, K),
\end{equation}
where $\mathcal{N}_{pk}$ is the peak number density
\begin{equation}\label{N_peak}
	\mathcal{N}_{pk} \left(\mu, K\right) \; d\mu \; dK = 2\left(\frac{1}{6 \pi}\right)^{3 / 2} \frac{\sigma_2^2}{\sigma_1^4} \frac{\sigma_4^3}{\sigma_3^3} \; \mu \; K \; f\left(\frac{\sigma_2}{\sigma_1^2} \mu K^2\right) \; P_1^{(3)}\left(\frac{\sigma_2}{\sigma_1^2} \mu, \frac{\sigma_2}{\sigma_1^2} \mu K^2\right) \; d\mu \; d K,
\end{equation}
and 
\begin{align}
	f(x) = & \;\frac{x^3-3 x}{2}\left[\operatorname{erf}\left(\sqrt{\frac{5}{2}}x\right)+\operatorname{erf}\left( \frac{1}{2}\sqrt{\frac{5}{2}}x\right)\right] + \sqrt{\frac{2}{5 \pi}}\left[\left(\frac{31 x^2}{4}+\frac{8}{5}\right) e^{-5 x^2 / 8}+\left(\frac{x^2}{2}-\frac{8}{5}\right) e^{-5 x^2 / 2}\right],\nonumber \\
	P_1^{(3)}(v, x) = & \; \frac{1}{2 \pi \sqrt{1-\gamma_3^2}} \exp \left[-\frac{1}{2}\left(v^2+\frac{\left(x-\gamma_3 v\right)^2}{1-\gamma_3^2}\right)\right].
\end{align}
The PBH mass function is defined as the PBH fraction of dark matter at the present time, i.e. $f_{PBH} = \Omega_{PBH,0}(M_{PBH}) / \Omega_{DM,0}$. Using Eq.\eqref{mass_mu} in \eqref{N_pbh}, the resulting mass function is obtained as
\begin{equation}\label{f_pbh}
	f_{PBH}(M) = \left( \frac{\Omega_{DM,0} \; h^2}{0.12} \right)^{-1} \; \left(\frac{M_{PBH}}{10^{20} \; {\rm gr} } \right) \; \left( \frac{k_\star}{1.56 \times 10^{13} {\rm Mpc^{-1}}} \right)^{3} \; \frac{\mathcal{N}_{PBH}}{1.74 \times 10^{-16} \; k_\star^3}
\end{equation}
and the total PBH abundance is achieved by integrating the above expression as 
\begin{equation}
	\label{f_pbh_tot}
	f_{PBH}^{tot} = \int f_{PBH}(M) \; d\ln M .
\end{equation}

\begin{figure}
	\centering
	\subfigure[\label{fpbh}]{\includegraphics[width=0.45\linewidth]{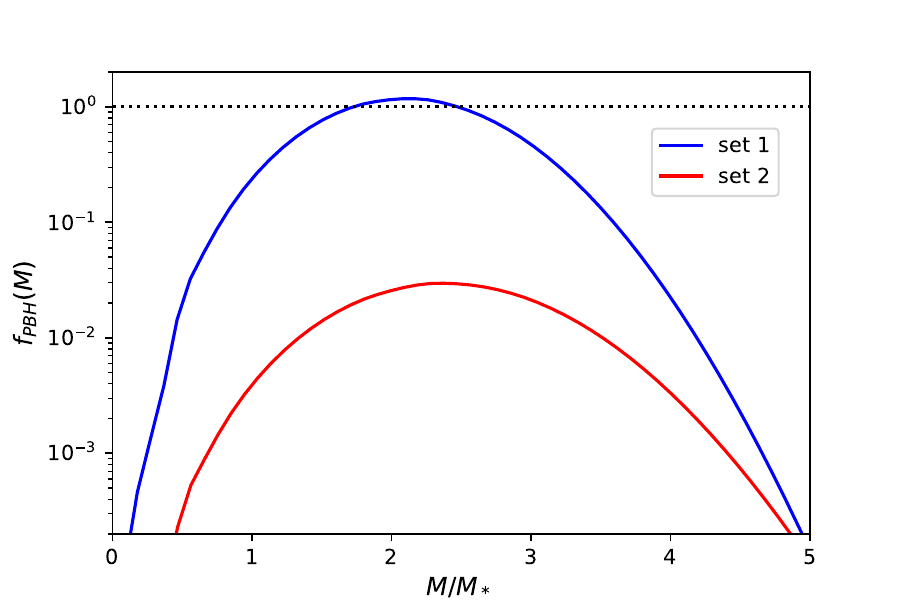}}
	\subfigure[\label{fpbh_tot}]{\includegraphics[width=0.45\linewidth]{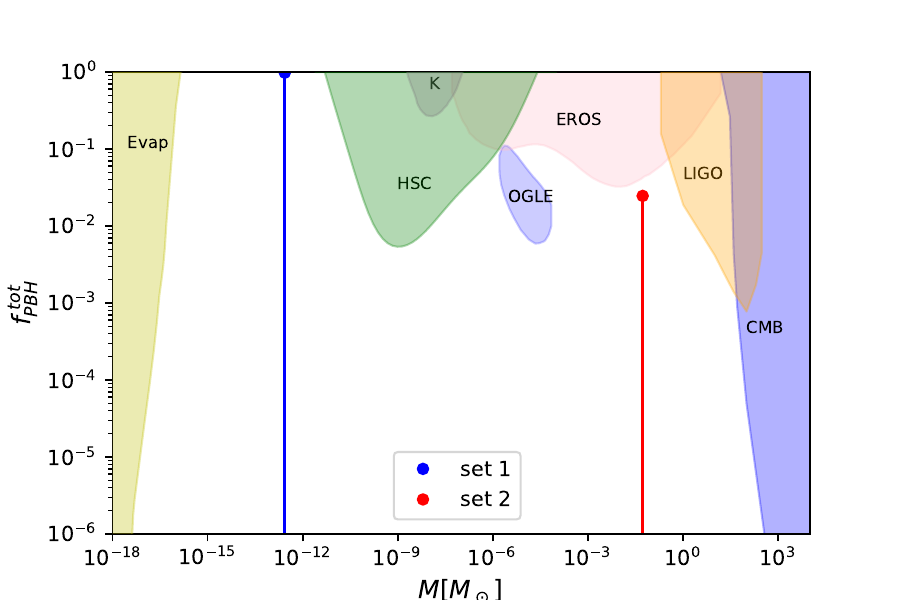}}
	\caption{The left side shows the PBH mass function versus the mass $M/M_H$ for two sets of the parameters introduced in Sec.\ref{model}: i) $(c, \phi_s, \delta) = (7.0043 \times 10^{-4}, 3.7, 1.26 \times 10^{-2})$ displayed by the blue color line, and ii)$(c, \phi_s, \delta) = (2.3658 \times 10^{-4}, 4.22, 7.2548 \times 10^{-3})$ as the red color line. The right side shows the total PBH abundance versus the mass $M/M_\odot$ for two sets of parameters. For the first set, the total PBH can account for the total DM, and for the second set, it can account for only $2.7\%$ of the total DM. }
	\label{fpbh_all}
\end{figure}

The total abundance of PBHs ($f^{\rm tot}_{\rm PBH}$) has been constrained severely by several observations. PBHs with $M_{PBH}< 2.5 \times 10^{-19}  M_\odot$ have been evaporated completely by the present epoch through the Hawking radiation, and they are not a DM candidate\cite{Hawking:1974rv}. PBHs with mass $M_{PBH} \leq   10^{-17} M_\odot$ are constrained by the observations from extra-galactic radiation background, Voyager, SPI/INTEGRAL   ~\cite{Siegert:2016ijv, Laha:2019ssq, Super-Kamiokande:2011lwo, Dasgupta:2019cae, Laha:2020ivk}. The mass window $10^{-16 } M_\odot \leq M_{PBH} \leq 5 \times 10^{-12} M_\odot$ shows the most promising possibility of explaining $100\%$ dark matter present in the universe~\cite{Carr:2016drx, Carr:2020xqk,Inomata:2017okj, Ballesteros:2017fsr, Bertone:2016nfn}. PBHs, with mass $10^{-11 } M_\odot \leq M_{PBH} \leq 10^{-1} M_\odot $, are bounded by their gravitational lensing by several observations like; HSC~\cite{Niikura:2017zjd}, EROS~\cite{Tisserand:2006zx} and OGLE~\cite{Niikura:2019kqi,Mroz:2024wia,Mroz:2024mse} where as PBHs contribution to DM above $1-10\%$ has been ruled out ~\cite{Smyth:2019whb,Tisserand:2006zx,Niikura:2017zjd,Niikura:2019kqi}. The present GW observations by LIGO/Virgo~\cite{Ali-Haimoud:2017rtz,Bird:2016dcv,Sasaki:2016jop,Cholis:2016kqi,Clesse:2016vqa,DeLuca:2020qqa,Andres-Carcasona:2024wqk,Liu:2018ess,Liu:2019rnx} have put constraints on the PBH with $0.2 M_\odot \leq M_{PBH} \leq 300 M_\odot $. Finally, more heavy mass PBHs $M_{PBH} \geq 100 M_\odot$ can impact the CMB spectrum, and anisotropies ~\cite{carr1981pregalactic,Ricotti:2007au,Serpico:2020ehh}.
The resulting $f_{PBH}$ for our model is displayed in Fig.\ref{fpbh} versus $M/M_\star$ for the aforementioned sets of parameters, so that the blue solid line is for the first set of parameters and the red solid line stands for the second set of parameters. The total PBH abundance is illustrated in Fig.\ref{fpbh_tot} for two sets of parameters, where blue and red dots indicate the values. For the first set of parameters, the mass is about $M_{PBH} \simeq 10^{-13} \; M_{\odot}$, where PBH can describe $100\%$ of DM. On the other side, for the second set of parameters, the resulting mass is about $M_{PBH} \simeq 10^{-2} \; M_{\odot}$, and PBHs only describe $2.4\%$ of the total DM. Fig.\ref{fpbh_all} shows the resulting mass spectrum and total PBH abundance for $K = \sqrt{\gamma_3}$; however, it should be mentioned that by decreasing the $K$ parameter, the mass spectrum rapidly grows, and the total abundance also increases. 

\section{Scalar Induced Gravitational Waves \label{sec_sigw}} 

In this section, first, we will briefly review the SIGW, compute the corresponding energy spectrum, and discuss the detectability of the resulting signal. The cosmological perturbations are generated during inflation, which are divided into scalar, vector, and tensor perturbations. The scalar perturbations are the primary seed of the universe's structure, and the generated tensor perturbations act as the source of the primordial gravitational waves. Up to the linear order, these perturbations are decoupled and evolve independently. At the first order, the gravitational waves propagate source-free (assuming no anisotropic stresses), and their power spectrum of the gravitational waves is given as \cite{Caprini:2018mtu, Christensen:2018iqi, Baumann:2007zm, Domenech:2024rks, Liu:2023hpw, Liu:2023pau, Chen:2024roo, Ananda:2006af, Kohri:2018awv, Domenech:2021ztg, Domenech:2019quo, Domenech:2020kqm, Witkowski:2022mtg, Balaji:2023ehk, Domenech:2021wkk}
\begin{equation}
	\Delta_h(k, \eta) = T(k, \eta) \; \mathcal{P}_{h, {\rm inf}}(k),
\end{equation}
where $\mathcal{P}_{h, {\rm inf}}(k)$ is the primordial tensor power spectrum at the horizon re-entry, 
\begin{equation}
	\mathcal{P}_{h, {\rm inf}}(k) = \frac{2}{\pi^2} \; \frac{H^2_\star}{M_p^2} \; \left( \frac{k}{k_\star} \right)^{n_t},
\end{equation}
in which $n_t$ is the spectral index, $k_\star$ is the pivot scale, and $H_\star$ is the Hubble parameter estimated at the time when $k_\star$ exits the horizon during the inflationary phase. The transfer function $T(k, \eta)$ encodes the background evolution from the moment a mode $k$ re-enters the horizon up to the time of observation. Assuming an instant reheating, the universe enters the radiation-dominant phase followed by a matter-dominant phase. Then, up to the first order of perturbation, the energy spectrum of the primordial GW is obtained as \cite{Figueroa:2019paj, Bernal:2019lpc, Bernal:2020ywq}
\begin{equation}
	\Omega_{\rm GW,0}^{(1)}(k) = \frac{\Omega_{\rm rad, 0}}{12\pi^2}\bigg(\frac{g_{*,k}}{g_{s,k}}\bigg)\bigg(\frac{g_{s,0}}{g_{s,k}}\bigg)^{4/3}\bigg(\frac{H_\star}{M_{\rm Pl}}\bigg)^2 \; \frac{\Gamma ^2 (5/6)}{2^{4/3} \;\Gamma ^2(3/2)}\mathcal{W}(\kappa),\label{GW_1_gen}
\end{equation}
where the dimensionless parameter $\kappa$ is defined as $\kappa=\frac{k}{k(T_1)}=\frac{f}{f(T_1)}$, with $T_1$ as the temperature at the start of the radiation-dominated phase. The function $\mathcal{W}(\kappa)$ is given as 
\begin{equation}
	\mathcal{W}(\kappa)= \frac{\pi}{2\kappa}\bigg[\bigg(\kappa J_{5/6}(\kappa)-J_{-1/6}(\kappa)\bigg)^2+\kappa^2J^2 _{-1/6}(\kappa)\bigg],\label{Bessel_GW1}
\end{equation}
and $J_{i}$ is the Bessel function of order $i$. Including second order, the scalar and tensor perturbations no longer evolve independently. At this order, the tensor perturbations are sourced by the first-order perturbations in which the evolution is described by the following equations \cite{Espinosa:2018eve, Kohri:2018awv}
\begin{equation}\label{sigw_equation}
	h_k'' + 2 {\cal H} h_k' + k^2 h_k=  {\cal S}(\mathbf{k}, \eta)\, ,
\end{equation}
where ${\cal S}(\mathbf{k}, \eta)$ is the source term that depends on the first-order perturbations. Enhanced scalar perturbations are expected to generate induced gravitational waves, which may be sufficiently strong to fall within the sensitivity range of present and upcoming detectors. Eq.\eqref{sigw_equation} is studied in \cite{Ananda:2006af,Baumann:2007zm,Kohri:2018awv,Domenech:2021ztg,Domenech:2019quo,Domenech:2020kqm,Witkowski:2022mtg,Balaji:2023ehk,Domenech:2021wkk} and the resulting energy density of SIGW can be written in the logarithmic interval of commoving wavenumber and normalized by total energy density present in the universe as: 
\begin{equation}
	\Omega_\text{GW}(k,\eta)=\frac1{\rho_\text{tot}}\frac{d\rho_\text{GW}}{d\ln k}=\frac{k^2}{12H^2a^2}\mathcal{P}_h(k,\eta),
\end{equation}
where $\mathcal{P}_h(k,\eta)$ is the power spectrum of the tensor perturbation. Following refs.~\cite{Espinosa:2018eve,Kohri:2018awv} we write 
\begin{eqnarray}\label{Omega0}
	\Omega_\text{GW,r}(k)
	& = & 3\int^\infty_0dv\int^{1+v}_{|1-v|}du\frac{\mathcal{T}(u,v)}{u^2v^2}\mathcal{P}_\mathcal{R}(vk)\mathcal{P}_\mathcal{R}(uk),\\
	\mathcal{T}(u,v)& = & \frac14\left[\frac{4v^2-(1+v^2-u^2)^2}{4uv}\right]^2\left(\frac{u^2+v^2-3}{2uv}\right)^4  \nonumber \\
	 & & \qquad \qquad \times \left[\left(\ln\left|\frac{3-(u+v)^2}{3-(u-v)^2}\right|-\frac{4uv}{u^2+v^2-3}\right)^2  +\pi^2\Theta\left(u+v-\sqrt3\right) \right], \nonumber
\end{eqnarray}
where, $\Theta$ is the step function and normalizing the $\mathcal{T}(u,v)$ in such a way that $u\approx v\rightarrow\infty$, this leads to $\mathcal{T}(u,v) \rightarrow(\ln(u+v))^2/4\sim(\ln v)^2$. The subscript ``r" in $\Omega_\text{GW,r}(k)$ in \eqref{Omega0} denotes the GW spectrum during the radiation-dominated epoch. Integrating Eq.~(\ref{Omega0}) numerically, one can obtain the spectrum of SIGWs for any given power spectrum. After the matter radiation equality, the energy density of the GWs starts to decay in comparison to the matter density, the present observational GW spectrum can be written as~\cite{Pi:2020otn}:  
\begin{align}\label{Omega6}
	\Omega_\text{GW}(f,\eta_0)h^2
	=\frac{g_*(\eta_0)^{4/3}}{g_*(\eta_0)g_{*s}(\eta_k)^{1/3}}\Omega_{r,0}\Omega_\text{GW,r}(f)
	=1.6\times10^{-5}\left(\frac{g_{*s}(\eta_k)}{106.75}\right)^{-1/3}\left(\frac{\Omega_{r,0}h^2}{4.1\times10^{-5}}\right)\Omega_\text{GW,r}(f).
\end{align}
Here $\Omega_{r,0}h^2$ is the present radiation energy density, and converting the co-moving wavenumber $k$ to frequency $f$\footnote{To compute the spectrum of the SIGWs we have utilized the publicly available code~\texttt{SIGWFast}~\cite{Witkowski:2022mtg}.}. 

\begin{figure}
	\centering
	\includegraphics[width=0.6\linewidth]{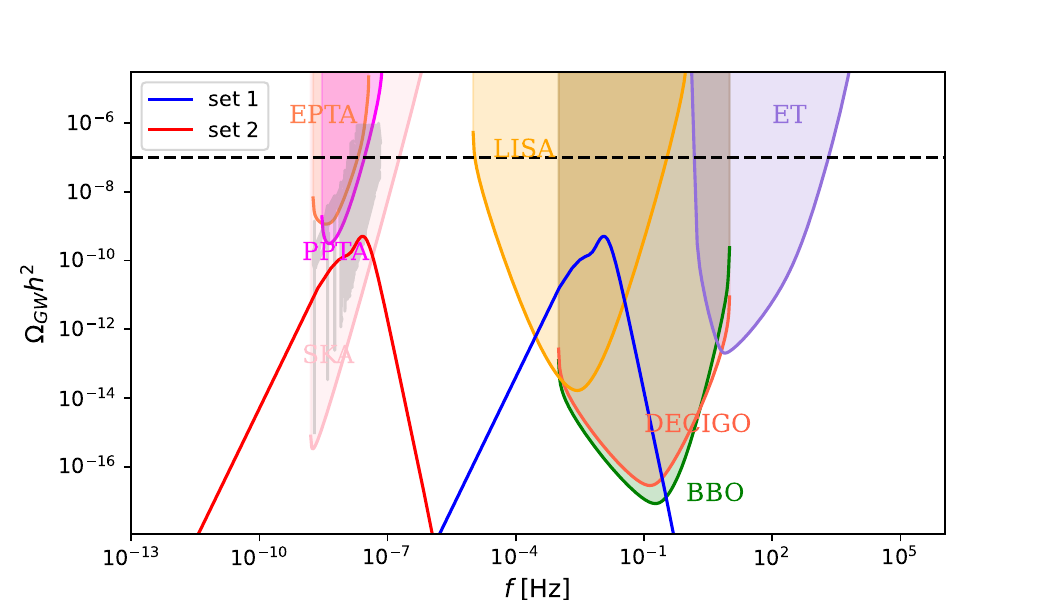}
	\caption{The plot shows the resulting SIGW versus the frequency for the two sets of parameters mentioned in Sec.\ref{model}: i) $(c, \phi_s, \delta) = (7.0043 \times 10^{-4}, 3.7, 1.26 \times 10^{-2})$ displayed by the blue color line, and ii)$(c, \phi_s, \delta) = (2.3658 \times 10^{-4}, 4.22, 7.2548 \times 10^{-3})$ as the red color line. For the first set, the peak of resulting SIGW stands in frequency around $f\simeq 10^{-2} \rm Hz$, and it crosses the observable range of the future experiments such as LISA, DECIGO, and BBO. For the second set, the peak of SIGW stands in frequency around $ f\simeq 10^{-8} \rm Hz$. It can explain NANOGrav, and it also crosses the observational range of the SKA observatory. }
	\label{sigw}
\end{figure}

The resulting induced GW spectrum is presented in Fig.~\ref{sigw} for two sets of parameters. The magnitudes of the GW spectra peaks are both of order $\mathcal{O}(10^{-8})$; however, they are located at different frequencies. For the first set of parameters, the peak of the GW spectrum is located at a frequency around $10^{-2} \; {\rm Hz}$. It crosses the sensitivity bounds of the future observatories such as LISA~\cite{Danzmann:1997hm, Harry:2010zz, LIGOScientific:2014qfs, LISA:2017pwj, LISACosmologyWorkingGroup:2022jok}, DECIGO~\cite{Seto:2001qf,Kawamura:2011zz, Kawamura:2006up}, and Big Bang Observer (BBO)~\cite{Crowder:2005nr, Corbin:2005ny, Harry:2006fi, Yagi:2011wg, Yagi:2017wg}. The obtained GW spectrum for the second set of parameters is displayed in solid red. The peak of its GW spectrum is located at a frequency around $10^{-8} \; {\rm Hz}$, and it crosses the sensitivity bounds of the SKA observatory~\cite{Carilli:2004nx, Moore:2014lga, Weltman:2018zrl, Correa:2023whf}. In addition, it can explain the NANOGrav result ~\cite{NANOGrav:2023gor, NANOGrav:2023hde, Gangopadhyay:2023qjr, Iovino:2024tyg,Correa:2023whf}. 


\section{Conclusion \label{conclusion}} 

In this work, we explored the initial spin of the PBHs along with their abundance and scalar-induced gravitational waves formed during the radiation-dominated epoch, using a power spectrum obtained numerically by solving the Mukhanov-Sasaki equation. We have considered the Mutated Hiltop inflation model as our working model. It is well known that for the formation of PBHs, an enhancement in the power spectrum is required. To obtain the desired amplitude of the power spectrum at small scales, we have added a step to the potential~(Eq.~\ref{pot_step}). We have chosen two different benchmark values for the step parameters. 
Evolution of the $\phi$, $H$, and $\epsilon_1$ is depicted in the Fig.~\ref{background_plot}, and the resulting power spectrum is displayed in Fig.~\ref{ps}, where the peak of the power spectrum is around $\mathcal{P_\zeta}(k)=0.0285$ at $k_{\rm peak}=6.708 \times 10^{12}$ for the first set, and for the second set it is $\mathcal{P_\zeta} (k)=0.0385$ at $k_{\rm peak}=1.523 \times 10^{7}$. To compute the profile of the curvature perturbation, we have followed the method mentioned in ~\cite{Pi:2024ert}, where, instead of $\zeta$, the authors used $\nabla^2 \zeta$ as the Gaussian random field. To compute the threshold for PBH formation, various methods have been employed to estimate $\mu_{\rm th}$. In some previous literature, the authors have considered $\mathcal{C}(r_m)=2/5$, whereas in others, $\bar{\mathcal{C}}_m(r_m)=2/5$ has been utilized. However, in this article, we have used the $q-$function approach to find the threshold value of $\mu$. The $q-$function method inherits the contribution from the shape of the curvature perturbation. Another advantage of using the $q-$ function method is that this method is accurate for different shapes of the profile within the $2 \%$ accuracy~\cite{Escriva:2022pnz}. While computing $f_{\rm PBH}$ and $f^{\rm tot}_{\rm PBH}$, we keep $K=\sqrt{\gamma_3}$; however, a similar analysis can be done for the different values of $K=(0.9\sqrt{\gamma_3}, 1.1\sqrt{\gamma_3})$. For set-1 we have obtained the  $M_{\rm PBH}=2.719 \times 10^{-13} M_\odot$ with $f^{\rm tot}_{\rm PBH} \approx 1$ where in case of set-2  $M_{\rm PBH}=5.2 \times 10^{-2} M_\odot$ with $f^{\rm tot}_{\rm PBH} \approx 0.024$ (see Fig.~\ref{fpbh_all}).  It is worth mentioning that by decreasing the parameter $K$, the $f^{\rm tot}_{\rm PBH}$ increases and may go beyond 1. To circumvent this, one needs to choose different benchmark points for the parameters ($c, \phi_s, \delta$) of Eq.~\ref{pot_step}.  \\

To calculate the initial spin of the PBHs, we first estimate the angular momentum of the collapsing region. In the CMC gauge, we assume a linear relation between the curvature and the density perturbation~(Eq.~\ref{density_curvature}). Eq.~\ref{zeta_profile} can be used to define the density profile, with the modified spectral momenta Eq.~\ref{sigma_d} and two-point correlation function Eq.~\ref{twopoint_d}. In Fig.~\ref{d_profile}, the behavior of the density profile can be found for the different values of the parameter $K_\delta$. Calculating the average of the density perturbation clarified that only for $k_\delta = [0.9, 1.07]$, the average of the density perturbation remains within the range $\delta_H = [0.63, 0.84]$. Determining the turnaround point as the moment when the velocity of the region reaches the minimum, the spin of the PBH was calculated. The resulting $\sqrt{\langle a_\star^2 \rangle}$ is plotted in terms of mass $M/M_H$ for different values of $K_\delta$, indicating that the magnitude of the spin is of the order of $10^{-3}$. This magnitude increases by decreasing mass, and it can reach the order of $10^{-2}$ for $M \ll M_H$.\\

Finally, in Section~\ref{sec_sigw} we compute the spectrum of scalar-induced gravitational waves, at second order in perturbation theory, where the scalar and tensor modes are coupled. An increment in the scalar perturbation can lead to the enhancement in the tensor perturbations which results in the SIGWs~\cite{Baumann:2007zm,Domenech:2024rks,Domenech:2020kqm}. We have computed the GW spectrum for two different sets ($=1,2$), and the results are presented in Fig.~\ref{sigw}. For set-1, the obtained spectrum peaks around $f \approx 10^{-2} \; {\rm Hz}$ and can be detected by the future GW detectors like; LISA, DECIGO, and BBO. In the case of set-2, the GW spectrum peaks at $f \approx 10^{-8} \; {\rm Hz}$, which can explain the stochastic GW measured by NANOGrav and can also be detected by the SKA.

\section*{Acknowledgments}

The authors would like to thank Tomohiro Harada, Chul-Moon Yoo, Kazunori Kohri, Albert Escriv\`a, and Cristian Joana for the fruitful discussions. This work is supported by the National Natural Science Foundation of China under Grants No. 12275238, 11675143 and W2433018, the National Key Research and Development Program under Grant No. 2020YFC2201503, and the Zhejiang Provincial Natural Science Foundation of China under Grants No. LR21A050001 and No. LY20A050002, and the Fundamental Research Funds for the Provincial Universities of Zhejiang in China under Grant No. RF-A2019015.

\bibliography{pbh.bib}

\begin{thebibliography}{182}%
\makeatletter
\providecommand \@ifxundefined [1]{%
 \@ifx{#1\undefined}
}%
\providecommand \@ifnum [1]{%
 \ifnum #1\expandafter \@firstoftwo
 \else \expandafter \@secondoftwo
 \fi
}%
\providecommand \@ifx [1]{%
 \ifx #1\expandafter \@firstoftwo
 \else \expandafter \@secondoftwo
 \fi
}%
\providecommand \natexlab [1]{#1}%
\providecommand \enquote  [1]{``#1''}%
\providecommand \bibnamefont  [1]{#1}%
\providecommand \bibfnamefont [1]{#1}%
\providecommand \citenamefont [1]{#1}%
\providecommand \href@noop [0]{\@secondoftwo}%
\providecommand \href [0]{\begingroup \@sanitize@url \@href}%
\providecommand \@href[1]{\@@startlink{#1}\@@href}%
\providecommand \@@href[1]{\endgroup#1\@@endlink}%
\providecommand \@sanitize@url [0]{\catcode `\\12\catcode `\$12\catcode
  `\&12\catcode `\#12\catcode `\^12\catcode `\_12\catcode `\%12\relax}%
\providecommand \@@startlink[1]{}%
\providecommand \@@endlink[0]{}%
\providecommand \url  [0]{\begingroup\@sanitize@url \@url }%
\providecommand \@url [1]{\endgroup\@href {#1}{\urlprefix }}%
\providecommand \urlprefix  [0]{URL }%
\providecommand \Eprint [0]{\href }%
\providecommand \doibase [0]{https://doi.org/}%
\providecommand \selectlanguage [0]{\@gobble}%
\providecommand \bibinfo  [0]{\@secondoftwo}%
\providecommand \bibfield  [0]{\@secondoftwo}%
\providecommand \translation [1]{[#1]}%
\providecommand \BibitemOpen [0]{}%
\providecommand \bibitemStop [0]{}%
\providecommand \bibitemNoStop [0]{.\EOS\space}%
\providecommand \EOS [0]{\spacefactor3000\relax}%
\providecommand \BibitemShut  [1]{\csname bibitem#1\endcsname}%
\let\auto@bib@innerbib\@empty
\bibitem [{\citenamefont {Abbott}\ \emph
  {et~al.}(2016{\natexlab{a}})\citenamefont {Abbott} \emph
  {et~al.}}]{LIGOScientific:2016dsl}%
  \BibitemOpen
  \bibfield  {author} {\bibinfo {author} {\bibfnamefont {B.~P.}\ \bibnamefont
  {Abbott}} \emph {et~al.} (\bibinfo {collaboration} {LIGO Scientific,
  Virgo}),\ }\bibfield  {title} {\bibinfo {title} {{Binary Black Hole Mergers
  in the first Advanced LIGO Observing Run}},\ }\href@noop {} {\bibfield
  {journal} {\bibinfo  {journal} {Phys. Rev. X}\ }\textbf {\bibinfo {volume}
  {6}},\ \bibinfo {pages} {041015} (\bibinfo {year} {2016}{\natexlab{a}})},\
  \bibinfo {note} {[Erratum: Phys.Rev.X 8, 039903 (2018)]}\BibitemShut
  {NoStop}%
\bibitem [{\citenamefont {Abbott}\ \emph
  {et~al.}(2016{\natexlab{b}})\citenamefont {Abbott} \emph
  {et~al.}}]{LIGOScientific:2016aoc}%
  \BibitemOpen
  \bibfield  {author} {\bibinfo {author} {\bibfnamefont {B.~P.}\ \bibnamefont
  {Abbott}} \emph {et~al.} (\bibinfo {collaboration} {LIGO Scientific,
  Virgo}),\ }\bibfield  {title} {\bibinfo {title} {{Observation of
  Gravitational Waves from a Binary Black Hole Merger}},\ }\href
  {https://doi.org/10.1103/PhysRevLett.116.061102} {\bibfield  {journal}
  {\bibinfo  {journal} {Phys. Rev. Lett.}\ }\textbf {\bibinfo {volume} {116}},\
  \bibinfo {pages} {061102} (\bibinfo {year} {2016}{\natexlab{b}})},\ \Eprint
  {https://arxiv.org/abs/1602.03837} {arXiv:1602.03837 [gr-qc]} \BibitemShut
  {NoStop}%
\bibitem [{\citenamefont {Abbott}\ \emph
  {et~al.}(2016{\natexlab{c}})\citenamefont {Abbott} \emph
  {et~al.}}]{LIGOScientific:2016sjg}%
  \BibitemOpen
  \bibfield  {author} {\bibinfo {author} {\bibfnamefont {B.~P.}\ \bibnamefont
  {Abbott}} \emph {et~al.} (\bibinfo {collaboration} {LIGO Scientific,
  Virgo}),\ }\bibfield  {title} {\bibinfo {title} {{GW151226: Observation of
  Gravitational Waves from a 22-Solar-Mass Binary Black Hole Coalescence}},\
  }\href {https://doi.org/10.1103/PhysRevLett.116.241103} {\bibfield  {journal}
  {\bibinfo  {journal} {Phys. Rev. Lett.}\ }\textbf {\bibinfo {volume} {116}},\
  \bibinfo {pages} {241103} (\bibinfo {year} {2016}{\natexlab{c}})},\ \Eprint
  {https://arxiv.org/abs/1606.04855} {arXiv:1606.04855 [gr-qc]} \BibitemShut
  {NoStop}%
\bibitem [{\citenamefont {Abbott}\ \emph
  {et~al.}(2017{\natexlab{a}})\citenamefont {Abbott} \emph
  {et~al.}}]{LIGOScientific:2016wyt}%
  \BibitemOpen
  \bibfield  {author} {\bibinfo {author} {\bibfnamefont {B.~P.}\ \bibnamefont
  {Abbott}} \emph {et~al.} (\bibinfo {collaboration} {LIGO Scientific,
  Virgo}),\ }\bibfield  {title} {\bibinfo {title} {{The basic physics of the
  binary black hole merger GW150914}},\ }\href
  {https://doi.org/10.1002/andp.201600209} {\bibfield  {journal} {\bibinfo
  {journal} {Annalen Phys.}\ }\textbf {\bibinfo {volume} {529}},\ \bibinfo
  {pages} {1600209} (\bibinfo {year} {2017}{\natexlab{a}})},\ \Eprint
  {https://arxiv.org/abs/1608.01940} {arXiv:1608.01940 [gr-qc]} \BibitemShut
  {NoStop}%
\bibitem [{\citenamefont {Abbott}\ \emph
  {et~al.}(2017{\natexlab{b}})\citenamefont {Abbott} \emph
  {et~al.}}]{LIGOScientific:2017bnn}%
  \BibitemOpen
  \bibfield  {author} {\bibinfo {author} {\bibfnamefont {B.~P.}\ \bibnamefont
  {Abbott}} \emph {et~al.} (\bibinfo {collaboration} {LIGO Scientific,
  VIRGO}),\ }\bibfield  {title} {\bibinfo {title} {{GW170104: Observation of a
  50-Solar-Mass Binary Black Hole Coalescence at Redshift 0.2}},\ }\href
  {https://doi.org/10.1103/PhysRevLett.118.221101} {\bibfield  {journal}
  {\bibinfo  {journal} {Phys. Rev. Lett.}\ }\textbf {\bibinfo {volume} {118}},\
  \bibinfo {pages} {221101} (\bibinfo {year} {2017}{\natexlab{b}})},\ \bibinfo
  {note} {[Erratum: Phys.Rev.Lett. 121, 129901 (2018)]},\ \Eprint
  {https://arxiv.org/abs/1706.01812} {arXiv:1706.01812 [gr-qc]} \BibitemShut
  {NoStop}%
\bibitem [{\citenamefont {Abbott}\ \emph
  {et~al.}(2017{\natexlab{c}})\citenamefont {Abbott} \emph
  {et~al.}}]{LIGOScientific:2017vox}%
  \BibitemOpen
  \bibfield  {author} {\bibinfo {author} {\bibfnamefont {B.~P.}\ \bibnamefont
  {Abbott}} \emph {et~al.} (\bibinfo {collaboration} {LIGO Scientific,
  Virgo}),\ }\bibfield  {title} {\bibinfo {title} {{GW170608: Observation of a
  19-solar-mass Binary Black Hole Coalescence}},\ }\href
  {https://doi.org/10.3847/2041-8213/aa9f0c} {\bibfield  {journal} {\bibinfo
  {journal} {Astrophys. J. Lett.}\ }\textbf {\bibinfo {volume} {851}},\
  \bibinfo {pages} {L35} (\bibinfo {year} {2017}{\natexlab{c}})},\ \Eprint
  {https://arxiv.org/abs/1711.05578} {arXiv:1711.05578 [astro-ph.HE]}
  \BibitemShut {NoStop}%
\bibitem [{\citenamefont {Abbott}\ \emph
  {et~al.}(2017{\natexlab{d}})\citenamefont {Abbott} \emph
  {et~al.}}]{LIGOScientific:2017ycc}%
  \BibitemOpen
  \bibfield  {author} {\bibinfo {author} {\bibfnamefont {B.~P.}\ \bibnamefont
  {Abbott}} \emph {et~al.} (\bibinfo {collaboration} {LIGO Scientific,
  Virgo}),\ }\bibfield  {title} {\bibinfo {title} {{GW170814: A Three-Detector
  Observation of Gravitational Waves from a Binary Black Hole Coalescence}},\
  }\href {https://doi.org/10.1103/PhysRevLett.119.141101} {\bibfield  {journal}
  {\bibinfo  {journal} {Phys. Rev. Lett.}\ }\textbf {\bibinfo {volume} {119}},\
  \bibinfo {pages} {141101} (\bibinfo {year} {2017}{\natexlab{d}})},\ \Eprint
  {https://arxiv.org/abs/1709.09660} {arXiv:1709.09660 [gr-qc]} \BibitemShut
  {NoStop}%
\bibitem [{\citenamefont {Zel'dovich}\ and\ \citenamefont
  {Novikov}(1967)}]{Zeldovich:1967lct}%
  \BibitemOpen
  \bibfield  {author} {\bibinfo {author} {\bibfnamefont {Y.~B.}\ \bibnamefont
  {Zel'dovich}}\ and\ \bibinfo {author} {\bibfnamefont {I.~D.}\ \bibnamefont
  {Novikov}},\ }\bibfield  {title} {\bibinfo {title} {{The Hypothesis of Cores
  Retarded during Expansion and the Hot Cosmological Model}},\ }\href@noop {}
  {\bibfield  {journal} {\bibinfo  {journal} {Sov. Astron.}\ }\textbf {\bibinfo
  {volume} {10}},\ \bibinfo {pages} {602} (\bibinfo {year} {1967})}\BibitemShut
  {NoStop}%
\bibitem [{\citenamefont {Hawking}(1971)}]{Hawking:1971ei}%
  \BibitemOpen
  \bibfield  {author} {\bibinfo {author} {\bibfnamefont {S.}~\bibnamefont
  {Hawking}},\ }\bibfield  {title} {\bibinfo {title} {{Gravitationally
  collapsed objects of very low mass}},\ }\href@noop {} {\bibfield  {journal}
  {\bibinfo  {journal} {Mon. Not. Roy. Astron. Soc.}\ }\textbf {\bibinfo
  {volume} {152}},\ \bibinfo {pages} {75} (\bibinfo {year} {1971})}\BibitemShut
  {NoStop}%
\bibitem [{\citenamefont {Carr}\ and\ \citenamefont
  {Hawking}(1974)}]{Carr:1974nx}%
  \BibitemOpen
  \bibfield  {author} {\bibinfo {author} {\bibfnamefont {B.~J.}\ \bibnamefont
  {Carr}}\ and\ \bibinfo {author} {\bibfnamefont {S.}~\bibnamefont {Hawking}},\
  }\bibfield  {title} {\bibinfo {title} {{Black holes in the early Universe}},\
  }\href@noop {} {\bibfield  {journal} {\bibinfo  {journal} {Mon. Not. Roy.
  Astron. Soc.}\ }\textbf {\bibinfo {volume} {168}},\ \bibinfo {pages} {399}
  (\bibinfo {year} {1974})}\BibitemShut {NoStop}%
\bibitem [{\citenamefont {Carr}(1975)}]{Carr:1975qj}%
  \BibitemOpen
  \bibfield  {author} {\bibinfo {author} {\bibfnamefont {B.~J.}\ \bibnamefont
  {Carr}},\ }\bibfield  {title} {\bibinfo {title} {{The Primordial black hole
  mass spectrum}},\ }\href {https://doi.org/10.1086/153853} {\bibfield
  {journal} {\bibinfo  {journal} {Astrophys. J.}\ }\textbf {\bibinfo {volume}
  {201}},\ \bibinfo {pages} {1} (\bibinfo {year} {1975})}\BibitemShut {NoStop}%
\bibitem [{\citenamefont {Fernandez}\ and\ \citenamefont
  {Profumo}(2019)}]{Fernandez:2019kyb}%
  \BibitemOpen
  \bibfield  {author} {\bibinfo {author} {\bibfnamefont {N.}~\bibnamefont
  {Fernandez}}\ and\ \bibinfo {author} {\bibfnamefont {S.}~\bibnamefont
  {Profumo}},\ }\bibfield  {title} {\bibinfo {title} {{Unraveling the origin of
  black holes from effective spin measurements with LIGO-Virgo}},\ }\href
  {https://doi.org/10.1088/1475-7516/2019/08/022} {\bibfield  {journal}
  {\bibinfo  {journal} {JCAP}\ }\textbf {\bibinfo {volume} {08}},\ \bibinfo
  {pages} {022}},\ \Eprint {https://arxiv.org/abs/1905.13019} {arXiv:1905.13019
  [astro-ph.HE]} \BibitemShut {NoStop}%
\bibitem [{\citenamefont {Carr}\ \emph {et~al.}(2021)\citenamefont {Carr},
  \citenamefont {Kohri}, \citenamefont {Sendouda},\ and\ \citenamefont
  {Yokoyama}}]{Carr:2020gox}%
  \BibitemOpen
  \bibfield  {author} {\bibinfo {author} {\bibfnamefont {B.}~\bibnamefont
  {Carr}}, \bibinfo {author} {\bibfnamefont {K.}~\bibnamefont {Kohri}},
  \bibinfo {author} {\bibfnamefont {Y.}~\bibnamefont {Sendouda}},\ and\
  \bibinfo {author} {\bibfnamefont {J.}~\bibnamefont {Yokoyama}},\ }\bibfield
  {title} {\bibinfo {title} {{Constraints on primordial black holes}},\ }\href
  {https://doi.org/10.1088/1361-6633/ac1e31} {\bibfield  {journal} {\bibinfo
  {journal} {Rept. Prog. Phys.}\ }\textbf {\bibinfo {volume} {84}},\ \bibinfo
  {pages} {116902} (\bibinfo {year} {2021})},\ \Eprint
  {https://arxiv.org/abs/2002.12778} {arXiv:2002.12778 [astro-ph.CO]}
  \BibitemShut {NoStop}%
\bibitem [{\citenamefont {Carr}\ and\ \citenamefont
  {Kuhnel}(2022)}]{Carr:2021bzv}%
  \BibitemOpen
  \bibfield  {author} {\bibinfo {author} {\bibfnamefont {B.}~\bibnamefont
  {Carr}}\ and\ \bibinfo {author} {\bibfnamefont {F.}~\bibnamefont {Kuhnel}},\
  }\bibfield  {title} {\bibinfo {title} {{Primordial black holes as dark matter
  candidates}},\ }\href {https://doi.org/10.21468/SciPostPhysLectNotes.48}
  {\bibfield  {journal} {\bibinfo  {journal} {SciPost Phys. Lect. Notes}\
  }\textbf {\bibinfo {volume} {48}},\ \bibinfo {pages} {1} (\bibinfo {year}
  {2022})},\ \Eprint {https://arxiv.org/abs/2110.02821} {arXiv:2110.02821
  [astro-ph.CO]} \BibitemShut {NoStop}%
\bibitem [{\citenamefont {Laha}(2019)}]{Laha:2019ssq}%
  \BibitemOpen
  \bibfield  {author} {\bibinfo {author} {\bibfnamefont {R.}~\bibnamefont
  {Laha}},\ }\bibfield  {title} {\bibinfo {title} {{Primordial Black Holes as a
  Dark Matter Candidate Are Severely Constrained by the Galactic Center 511 keV
  $\gamma$ -Ray Line}},\ }\href
  {https://doi.org/10.1103/PhysRevLett.123.251101} {\bibfield  {journal}
  {\bibinfo  {journal} {Phys. Rev. Lett.}\ }\textbf {\bibinfo {volume} {123}},\
  \bibinfo {pages} {251101} (\bibinfo {year} {2019})},\ \Eprint
  {https://arxiv.org/abs/1906.09994} {arXiv:1906.09994 [astro-ph.HE]}
  \BibitemShut {NoStop}%
\bibitem [{\citenamefont {Churazov}\ \emph {et~al.}(2011)\citenamefont
  {Churazov}, \citenamefont {Sazonov}, \citenamefont {Tsygankov}, \citenamefont
  {Sunyaev},\ and\ \citenamefont {Varshalovich}}]{2011MNRAS.411.1727C}%
  \BibitemOpen
  \bibfield  {author} {\bibinfo {author} {\bibfnamefont {E.}~\bibnamefont
  {Churazov}}, \bibinfo {author} {\bibfnamefont {S.}~\bibnamefont {Sazonov}},
  \bibinfo {author} {\bibfnamefont {S.}~\bibnamefont {Tsygankov}}, \bibinfo
  {author} {\bibfnamefont {R.}~\bibnamefont {Sunyaev}},\ and\ \bibinfo {author}
  {\bibfnamefont {D.}~\bibnamefont {Varshalovich}},\ }\bibfield  {title}
  {\bibinfo {title} {{Positron annihilation spectrum from the Galactic Centre
  region observed by SPI/INTEGRAL, revisited: annihilation in a cooling
  ISM?}},\ }\href {https://doi.org/10.1111/j.1365-2966.2010.17804.x} {\bibfield
   {journal} {\bibinfo  {journal} {Mon. Not. Roy. Astron. Soc.}\ }\textbf
  {\bibinfo {volume} {411}},\ \bibinfo {pages} {1727} (\bibinfo {year}
  {2011})},\ \Eprint {https://arxiv.org/abs/1010.0864} {arXiv:1010.0864
  [astro-ph.HE]} \BibitemShut {NoStop}%
\bibitem [{\citenamefont {Siegert}\ \emph {et~al.}(2016)\citenamefont
  {Siegert}, \citenamefont {Diehl}, \citenamefont {Vincent}, \citenamefont
  {Guglielmetti}, \citenamefont {Krause},\ and\ \citenamefont
  {Boehm}}]{Siegert:2016ijv}%
  \BibitemOpen
  \bibfield  {author} {\bibinfo {author} {\bibfnamefont {T.}~\bibnamefont
  {Siegert}}, \bibinfo {author} {\bibfnamefont {R.}~\bibnamefont {Diehl}},
  \bibinfo {author} {\bibfnamefont {A.~C.}\ \bibnamefont {Vincent}}, \bibinfo
  {author} {\bibfnamefont {F.}~\bibnamefont {Guglielmetti}}, \bibinfo {author}
  {\bibfnamefont {M.~G.~H.}\ \bibnamefont {Krause}},\ and\ \bibinfo {author}
  {\bibfnamefont {C.}~\bibnamefont {Boehm}},\ }\bibfield  {title} {\bibinfo
  {title} {{Search for 511 keV Emission in Satellite Galaxies of the Milky Way
  with INTEGRAL/SPI}},\ }\href {https://doi.org/10.1051/0004-6361/201629136}
  {\bibfield  {journal} {\bibinfo  {journal} {Astron. Astrophys.}\ }\textbf
  {\bibinfo {volume} {595}},\ \bibinfo {pages} {A25} (\bibinfo {year}
  {2016})},\ \Eprint {https://arxiv.org/abs/1608.00393} {arXiv:1608.00393
  [astro-ph.HE]} \BibitemShut {NoStop}%
\bibitem [{\citenamefont {Dasgupta}\ \emph {et~al.}(2020)\citenamefont
  {Dasgupta}, \citenamefont {Laha},\ and\ \citenamefont
  {Ray}}]{Dasgupta:2019cae}%
  \BibitemOpen
  \bibfield  {author} {\bibinfo {author} {\bibfnamefont {B.}~\bibnamefont
  {Dasgupta}}, \bibinfo {author} {\bibfnamefont {R.}~\bibnamefont {Laha}},\
  and\ \bibinfo {author} {\bibfnamefont {A.}~\bibnamefont {Ray}},\ }\bibfield
  {title} {\bibinfo {title} {{Neutrino and positron constraints on spinning
  primordial black hole dark matter}},\ }\href
  {https://doi.org/10.1103/PhysRevLett.125.101101} {\bibfield  {journal}
  {\bibinfo  {journal} {Phys. Rev. Lett.}\ }\textbf {\bibinfo {volume} {125}},\
  \bibinfo {pages} {101101} (\bibinfo {year} {2020})},\ \Eprint
  {https://arxiv.org/abs/1912.01014} {arXiv:1912.01014 [hep-ph]} \BibitemShut
  {NoStop}%
\bibitem [{\citenamefont {Laha}\ \emph {et~al.}(2020)\citenamefont {Laha},
  \citenamefont {Mu{\~n}oz},\ and\ \citenamefont {Slatyer}}]{Laha:2020ivk}%
  \BibitemOpen
  \bibfield  {author} {\bibinfo {author} {\bibfnamefont {R.}~\bibnamefont
  {Laha}}, \bibinfo {author} {\bibfnamefont {J.~B.}\ \bibnamefont
  {Mu{\~n}oz}},\ and\ \bibinfo {author} {\bibfnamefont {T.~R.}\ \bibnamefont
  {Slatyer}},\ }\bibfield  {title} {\bibinfo {title} {{INTEGRAL constraints on
  primordial black holes and particle dark matter}},\ }\href
  {https://doi.org/10.1103/PhysRevD.101.123514} {\bibfield  {journal} {\bibinfo
   {journal} {Phys. Rev. D}\ }\textbf {\bibinfo {volume} {101}},\ \bibinfo
  {pages} {123514} (\bibinfo {year} {2020})},\ \Eprint
  {https://arxiv.org/abs/2004.00627} {arXiv:2004.00627 [astro-ph.CO]}
  \BibitemShut {NoStop}%
\bibitem [{\citenamefont {Ivanov}\ \emph {et~al.}(1994)\citenamefont {Ivanov},
  \citenamefont {Naselsky},\ and\ \citenamefont {Novikov}}]{PhysRevD.50.7173}%
  \BibitemOpen
  \bibfield  {author} {\bibinfo {author} {\bibfnamefont {P.}~\bibnamefont
  {Ivanov}}, \bibinfo {author} {\bibfnamefont {P.}~\bibnamefont {Naselsky}},\
  and\ \bibinfo {author} {\bibfnamefont {I.}~\bibnamefont {Novikov}},\
  }\bibfield  {title} {\bibinfo {title} {Inflation and primordial black holes
  as dark matter},\ }\href {https://doi.org/10.1103/PhysRevD.50.7173}
  {\bibfield  {journal} {\bibinfo  {journal} {Phys. Rev. D}\ }\textbf {\bibinfo
  {volume} {50}},\ \bibinfo {pages} {7173} (\bibinfo {year}
  {1994})}\BibitemShut {NoStop}%
\bibitem [{\citenamefont {Yokoyama}(1998)}]{Yokoyama:1998pt}%
  \BibitemOpen
  \bibfield  {author} {\bibinfo {author} {\bibfnamefont {J.}~\bibnamefont
  {Yokoyama}},\ }\bibfield  {title} {\bibinfo {title} {{Chaotic new inflation
  and formation of primordial black holes}},\ }\href
  {https://doi.org/10.1103/PhysRevD.58.083510} {\bibfield  {journal} {\bibinfo
  {journal} {Phys. Rev. D}\ }\textbf {\bibinfo {volume} {58}},\ \bibinfo
  {pages} {083510} (\bibinfo {year} {1998})},\ \Eprint
  {https://arxiv.org/abs/astro-ph/9802357} {arXiv:astro-ph/9802357}
  \BibitemShut {NoStop}%
\bibitem [{\citenamefont {Garcia-Bellido}\ and\ \citenamefont
  {Ruiz~Morales}(2017)}]{Garcia-Bellido:2017mdw}%
  \BibitemOpen
  \bibfield  {author} {\bibinfo {author} {\bibfnamefont {J.}~\bibnamefont
  {Garcia-Bellido}}\ and\ \bibinfo {author} {\bibfnamefont {E.}~\bibnamefont
  {Ruiz~Morales}},\ }\bibfield  {title} {\bibinfo {title} {{Primordial black
  holes from single field models of inflation}},\ }\href
  {https://doi.org/10.1016/j.dark.2017.09.007} {\bibfield  {journal} {\bibinfo
  {journal} {Phys. Dark Univ.}\ }\textbf {\bibinfo {volume} {18}},\ \bibinfo
  {pages} {47} (\bibinfo {year} {2017})},\ \Eprint
  {https://arxiv.org/abs/1702.03901} {arXiv:1702.03901 [astro-ph.CO]}
  \BibitemShut {NoStop}%
\bibitem [{\citenamefont {Ballesteros}\ and\ \citenamefont
  {Taoso}(2018)}]{Ballesteros:2017fsr}%
  \BibitemOpen
  \bibfield  {author} {\bibinfo {author} {\bibfnamefont {G.}~\bibnamefont
  {Ballesteros}}\ and\ \bibinfo {author} {\bibfnamefont {M.}~\bibnamefont
  {Taoso}},\ }\bibfield  {title} {\bibinfo {title} {{Primordial black hole dark
  matter from single field inflation}},\ }\href
  {https://doi.org/10.1103/PhysRevD.97.023501} {\bibfield  {journal} {\bibinfo
  {journal} {Phys. Rev. D}\ }\textbf {\bibinfo {volume} {97}},\ \bibinfo
  {pages} {023501} (\bibinfo {year} {2018})},\ \Eprint
  {https://arxiv.org/abs/1709.05565} {arXiv:1709.05565 [hep-ph]} \BibitemShut
  {NoStop}%
\bibitem [{\citenamefont {Hertzberg}\ and\ \citenamefont
  {Yamada}(2018)}]{Hertzberg:2017dkh}%
  \BibitemOpen
  \bibfield  {author} {\bibinfo {author} {\bibfnamefont {M.~P.}\ \bibnamefont
  {Hertzberg}}\ and\ \bibinfo {author} {\bibfnamefont {M.}~\bibnamefont
  {Yamada}},\ }\bibfield  {title} {\bibinfo {title} {{Primordial Black Holes
  from Polynomial Potentials in Single Field Inflation}},\ }\href
  {https://doi.org/10.1103/PhysRevD.97.083509} {\bibfield  {journal} {\bibinfo
  {journal} {Phys. Rev. D}\ }\textbf {\bibinfo {volume} {97}},\ \bibinfo
  {pages} {083509} (\bibinfo {year} {2018})},\ \Eprint
  {https://arxiv.org/abs/1712.09750} {arXiv:1712.09750 [astro-ph.CO]}
  \BibitemShut {NoStop}%
\bibitem [{\citenamefont {Kinney}(2005)}]{Kinney:2005vj}%
  \BibitemOpen
  \bibfield  {author} {\bibinfo {author} {\bibfnamefont {W.~H.}\ \bibnamefont
  {Kinney}},\ }\bibfield  {title} {\bibinfo {title} {{Horizon crossing and
  inflation with large eta}},\ }\href
  {https://doi.org/10.1103/PhysRevD.72.023515} {\bibfield  {journal} {\bibinfo
  {journal} {Phys. Rev. D}\ }\textbf {\bibinfo {volume} {72}},\ \bibinfo
  {pages} {023515} (\bibinfo {year} {2005})},\ \Eprint
  {https://arxiv.org/abs/gr-qc/0503017} {arXiv:gr-qc/0503017} \BibitemShut
  {NoStop}%
\bibitem [{\citenamefont {Germani}\ and\ \citenamefont
  {Prokopec}(2017)}]{Germani:2017bcs}%
  \BibitemOpen
  \bibfield  {author} {\bibinfo {author} {\bibfnamefont {C.}~\bibnamefont
  {Germani}}\ and\ \bibinfo {author} {\bibfnamefont {T.}~\bibnamefont
  {Prokopec}},\ }\bibfield  {title} {\bibinfo {title} {{On primordial black
  holes from an inflection point}},\ }\href
  {https://doi.org/10.1016/j.dark.2017.09.001} {\bibfield  {journal} {\bibinfo
  {journal} {Phys. Dark Univ.}\ }\textbf {\bibinfo {volume} {18}},\ \bibinfo
  {pages} {6} (\bibinfo {year} {2017})},\ \Eprint
  {https://arxiv.org/abs/1706.04226} {arXiv:1706.04226 [astro-ph.CO]}
  \BibitemShut {NoStop}%
\bibitem [{\citenamefont {Pattison}\ \emph {et~al.}(2017)\citenamefont
  {Pattison}, \citenamefont {Vennin}, \citenamefont {Assadullahi},\ and\
  \citenamefont {Wands}}]{Pattison:2017mbe}%
  \BibitemOpen
  \bibfield  {author} {\bibinfo {author} {\bibfnamefont {C.}~\bibnamefont
  {Pattison}}, \bibinfo {author} {\bibfnamefont {V.}~\bibnamefont {Vennin}},
  \bibinfo {author} {\bibfnamefont {H.}~\bibnamefont {Assadullahi}},\ and\
  \bibinfo {author} {\bibfnamefont {D.}~\bibnamefont {Wands}},\ }\bibfield
  {title} {\bibinfo {title} {{Quantum diffusion during inflation and primordial
  black holes}},\ }\href {https://doi.org/10.1088/1475-7516/2017/10/046}
  {\bibfield  {journal} {\bibinfo  {journal} {JCAP}\ }\textbf {\bibinfo
  {volume} {10}},\ \bibinfo {pages} {046}},\ \Eprint
  {https://arxiv.org/abs/1707.00537} {arXiv:1707.00537 [hep-th]} \BibitemShut
  {NoStop}%
\bibitem [{\citenamefont {Ezquiaga}\ and\ \citenamefont
  {Garc\'\i{}a-Bellido}(2018)}]{Ezquiaga:2018gbw}%
  \BibitemOpen
  \bibfield  {author} {\bibinfo {author} {\bibfnamefont {J.~M.}\ \bibnamefont
  {Ezquiaga}}\ and\ \bibinfo {author} {\bibfnamefont {J.}~\bibnamefont
  {Garc\'\i{}a-Bellido}},\ }\bibfield  {title} {\bibinfo {title} {{Quantum
  diffusion beyond slow-roll: implications for primordial black-hole
  production}},\ }\href {https://doi.org/10.1088/1475-7516/2018/08/018}
  {\bibfield  {journal} {\bibinfo  {journal} {JCAP}\ }\textbf {\bibinfo
  {volume} {08}},\ \bibinfo {pages} {018}},\ \Eprint
  {https://arxiv.org/abs/1805.06731} {arXiv:1805.06731 [astro-ph.CO]}
  \BibitemShut {NoStop}%
\bibitem [{\citenamefont {Biagetti}\ \emph {et~al.}(2018)\citenamefont
  {Biagetti}, \citenamefont {Franciolini}, \citenamefont {Kehagias},\ and\
  \citenamefont {Riotto}}]{Biagetti:2018pjj}%
  \BibitemOpen
  \bibfield  {author} {\bibinfo {author} {\bibfnamefont {M.}~\bibnamefont
  {Biagetti}}, \bibinfo {author} {\bibfnamefont {G.}~\bibnamefont
  {Franciolini}}, \bibinfo {author} {\bibfnamefont {A.}~\bibnamefont
  {Kehagias}},\ and\ \bibinfo {author} {\bibfnamefont {A.}~\bibnamefont
  {Riotto}},\ }\bibfield  {title} {\bibinfo {title} {{Primordial Black Holes
  from Inflation and Quantum Diffusion}},\ }\href
  {https://doi.org/10.1088/1475-7516/2018/07/032} {\bibfield  {journal}
  {\bibinfo  {journal} {JCAP}\ }\textbf {\bibinfo {volume} {07}},\ \bibinfo
  {pages} {032}},\ \Eprint {https://arxiv.org/abs/1804.07124} {arXiv:1804.07124
  [astro-ph.CO]} \BibitemShut {NoStop}%
\bibitem [{\citenamefont {Kohri}\ \emph {et~al.}(2008)\citenamefont {Kohri},
  \citenamefont {Lyth},\ and\ \citenamefont {Melchiorri}}]{Kohri:2007qn}%
  \BibitemOpen
  \bibfield  {author} {\bibinfo {author} {\bibfnamefont {K.}~\bibnamefont
  {Kohri}}, \bibinfo {author} {\bibfnamefont {D.~H.}\ \bibnamefont {Lyth}},\
  and\ \bibinfo {author} {\bibfnamefont {A.}~\bibnamefont {Melchiorri}},\
  }\bibfield  {title} {\bibinfo {title} {{Black hole formation and slow-roll
  inflation}},\ }\href {https://doi.org/10.1088/1475-7516/2008/04/038}
  {\bibfield  {journal} {\bibinfo  {journal} {JCAP}\ }\textbf {\bibinfo
  {volume} {04}},\ \bibinfo {pages} {038}},\ \Eprint
  {https://arxiv.org/abs/0711.5006} {arXiv:0711.5006 [hep-ph]} \BibitemShut
  {NoStop}%
\bibitem [{\citenamefont {Garcia-Bellido}\ \emph {et~al.}(1996)\citenamefont
  {Garcia-Bellido}, \citenamefont {Linde},\ and\ \citenamefont
  {Wands}}]{Garcia-Bellido:1996mdl}%
  \BibitemOpen
  \bibfield  {author} {\bibinfo {author} {\bibfnamefont {J.}~\bibnamefont
  {Garcia-Bellido}}, \bibinfo {author} {\bibfnamefont {A.~D.}\ \bibnamefont
  {Linde}},\ and\ \bibinfo {author} {\bibfnamefont {D.}~\bibnamefont {Wands}},\
  }\bibfield  {title} {\bibinfo {title} {{Density perturbations and black hole
  formation in hybrid inflation}},\ }\href
  {https://doi.org/10.1103/PhysRevD.54.6040} {\bibfield  {journal} {\bibinfo
  {journal} {Phys. Rev. D}\ }\textbf {\bibinfo {volume} {54}},\ \bibinfo
  {pages} {6040} (\bibinfo {year} {1996})},\ \Eprint
  {https://arxiv.org/abs/astro-ph/9605094} {arXiv:astro-ph/9605094}
  \BibitemShut {NoStop}%
\bibitem [{\citenamefont {Pi}\ and\ \citenamefont {Wang}(2023)}]{Pi:2022zxs}%
  \BibitemOpen
  \bibfield  {author} {\bibinfo {author} {\bibfnamefont {S.}~\bibnamefont
  {Pi}}\ and\ \bibinfo {author} {\bibfnamefont {J.}~\bibnamefont {Wang}},\
  }\bibfield  {title} {\bibinfo {title} {{Primordial black hole formation in
  Starobinsky's linear potential model}},\ }\href
  {https://doi.org/10.1088/1475-7516/2023/06/018} {\bibfield  {journal}
  {\bibinfo  {journal} {JCAP}\ }\textbf {\bibinfo {volume} {06}},\ \bibinfo
  {pages} {018}},\ \Eprint {https://arxiv.org/abs/2209.14183} {arXiv:2209.14183
  [astro-ph.CO]} \BibitemShut {NoStop}%
\bibitem [{\citenamefont {Zhou}\ \emph {et~al.}(2020)\citenamefont {Zhou},
  \citenamefont {Jiang}, \citenamefont {Cai}, \citenamefont {Sasaki},\ and\
  \citenamefont {Pi}}]{Zhou:2020kkf}%
  \BibitemOpen
  \bibfield  {author} {\bibinfo {author} {\bibfnamefont {Z.}~\bibnamefont
  {Zhou}}, \bibinfo {author} {\bibfnamefont {J.}~\bibnamefont {Jiang}},
  \bibinfo {author} {\bibfnamefont {Y.-F.}\ \bibnamefont {Cai}}, \bibinfo
  {author} {\bibfnamefont {M.}~\bibnamefont {Sasaki}},\ and\ \bibinfo {author}
  {\bibfnamefont {S.}~\bibnamefont {Pi}},\ }\bibfield  {title} {\bibinfo
  {title} {{Primordial black holes and gravitational waves from resonant
  amplification during inflation}},\ }\href
  {https://doi.org/10.1103/PhysRevD.102.103527} {\bibfield  {journal} {\bibinfo
   {journal} {Phys. Rev. D}\ }\textbf {\bibinfo {volume} {102}},\ \bibinfo
  {pages} {103527} (\bibinfo {year} {2020})},\ \Eprint
  {https://arxiv.org/abs/2010.03537} {arXiv:2010.03537 [astro-ph.CO]}
  \BibitemShut {NoStop}%
\bibitem [{\citenamefont {Cacciapaglia}\ \emph {et~al.}(2025)\citenamefont
  {Cacciapaglia}, \citenamefont {Cheong}, \citenamefont {Deandrea},
  \citenamefont {Isnard}, \citenamefont {Park}, \citenamefont {Wang},\ and\
  \citenamefont {Zhang}}]{Cacciapaglia:2025xqd}%
  \BibitemOpen
  \bibfield  {author} {\bibinfo {author} {\bibfnamefont {G.}~\bibnamefont
  {Cacciapaglia}}, \bibinfo {author} {\bibfnamefont {D.~Y.}\ \bibnamefont
  {Cheong}}, \bibinfo {author} {\bibfnamefont {A.}~\bibnamefont {Deandrea}},
  \bibinfo {author} {\bibfnamefont {W.}~\bibnamefont {Isnard}}, \bibinfo
  {author} {\bibfnamefont {S.~C.}\ \bibnamefont {Park}}, \bibinfo {author}
  {\bibfnamefont {X.}~\bibnamefont {Wang}},\ and\ \bibinfo {author}
  {\bibfnamefont {Y.-l.}\ \bibnamefont {Zhang}},\ }\bibfield  {title} {\bibinfo
  {title} {{Composite Hybrid Inflation : Primordial Black Holes and Stochastic
  Gravitational Waves}},\ }\href@noop {} {\  (\bibinfo {year} {2025})},\
  \Eprint {https://arxiv.org/abs/2506.06655} {arXiv:2506.06655 [hep-ph]}
  \BibitemShut {NoStop}%
\bibitem [{\citenamefont {Wang}\ \emph {et~al.}(2025)\citenamefont {Wang},
  \citenamefont {Sasaki},\ and\ \citenamefont {Zhang}}]{Wang:2025lti}%
  \BibitemOpen
  \bibfield  {author} {\bibinfo {author} {\bibfnamefont {X.}~\bibnamefont
  {Wang}}, \bibinfo {author} {\bibfnamefont {M.}~\bibnamefont {Sasaki}},\ and\
  \bibinfo {author} {\bibfnamefont {Y.-l.}\ \bibnamefont {Zhang}},\ }\bibfield
  {title} {\bibinfo {title} {{The Dual Primordial Black Hole Formation
  Scenario}},\ }\href@noop {} {\  (\bibinfo {year} {2025})},\ \Eprint
  {https://arxiv.org/abs/2505.09337} {arXiv:2505.09337 [astro-ph.CO]}
  \BibitemShut {NoStop}%
\bibitem [{\citenamefont {Kim}\ \emph {et~al.}(2025)\citenamefont {Kim},
  \citenamefont {Wang}, \citenamefont {Zhang},\ and\ \citenamefont
  {Ren}}]{Kim:2025dyi}%
  \BibitemOpen
  \bibfield  {author} {\bibinfo {author} {\bibfnamefont {J.}~\bibnamefont
  {Kim}}, \bibinfo {author} {\bibfnamefont {X.}~\bibnamefont {Wang}}, \bibinfo
  {author} {\bibfnamefont {Y.-l.}\ \bibnamefont {Zhang}},\ and\ \bibinfo
  {author} {\bibfnamefont {Z.}~\bibnamefont {Ren}},\ }\bibfield  {title}
  {\bibinfo {title} {{Enhancement of primordial curvature perturbations in R
  $^{3}$-corrected Starobinsky-Higgs inflation}},\ }\href
  {https://doi.org/10.1088/1475-7516/2025/09/011} {\bibfield  {journal}
  {\bibinfo  {journal} {JCAP}\ }\textbf {\bibinfo {volume} {09}},\ \bibinfo
  {pages} {011}},\ \Eprint {https://arxiv.org/abs/2504.12035} {arXiv:2504.12035
  [astro-ph.CO]} \BibitemShut {NoStop}%
\bibitem [{\citenamefont {Cai}\ \emph {et~al.}(2018)\citenamefont {Cai},
  \citenamefont {Tong}, \citenamefont {Wang},\ and\ \citenamefont
  {Yan}}]{Cai:2018tuh}%
  \BibitemOpen
  \bibfield  {author} {\bibinfo {author} {\bibfnamefont {Y.-F.}\ \bibnamefont
  {Cai}}, \bibinfo {author} {\bibfnamefont {X.}~\bibnamefont {Tong}}, \bibinfo
  {author} {\bibfnamefont {D.-G.}\ \bibnamefont {Wang}},\ and\ \bibinfo
  {author} {\bibfnamefont {S.-F.}\ \bibnamefont {Yan}},\ }\bibfield  {title}
  {\bibinfo {title} {{Primordial Black Holes from Sound Speed Resonance during
  Inflation}},\ }\href {https://doi.org/10.1103/PhysRevLett.121.081306}
  {\bibfield  {journal} {\bibinfo  {journal} {Phys. Rev. Lett.}\ }\textbf
  {\bibinfo {volume} {121}},\ \bibinfo {pages} {081306} (\bibinfo {year}
  {2018})},\ \Eprint {https://arxiv.org/abs/1805.03639} {arXiv:1805.03639
  [astro-ph.CO]} \BibitemShut {NoStop}%
\bibitem [{\citenamefont {Kawasaki}\ \emph {et~al.}(1998)\citenamefont
  {Kawasaki}, \citenamefont {Sugiyama},\ and\ \citenamefont
  {Yanagida}}]{Kawasaki:1997ju}%
  \BibitemOpen
  \bibfield  {author} {\bibinfo {author} {\bibfnamefont {M.}~\bibnamefont
  {Kawasaki}}, \bibinfo {author} {\bibfnamefont {N.}~\bibnamefont {Sugiyama}},\
  and\ \bibinfo {author} {\bibfnamefont {T.}~\bibnamefont {Yanagida}},\
  }\bibfield  {title} {\bibinfo {title} {{Primordial black hole formation in a
  double inflation model in supergravity}},\ }\href
  {https://doi.org/10.1103/PhysRevD.57.6050} {\bibfield  {journal} {\bibinfo
  {journal} {Phys. Rev. D}\ }\textbf {\bibinfo {volume} {57}},\ \bibinfo
  {pages} {6050} (\bibinfo {year} {1998})},\ \Eprint
  {https://arxiv.org/abs/hep-ph/9710259} {arXiv:hep-ph/9710259} \BibitemShut
  {NoStop}%
\bibitem [{\citenamefont {Lyth}\ and\ \citenamefont
  {Wands}(2002)}]{Lyth:2001nq}%
  \BibitemOpen
  \bibfield  {author} {\bibinfo {author} {\bibfnamefont {D.~H.}\ \bibnamefont
  {Lyth}}\ and\ \bibinfo {author} {\bibfnamefont {D.}~\bibnamefont {Wands}},\
  }\bibfield  {title} {\bibinfo {title} {{Generating the curvature perturbation
  without an inflaton}},\ }\href
  {https://doi.org/10.1016/S0370-2693(01)01366-1} {\bibfield  {journal}
  {\bibinfo  {journal} {Phys. Lett. B}\ }\textbf {\bibinfo {volume} {524}},\
  \bibinfo {pages} {5} (\bibinfo {year} {2002})},\ \Eprint
  {https://arxiv.org/abs/hep-ph/0110002} {arXiv:hep-ph/0110002} \BibitemShut
  {NoStop}%
\bibitem [{\citenamefont {Kawasaki}\ \emph {et~al.}(2013)\citenamefont
  {Kawasaki}, \citenamefont {Kitajima},\ and\ \citenamefont
  {Yanagida}}]{Kawasaki:2012wr}%
  \BibitemOpen
  \bibfield  {author} {\bibinfo {author} {\bibfnamefont {M.}~\bibnamefont
  {Kawasaki}}, \bibinfo {author} {\bibfnamefont {N.}~\bibnamefont {Kitajima}},\
  and\ \bibinfo {author} {\bibfnamefont {T.~T.}\ \bibnamefont {Yanagida}},\
  }\bibfield  {title} {\bibinfo {title} {{Primordial black hole formation from
  an axionlike curvaton model}},\ }\href
  {https://doi.org/10.1103/PhysRevD.87.063519} {\bibfield  {journal} {\bibinfo
  {journal} {Phys. Rev. D}\ }\textbf {\bibinfo {volume} {87}},\ \bibinfo
  {pages} {063519} (\bibinfo {year} {2013})},\ \Eprint
  {https://arxiv.org/abs/1207.2550} {arXiv:1207.2550 [hep-ph]} \BibitemShut
  {NoStop}%
\bibitem [{\citenamefont {Kohri}\ \emph {et~al.}(2013)\citenamefont {Kohri},
  \citenamefont {Lin},\ and\ \citenamefont {Matsuda}}]{Kohri:2012yw}%
  \BibitemOpen
  \bibfield  {author} {\bibinfo {author} {\bibfnamefont {K.}~\bibnamefont
  {Kohri}}, \bibinfo {author} {\bibfnamefont {C.-M.}\ \bibnamefont {Lin}},\
  and\ \bibinfo {author} {\bibfnamefont {T.}~\bibnamefont {Matsuda}},\
  }\bibfield  {title} {\bibinfo {title} {{Primordial black holes from the
  inflating curvaton}},\ }\href {https://doi.org/10.1103/PhysRevD.87.103527}
  {\bibfield  {journal} {\bibinfo  {journal} {Phys. Rev. D}\ }\textbf {\bibinfo
  {volume} {87}},\ \bibinfo {pages} {103527} (\bibinfo {year} {2013})},\
  \Eprint {https://arxiv.org/abs/1211.2371} {arXiv:1211.2371 [hep-ph]}
  \BibitemShut {NoStop}%
\bibitem [{\citenamefont {Yokoyama}(1997)}]{Yokoyama:1995ex}%
  \BibitemOpen
  \bibfield  {author} {\bibinfo {author} {\bibfnamefont {J.}~\bibnamefont
  {Yokoyama}},\ }\bibfield  {title} {\bibinfo {title} {{Formation of MACHO
  primordial black holes in inflationary cosmology}},\ }\href@noop {}
  {\bibfield  {journal} {\bibinfo  {journal} {Astron. Astrophys.}\ }\textbf
  {\bibinfo {volume} {318}},\ \bibinfo {pages} {673} (\bibinfo {year}
  {1997})},\ \Eprint {https://arxiv.org/abs/astro-ph/9509027}
  {arXiv:astro-ph/9509027} \BibitemShut {NoStop}%
\bibitem [{\citenamefont {Kristiano}\ and\ \citenamefont
  {Yokoyama}(2024)}]{PhysRevLett.132.221003}%
  \BibitemOpen
  \bibfield  {author} {\bibinfo {author} {\bibfnamefont {J.}~\bibnamefont
  {Kristiano}}\ and\ \bibinfo {author} {\bibfnamefont {J.}~\bibnamefont
  {Yokoyama}},\ }\bibfield  {title} {\bibinfo {title} {Constraining primordial
  black hole formation from single-field inflation},\ }\href
  {https://doi.org/10.1103/PhysRevLett.132.221003} {\bibfield  {journal}
  {\bibinfo  {journal} {Phys. Rev. Lett.}\ }\textbf {\bibinfo {volume} {132}},\
  \bibinfo {pages} {221003} (\bibinfo {year} {2024})}\BibitemShut {NoStop}%
\bibitem [{\citenamefont {Bhattacharya}\ \emph {et~al.}(2024)\citenamefont
  {Bhattacharya}, \citenamefont {Choudhury}, \citenamefont {Dey}, \citenamefont
  {Ghosh}, \citenamefont {Karde},\ and\ \citenamefont
  {Mishra}}]{Bhattacharya:2023ysp}%
  \BibitemOpen
  \bibfield  {author} {\bibinfo {author} {\bibfnamefont {G.}~\bibnamefont
  {Bhattacharya}}, \bibinfo {author} {\bibfnamefont {S.}~\bibnamefont
  {Choudhury}}, \bibinfo {author} {\bibfnamefont {K.}~\bibnamefont {Dey}},
  \bibinfo {author} {\bibfnamefont {S.}~\bibnamefont {Ghosh}}, \bibinfo
  {author} {\bibfnamefont {A.}~\bibnamefont {Karde}},\ and\ \bibinfo {author}
  {\bibfnamefont {N.~S.}\ \bibnamefont {Mishra}},\ }\bibfield  {title}
  {\bibinfo {title} {{Evading no-go for PBH formation and production of SIGWs
  using Multiple Sharp Transitions in EFT of single field inflation}},\ }\href
  {https://doi.org/10.1016/j.dark.2024.101602} {\bibfield  {journal} {\bibinfo
  {journal} {Phys. Dark Univ.}\ }\textbf {\bibinfo {volume} {46}},\ \bibinfo
  {pages} {101602} (\bibinfo {year} {2024})},\ \Eprint
  {https://arxiv.org/abs/2309.00973} {arXiv:2309.00973 [astro-ph.CO]}
  \BibitemShut {NoStop}%
\bibitem [{\citenamefont {Sharma}\ \emph {et~al.}(2024)\citenamefont {Sharma},
  \citenamefont {Sami},\ and\ \citenamefont {Mota}}]{Sharma:2024whg}%
  \BibitemOpen
  \bibfield  {author} {\bibinfo {author} {\bibfnamefont {M.~K.}\ \bibnamefont
  {Sharma}}, \bibinfo {author} {\bibfnamefont {M.}~\bibnamefont {Sami}},\ and\
  \bibinfo {author} {\bibfnamefont {D.~F.}\ \bibnamefont {Mota}},\ }\bibfield
  {title} {\bibinfo {title} {{Generic predictions for primordial perturbations
  and their implications}},\ }\href
  {https://doi.org/10.1016/j.physletb.2024.138956} {\bibfield  {journal}
  {\bibinfo  {journal} {Phys. Lett. B}\ }\textbf {\bibinfo {volume} {856}},\
  \bibinfo {pages} {138956} (\bibinfo {year} {2024})},\ \Eprint
  {https://arxiv.org/abs/2401.11142} {arXiv:2401.11142 [astro-ph.CO]}
  \BibitemShut {NoStop}%
\bibitem [{\citenamefont {Gangopadhyay}\ \emph {et~al.}(2022)\citenamefont
  {Gangopadhyay}, \citenamefont {Jain}, \citenamefont {Sharma},\ and\
  \citenamefont {Yogesh}}]{Gangopadhyay:2021kmf}%
  \BibitemOpen
  \bibfield  {author} {\bibinfo {author} {\bibfnamefont {M.~R.}\ \bibnamefont
  {Gangopadhyay}}, \bibinfo {author} {\bibfnamefont {J.~C.}\ \bibnamefont
  {Jain}}, \bibinfo {author} {\bibfnamefont {D.}~\bibnamefont {Sharma}},\ and\
  \bibinfo {author} {\bibnamefont {Yogesh}},\ }\bibfield  {title} {\bibinfo
  {title} {{Production of primordial black holes via single field inflation and
  observational constraints}},\ }\href
  {https://doi.org/10.1140/epjc/s10052-022-10796-x} {\bibfield  {journal}
  {\bibinfo  {journal} {Eur. Phys. J. C}\ }\textbf {\bibinfo {volume} {82}},\
  \bibinfo {pages} {849} (\bibinfo {year} {2022})},\ \Eprint
  {https://arxiv.org/abs/2108.13839} {arXiv:2108.13839 [astro-ph.CO]}
  \BibitemShut {NoStop}%
\bibitem [{\citenamefont {Correa}\ \emph {et~al.}(2022)\citenamefont {Correa},
  \citenamefont {Gangopadhyay}, \citenamefont {Jaman},\ and\ \citenamefont
  {Mathews}}]{Correa:2022ngq}%
  \BibitemOpen
  \bibfield  {author} {\bibinfo {author} {\bibfnamefont {M.}~\bibnamefont
  {Correa}}, \bibinfo {author} {\bibfnamefont {M.~R.}\ \bibnamefont
  {Gangopadhyay}}, \bibinfo {author} {\bibfnamefont {N.}~\bibnamefont
  {Jaman}},\ and\ \bibinfo {author} {\bibfnamefont {G.~J.}\ \bibnamefont
  {Mathews}},\ }\bibfield  {title} {\bibinfo {title} {{Primordial black-hole
  dark matter via warm natural inflation}},\ }\href
  {https://doi.org/10.1016/j.physletb.2022.137510} {\bibfield  {journal}
  {\bibinfo  {journal} {Phys. Lett. B}\ }\textbf {\bibinfo {volume} {835}},\
  \bibinfo {pages} {137510} (\bibinfo {year} {2022})},\ \Eprint
  {https://arxiv.org/abs/2207.10394} {arXiv:2207.10394 [gr-qc]} \BibitemShut
  {NoStop}%
\bibitem [{\citenamefont {Braglia}\ \emph {et~al.}(2020)\citenamefont
  {Braglia}, \citenamefont {Hazra}, \citenamefont {Finelli}, \citenamefont
  {Smoot}, \citenamefont {Sriramkumar},\ and\ \citenamefont
  {Starobinsky}}]{Braglia:2020eai}%
  \BibitemOpen
  \bibfield  {author} {\bibinfo {author} {\bibfnamefont {M.}~\bibnamefont
  {Braglia}}, \bibinfo {author} {\bibfnamefont {D.~K.}\ \bibnamefont {Hazra}},
  \bibinfo {author} {\bibfnamefont {F.}~\bibnamefont {Finelli}}, \bibinfo
  {author} {\bibfnamefont {G.~F.}\ \bibnamefont {Smoot}}, \bibinfo {author}
  {\bibfnamefont {L.}~\bibnamefont {Sriramkumar}},\ and\ \bibinfo {author}
  {\bibfnamefont {A.~A.}\ \bibnamefont {Starobinsky}},\ }\bibfield  {title}
  {\bibinfo {title} {{Generating PBHs and small-scale GWs in two-field models
  of inflation}},\ }\href {https://doi.org/10.1088/1475-7516/2020/08/001}
  {\bibfield  {journal} {\bibinfo  {journal} {JCAP}\ }\textbf {\bibinfo
  {volume} {08}},\ \bibinfo {pages} {001}},\ \Eprint
  {https://arxiv.org/abs/2005.02895} {arXiv:2005.02895 [astro-ph.CO]}
  \BibitemShut {NoStop}%
\bibitem [{\citenamefont {Bhattacharya}\ and\ \citenamefont
  {Zavala}(2023)}]{Bhattacharya:2022fze}%
  \BibitemOpen
  \bibfield  {author} {\bibinfo {author} {\bibfnamefont {S.}~\bibnamefont
  {Bhattacharya}}\ and\ \bibinfo {author} {\bibfnamefont {I.}~\bibnamefont
  {Zavala}},\ }\bibfield  {title} {\bibinfo {title} {{Sharp turns in axion
  monodromy: primordial black holes and gravitational waves}},\ }\href
  {https://doi.org/10.1088/1475-7516/2023/04/065} {\bibfield  {journal}
  {\bibinfo  {journal} {JCAP}\ }\textbf {\bibinfo {volume} {04}},\ \bibinfo
  {pages} {065}},\ \Eprint {https://arxiv.org/abs/2205.06065} {arXiv:2205.06065
  [astro-ph.CO]} \BibitemShut {NoStop}%
\bibitem [{\citenamefont {Papanikolaou}\ \emph {et~al.}(2024)\citenamefont
  {Papanikolaou}, \citenamefont {Banerjee}, \citenamefont {Cai}, \citenamefont
  {Capozziello},\ and\ \citenamefont {Saridakis}}]{Papanikolaou:2024fzf}%
  \BibitemOpen
  \bibfield  {author} {\bibinfo {author} {\bibfnamefont {T.}~\bibnamefont
  {Papanikolaou}}, \bibinfo {author} {\bibfnamefont {S.}~\bibnamefont
  {Banerjee}}, \bibinfo {author} {\bibfnamefont {Y.-F.}\ \bibnamefont {Cai}},
  \bibinfo {author} {\bibfnamefont {S.}~\bibnamefont {Capozziello}},\ and\
  \bibinfo {author} {\bibfnamefont {E.~N.}\ \bibnamefont {Saridakis}},\
  }\bibfield  {title} {\bibinfo {title} {{Primordial black holes and induced
  gravitational waves in non-singular matter bouncing cosmology}},\ }\href
  {https://doi.org/10.1088/1475-7516/2024/06/066} {\bibfield  {journal}
  {\bibinfo  {journal} {JCAP}\ }\textbf {\bibinfo {volume} {06}},\ \bibinfo
  {pages} {066}},\ \Eprint {https://arxiv.org/abs/2404.03779} {arXiv:2404.03779
  [gr-qc]} \BibitemShut {NoStop}%
\bibitem [{\citenamefont {Teimoori}\ \emph {et~al.}(2021)\citenamefont
  {Teimoori}, \citenamefont {Rezazadeh},\ and\ \citenamefont
  {Karami}}]{Teimoori:2021thk}%
  \BibitemOpen
  \bibfield  {author} {\bibinfo {author} {\bibfnamefont {Z.}~\bibnamefont
  {Teimoori}}, \bibinfo {author} {\bibfnamefont {K.}~\bibnamefont
  {Rezazadeh}},\ and\ \bibinfo {author} {\bibfnamefont {K.}~\bibnamefont
  {Karami}},\ }\bibfield  {title} {\bibinfo {title} {{Primordial Black Holes
  Formation and Secondary Gravitational Waves in Nonminimal Derivative Coupling
  Inflation}},\ }\href {https://doi.org/10.3847/1538-4357/ac01cf} {\bibfield
  {journal} {\bibinfo  {journal} {Astrophys. J.}\ }\textbf {\bibinfo {volume}
  {915}},\ \bibinfo {pages} {118} (\bibinfo {year} {2021})},\ \Eprint
  {https://arxiv.org/abs/2107.08048} {arXiv:2107.08048 [gr-qc]} \BibitemShut
  {NoStop}%
\bibitem [{\citenamefont {Solbi}\ and\ \citenamefont
  {Karami}(2021)}]{Solbi:2021rse}%
  \BibitemOpen
  \bibfield  {author} {\bibinfo {author} {\bibfnamefont {M.}~\bibnamefont
  {Solbi}}\ and\ \bibinfo {author} {\bibfnamefont {K.}~\bibnamefont {Karami}},\
  }\bibfield  {title} {\bibinfo {title} {{Primordial black holes formation in
  the inflationary model with field-dependent kinetic term for quartic and
  natural potentials}},\ }\href
  {https://doi.org/10.1140/epjc/s10052-021-09690-9} {\bibfield  {journal}
  {\bibinfo  {journal} {Eur. Phys. J. C}\ }\textbf {\bibinfo {volume} {81}},\
  \bibinfo {pages} {884} (\bibinfo {year} {2021})},\ \Eprint
  {https://arxiv.org/abs/2106.02863} {arXiv:2106.02863 [astro-ph.CO]}
  \BibitemShut {NoStop}%
\bibitem [{\citenamefont {Heydari}\ and\ \citenamefont
  {Karami}(2024)}]{Heydari:2023rmq}%
  \BibitemOpen
  \bibfield  {author} {\bibinfo {author} {\bibfnamefont {S.}~\bibnamefont
  {Heydari}}\ and\ \bibinfo {author} {\bibfnamefont {K.}~\bibnamefont
  {Karami}},\ }\bibfield  {title} {\bibinfo {title} {{Primordial black holes
  and secondary gravitational waves from generalized power-law non-canonical
  inflation with quartic potential}},\ }\href
  {https://doi.org/10.1140/epjc/s10052-024-12489-z} {\bibfield  {journal}
  {\bibinfo  {journal} {Eur. Phys. J. C}\ }\textbf {\bibinfo {volume} {84}},\
  \bibinfo {pages} {127} (\bibinfo {year} {2024})},\ \Eprint
  {https://arxiv.org/abs/2310.11030} {arXiv:2310.11030 [gr-qc]} \BibitemShut
  {NoStop}%
\bibitem [{\citenamefont {Kawai}\ and\ \citenamefont
  {Kim}(2021)}]{Kawai:2021edk}%
  \BibitemOpen
  \bibfield  {author} {\bibinfo {author} {\bibfnamefont {S.}~\bibnamefont
  {Kawai}}\ and\ \bibinfo {author} {\bibfnamefont {J.}~\bibnamefont {Kim}},\
  }\bibfield  {title} {\bibinfo {title} {{Primordial black holes from
  Gauss-Bonnet-corrected single field inflation}},\ }\href
  {https://doi.org/10.1103/PhysRevD.104.083545} {\bibfield  {journal} {\bibinfo
   {journal} {Phys. Rev. D}\ }\textbf {\bibinfo {volume} {104}},\ \bibinfo
  {pages} {083545} (\bibinfo {year} {2021})},\ \Eprint
  {https://arxiv.org/abs/2108.01340} {arXiv:2108.01340 [astro-ph.CO]}
  \BibitemShut {NoStop}%
\bibitem [{\citenamefont {Ragavendra}\ \emph {et~al.}(2021)\citenamefont
  {Ragavendra}, \citenamefont {Saha}, \citenamefont {Sriramkumar},\ and\
  \citenamefont {Silk}}]{Ragavendra:2020sop}%
  \BibitemOpen
  \bibfield  {author} {\bibinfo {author} {\bibfnamefont {H.~V.}\ \bibnamefont
  {Ragavendra}}, \bibinfo {author} {\bibfnamefont {P.}~\bibnamefont {Saha}},
  \bibinfo {author} {\bibfnamefont {L.}~\bibnamefont {Sriramkumar}},\ and\
  \bibinfo {author} {\bibfnamefont {J.}~\bibnamefont {Silk}},\ }\bibfield
  {title} {\bibinfo {title} {{Primordial black holes and secondary
  gravitational waves from ultraslow roll and punctuated inflation}},\ }\href
  {https://doi.org/10.1103/PhysRevD.103.083510} {\bibfield  {journal} {\bibinfo
   {journal} {Phys. Rev. D}\ }\textbf {\bibinfo {volume} {103}},\ \bibinfo
  {pages} {083510} (\bibinfo {year} {2021})},\ \Eprint
  {https://arxiv.org/abs/2008.12202} {arXiv:2008.12202 [astro-ph.CO]}
  \BibitemShut {NoStop}%
\bibitem [{\citenamefont {Ragavendra}\ and\ \citenamefont
  {Sriramkumar}(2023)}]{Ragavendra:2023ret}%
  \BibitemOpen
  \bibfield  {author} {\bibinfo {author} {\bibfnamefont {H.~V.}\ \bibnamefont
  {Ragavendra}}\ and\ \bibinfo {author} {\bibfnamefont {L.}~\bibnamefont
  {Sriramkumar}},\ }\bibfield  {title} {\bibinfo {title} {{Observational
  Imprints of Enhanced Scalar Power on Small Scales in Ultra Slow Roll
  Inflation and Associated Non-Gaussianities}},\ }\href
  {https://doi.org/10.3390/galaxies11010034} {\bibfield  {journal} {\bibinfo
  {journal} {Galaxies}\ }\textbf {\bibinfo {volume} {11}},\ \bibinfo {pages}
  {34} (\bibinfo {year} {2023})},\ \Eprint {https://arxiv.org/abs/2301.08887}
  {arXiv:2301.08887 [astro-ph.CO]} \BibitemShut {NoStop}%
\bibitem [{\citenamefont {Choudhury}\ \emph {et~al.}(2024)\citenamefont
  {Choudhury}, \citenamefont {Gangopadhyay},\ and\ \citenamefont
  {Sami}}]{Choudhury:2023vuj}%
  \BibitemOpen
  \bibfield  {author} {\bibinfo {author} {\bibfnamefont {S.}~\bibnamefont
  {Choudhury}}, \bibinfo {author} {\bibfnamefont {M.~R.}\ \bibnamefont
  {Gangopadhyay}},\ and\ \bibinfo {author} {\bibfnamefont {M.}~\bibnamefont
  {Sami}},\ }\bibfield  {title} {\bibinfo {title} {{No-go for the formation of
  heavy mass Primordial Black Holes in Single Field Inflation}},\ }\href
  {https://doi.org/10.1140/epjc/s10052-024-13218-2} {\bibfield  {journal}
  {\bibinfo  {journal} {Eur. Phys. J. C}\ }\textbf {\bibinfo {volume} {84}},\
  \bibinfo {pages} {884} (\bibinfo {year} {2024})},\ \Eprint
  {https://arxiv.org/abs/2301.10000} {arXiv:2301.10000 [astro-ph.CO]}
  \BibitemShut {NoStop}%
\bibitem [{\citenamefont {Riotto}(2023)}]{Riotto:2023hoz}%
  \BibitemOpen
  \bibfield  {author} {\bibinfo {author} {\bibfnamefont {A.}~\bibnamefont
  {Riotto}},\ }\bibfield  {title} {\bibinfo {title} {{The Primordial Black Hole
  Formation from Single-Field Inflation is Not Ruled Out}},\ }\href@noop {} {\
  (\bibinfo {year} {2023})},\ \Eprint {https://arxiv.org/abs/2301.00599}
  {arXiv:2301.00599 [astro-ph.CO]} \BibitemShut {NoStop}%
\bibitem [{\citenamefont {Choudhury}\ \emph {et~al.}(2023)\citenamefont
  {Choudhury}, \citenamefont {Panda},\ and\ \citenamefont
  {Sami}}]{Choudhury:2023hvf}%
  \BibitemOpen
  \bibfield  {author} {\bibinfo {author} {\bibfnamefont {S.}~\bibnamefont
  {Choudhury}}, \bibinfo {author} {\bibfnamefont {S.}~\bibnamefont {Panda}},\
  and\ \bibinfo {author} {\bibfnamefont {M.}~\bibnamefont {Sami}},\ }\bibfield
  {title} {\bibinfo {title} {{Galileon inflation evades the no-go for PBH
  formation in the single-field framework}},\ }\href
  {https://doi.org/10.1088/1475-7516/2023/08/078} {\bibfield  {journal}
  {\bibinfo  {journal} {JCAP}\ }\textbf {\bibinfo {volume} {08}},\ \bibinfo
  {pages} {078}},\ \Eprint {https://arxiv.org/abs/2304.04065} {arXiv:2304.04065
  [astro-ph.CO]} \BibitemShut {NoStop}%
\bibitem [{\citenamefont {Harada}\ \emph {et~al.}(2013)\citenamefont {Harada},
  \citenamefont {Yoo},\ and\ \citenamefont {Kohri}}]{Harada:2013epa}%
  \BibitemOpen
  \bibfield  {author} {\bibinfo {author} {\bibfnamefont {T.}~\bibnamefont
  {Harada}}, \bibinfo {author} {\bibfnamefont {C.-M.}\ \bibnamefont {Yoo}},\
  and\ \bibinfo {author} {\bibfnamefont {K.}~\bibnamefont {Kohri}},\ }\bibfield
   {title} {\bibinfo {title} {{Threshold of primordial black hole formation}},\
  }\href {https://doi.org/10.1103/PhysRevD.88.084051} {\bibfield  {journal}
  {\bibinfo  {journal} {Phys. Rev. D}\ }\textbf {\bibinfo {volume} {88}},\
  \bibinfo {pages} {084051} (\bibinfo {year} {2013})},\ \bibinfo {note}
  {[Erratum: Phys.Rev.D 89, 029903 (2014)]},\ \Eprint
  {https://arxiv.org/abs/1309.4201} {arXiv:1309.4201 [astro-ph.CO]}
  \BibitemShut {NoStop}%
\bibitem [{\citenamefont {Escriv\`a}\ \emph {et~al.}(2021)\citenamefont
  {Escriv\`a}, \citenamefont {Germani},\ and\ \citenamefont
  {Sheth}}]{Escriva:2020tak}%
  \BibitemOpen
  \bibfield  {author} {\bibinfo {author} {\bibfnamefont {A.}~\bibnamefont
  {Escriv\`a}}, \bibinfo {author} {\bibfnamefont {C.}~\bibnamefont {Germani}},\
  and\ \bibinfo {author} {\bibfnamefont {R.~K.}\ \bibnamefont {Sheth}},\
  }\bibfield  {title} {\bibinfo {title} {{Analytical thresholds for black hole
  formation in general cosmological backgrounds}},\ }\href
  {https://doi.org/10.1088/1475-7516/2021/01/030} {\bibfield  {journal}
  {\bibinfo  {journal} {JCAP}\ }\textbf {\bibinfo {volume} {01}},\ \bibinfo
  {pages} {030}},\ \Eprint {https://arxiv.org/abs/2007.05564} {arXiv:2007.05564
  [gr-qc]} \BibitemShut {NoStop}%
\bibitem [{\citenamefont {{Press}}\ and\ \citenamefont
  {{Schechter}}(1974)}]{1974ApJ...187..425P}%
  \BibitemOpen
  \bibfield  {author} {\bibinfo {author} {\bibfnamefont {W.~H.}\ \bibnamefont
  {{Press}}}\ and\ \bibinfo {author} {\bibfnamefont {P.}~\bibnamefont
  {{Schechter}}},\ }\bibfield  {title} {\bibinfo {title} {{Formation of
  Galaxies and Clusters of Galaxies by Self-Similar Gravitational
  Condensation}},\ }\href {https://doi.org/10.1086/152650} {\bibfield
  {journal} {\bibinfo  {journal} {\apj}\ }\textbf {\bibinfo {volume} {187}},\
  \bibinfo {pages} {425} (\bibinfo {year} {1974})}\BibitemShut {NoStop}%
\bibitem [{\citenamefont {Shibata}\ and\ \citenamefont
  {Sasaki}(1999)}]{Shibata:1999zs}%
  \BibitemOpen
  \bibfield  {author} {\bibinfo {author} {\bibfnamefont {M.}~\bibnamefont
  {Shibata}}\ and\ \bibinfo {author} {\bibfnamefont {M.}~\bibnamefont
  {Sasaki}},\ }\bibfield  {title} {\bibinfo {title} {{Black hole formation in
  the Friedmann universe: Formulation and computation in numerical
  relativity}},\ }\href {https://doi.org/10.1103/PhysRevD.60.084002} {\bibfield
   {journal} {\bibinfo  {journal} {Phys. Rev. D}\ }\textbf {\bibinfo {volume}
  {60}},\ \bibinfo {pages} {084002} (\bibinfo {year} {1999})},\ \Eprint
  {https://arxiv.org/abs/gr-qc/9905064} {arXiv:gr-qc/9905064} \BibitemShut
  {NoStop}%
\bibitem [{\citenamefont {Escriv\`a}\ \emph {et~al.}(2022)\citenamefont
  {Escriv\`a}, \citenamefont {Kuhnel},\ and\ \citenamefont
  {Tada}}]{Escriva:2022duf}%
  \BibitemOpen
  \bibfield  {author} {\bibinfo {author} {\bibfnamefont {A.}~\bibnamefont
  {Escriv\`a}}, \bibinfo {author} {\bibfnamefont {F.}~\bibnamefont {Kuhnel}},\
  and\ \bibinfo {author} {\bibfnamefont {Y.}~\bibnamefont {Tada}},\ }\bibfield
  {title} {\bibinfo {title} {{Primordial Black Holes}}\ }\href
  {https://doi.org/10.1016/B978-0-32-395636-9.00012-8}
  {10.1016/B978-0-32-395636-9.00012-8} (\bibinfo {year} {2022}),\ \Eprint
  {https://arxiv.org/abs/2211.05767} {arXiv:2211.05767 [astro-ph.CO]}
  \BibitemShut {NoStop}%
\bibitem [{\citenamefont {Harada}\ \emph {et~al.}(2023)\citenamefont {Harada},
  \citenamefont {Yoo},\ and\ \citenamefont {Koga}}]{Harada:2023ffo}%
  \BibitemOpen
  \bibfield  {author} {\bibinfo {author} {\bibfnamefont {T.}~\bibnamefont
  {Harada}}, \bibinfo {author} {\bibfnamefont {C.-M.}\ \bibnamefont {Yoo}},\
  and\ \bibinfo {author} {\bibfnamefont {Y.}~\bibnamefont {Koga}},\ }\bibfield
  {title} {\bibinfo {title} {{Revisiting compaction functions for primordial
  black hole formation}},\ }\href {https://doi.org/10.1103/PhysRevD.108.043515}
  {\bibfield  {journal} {\bibinfo  {journal} {Phys. Rev. D}\ }\textbf {\bibinfo
  {volume} {108}},\ \bibinfo {pages} {043515} (\bibinfo {year} {2023})},\
  \Eprint {https://arxiv.org/abs/2304.13284} {arXiv:2304.13284 [gr-qc]}
  \BibitemShut {NoStop}%
\bibitem [{\citenamefont {Yoo}\ \emph {et~al.}(2018)\citenamefont {Yoo},
  \citenamefont {Harada}, \citenamefont {Garriga},\ and\ \citenamefont
  {Kohri}}]{Yoo:2018kvb}%
  \BibitemOpen
  \bibfield  {author} {\bibinfo {author} {\bibfnamefont {C.-M.}\ \bibnamefont
  {Yoo}}, \bibinfo {author} {\bibfnamefont {T.}~\bibnamefont {Harada}},
  \bibinfo {author} {\bibfnamefont {J.}~\bibnamefont {Garriga}},\ and\ \bibinfo
  {author} {\bibfnamefont {K.}~\bibnamefont {Kohri}},\ }\bibfield  {title}
  {\bibinfo {title} {{Primordial black hole abundance from random Gaussian
  curvature perturbations and a local density threshold}},\ }\href
  {https://doi.org/10.1093/ptep/pty120} {\bibfield  {journal} {\bibinfo
  {journal} {PTEP}\ }\textbf {\bibinfo {volume} {2018}},\ \bibinfo {pages}
  {123E01} (\bibinfo {year} {2018})},\ \bibinfo {note} {[Erratum: PTEP 2024,
  049202 (2024)]},\ \Eprint {https://arxiv.org/abs/1805.03946}
  {arXiv:1805.03946 [astro-ph.CO]} \BibitemShut {NoStop}%
\bibitem [{\citenamefont {Atal}\ \emph {et~al.}(2019)\citenamefont {Atal},
  \citenamefont {Garriga},\ and\ \citenamefont
  {Marcos-Caballero}}]{Atal:2019cdz}%
  \BibitemOpen
  \bibfield  {author} {\bibinfo {author} {\bibfnamefont {V.}~\bibnamefont
  {Atal}}, \bibinfo {author} {\bibfnamefont {J.}~\bibnamefont {Garriga}},\ and\
  \bibinfo {author} {\bibfnamefont {A.}~\bibnamefont {Marcos-Caballero}},\
  }\bibfield  {title} {\bibinfo {title} {{Primordial black hole formation with
  non-Gaussian curvature perturbations}},\ }\href
  {https://doi.org/10.1088/1475-7516/2019/09/073} {\bibfield  {journal}
  {\bibinfo  {journal} {JCAP}\ }\textbf {\bibinfo {volume} {09}},\ \bibinfo
  {pages} {073}},\ \Eprint {https://arxiv.org/abs/1905.13202} {arXiv:1905.13202
  [astro-ph.CO]} \BibitemShut {NoStop}%
\bibitem [{\citenamefont {Yoo}\ \emph {et~al.}(2019)\citenamefont {Yoo},
  \citenamefont {Gong},\ and\ \citenamefont {Yokoyama}}]{Yoo:2019pma}%
  \BibitemOpen
  \bibfield  {author} {\bibinfo {author} {\bibfnamefont {C.-M.}\ \bibnamefont
  {Yoo}}, \bibinfo {author} {\bibfnamefont {J.-O.}\ \bibnamefont {Gong}},\ and\
  \bibinfo {author} {\bibfnamefont {S.}~\bibnamefont {Yokoyama}},\ }\bibfield
  {title} {\bibinfo {title} {{Abundance of primordial black holes with local
  non-Gaussianity in peak theory}},\ }\href
  {https://doi.org/10.1088/1475-7516/2019/09/033} {\bibfield  {journal}
  {\bibinfo  {journal} {JCAP}\ }\textbf {\bibinfo {volume} {09}},\ \bibinfo
  {pages} {033}},\ \Eprint {https://arxiv.org/abs/1906.06790} {arXiv:1906.06790
  [astro-ph.CO]} \BibitemShut {NoStop}%
\bibitem [{\citenamefont {Atal}\ \emph {et~al.}(2020)\citenamefont {Atal},
  \citenamefont {Cid}, \citenamefont {Escriv{\`a}},\ and\ \citenamefont
  {Garriga}}]{Atal:2019erb}%
  \BibitemOpen
  \bibfield  {author} {\bibinfo {author} {\bibfnamefont {V.}~\bibnamefont
  {Atal}}, \bibinfo {author} {\bibfnamefont {J.}~\bibnamefont {Cid}}, \bibinfo
  {author} {\bibfnamefont {A.}~\bibnamefont {Escriv{\`a}}},\ and\ \bibinfo
  {author} {\bibfnamefont {J.}~\bibnamefont {Garriga}},\ }\bibfield  {title}
  {\bibinfo {title} {{PBH in single field inflation: the effect of shape
  dispersion and non-Gaussianities}},\ }\href
  {https://doi.org/10.1088/1475-7516/2020/05/022} {\bibfield  {journal}
  {\bibinfo  {journal} {JCAP}\ }\textbf {\bibinfo {volume} {05}},\ \bibinfo
  {pages} {022}},\ \Eprint {https://arxiv.org/abs/1908.11357} {arXiv:1908.11357
  [astro-ph.CO]} \BibitemShut {NoStop}%
\bibitem [{\citenamefont {Germani}\ and\ \citenamefont
  {Sheth}(2020)}]{Germani:2019zez}%
  \BibitemOpen
  \bibfield  {author} {\bibinfo {author} {\bibfnamefont {C.}~\bibnamefont
  {Germani}}\ and\ \bibinfo {author} {\bibfnamefont {R.~K.}\ \bibnamefont
  {Sheth}},\ }\bibfield  {title} {\bibinfo {title} {{Nonlinear statistics of
  primordial black holes from Gaussian curvature perturbations}},\ }\href
  {https://doi.org/10.1103/PhysRevD.101.063520} {\bibfield  {journal} {\bibinfo
   {journal} {Phys. Rev. D}\ }\textbf {\bibinfo {volume} {101}},\ \bibinfo
  {pages} {063520} (\bibinfo {year} {2020})},\ \Eprint
  {https://arxiv.org/abs/1912.07072} {arXiv:1912.07072 [astro-ph.CO]}
  \BibitemShut {NoStop}%
\bibitem [{\citenamefont {Young}(2022)}]{Young:2022phe}%
  \BibitemOpen
  \bibfield  {author} {\bibinfo {author} {\bibfnamefont {S.}~\bibnamefont
  {Young}},\ }\bibfield  {title} {\bibinfo {title} {{Peaks and primordial black
  holes: the~effect of non-Gaussianity}},\ }\href
  {https://doi.org/10.1088/1475-7516/2022/05/037} {\bibfield  {journal}
  {\bibinfo  {journal} {JCAP}\ }\textbf {\bibinfo {volume} {05}}\bibfield
  {number} {\bibinfo  {number} { (05)},\ \bibinfo {pages} {037}},\ }\Eprint
  {https://arxiv.org/abs/2201.13345} {arXiv:2201.13345 [astro-ph.CO]}
  \BibitemShut {NoStop}%
\bibitem [{\citenamefont {Yoo}\ \emph {et~al.}(2021)\citenamefont {Yoo},
  \citenamefont {Harada}, \citenamefont {Hirano},\ and\ \citenamefont
  {Kohri}}]{Yoo:2020dkz}%
  \BibitemOpen
  \bibfield  {author} {\bibinfo {author} {\bibfnamefont {C.-M.}\ \bibnamefont
  {Yoo}}, \bibinfo {author} {\bibfnamefont {T.}~\bibnamefont {Harada}},
  \bibinfo {author} {\bibfnamefont {S.}~\bibnamefont {Hirano}},\ and\ \bibinfo
  {author} {\bibfnamefont {K.}~\bibnamefont {Kohri}},\ }\bibfield  {title}
  {\bibinfo {title} {{Abundance of Primordial Black Holes in Peak Theory for an
  Arbitrary Power Spectrum}},\ }\href {https://doi.org/10.1093/ptep/ptaa155}
  {\bibfield  {journal} {\bibinfo  {journal} {PTEP}\ }\textbf {\bibinfo
  {volume} {2021}},\ \bibinfo {pages} {013E02} (\bibinfo {year} {2021})},\
  \bibinfo {note} {[Erratum: PTEP 2024, 049203 (2024)]},\ \Eprint
  {https://arxiv.org/abs/2008.02425} {arXiv:2008.02425 [astro-ph.CO]}
  \BibitemShut {NoStop}%
\bibitem [{\citenamefont {Kitajima}\ \emph {et~al.}(2021)\citenamefont
  {Kitajima}, \citenamefont {Tada}, \citenamefont {Yokoyama},\ and\
  \citenamefont {Yoo}}]{Kitajima:2021fpq}%
  \BibitemOpen
  \bibfield  {author} {\bibinfo {author} {\bibfnamefont {N.}~\bibnamefont
  {Kitajima}}, \bibinfo {author} {\bibfnamefont {Y.}~\bibnamefont {Tada}},
  \bibinfo {author} {\bibfnamefont {S.}~\bibnamefont {Yokoyama}},\ and\
  \bibinfo {author} {\bibfnamefont {C.-M.}\ \bibnamefont {Yoo}},\ }\bibfield
  {title} {\bibinfo {title} {{Primordial black holes in peak theory with a
  non-Gaussian tail}},\ }\href {https://doi.org/10.1088/1475-7516/2021/10/053}
  {\bibfield  {journal} {\bibinfo  {journal} {JCAP}\ }\textbf {\bibinfo
  {volume} {10}},\ \bibinfo {pages} {053}},\ \Eprint
  {https://arxiv.org/abs/2109.00791} {arXiv:2109.00791 [astro-ph.CO]}
  \BibitemShut {NoStop}%
\bibitem [{\citenamefont {De~Luca}\ \emph {et~al.}(2019)\citenamefont
  {De~Luca}, \citenamefont {Desjacques}, \citenamefont {Franciolini},
  \citenamefont {Malhotra},\ and\ \citenamefont {Riotto}}]{DeLuca:2019buf}%
  \BibitemOpen
  \bibfield  {author} {\bibinfo {author} {\bibfnamefont {V.}~\bibnamefont
  {De~Luca}}, \bibinfo {author} {\bibfnamefont {V.}~\bibnamefont {Desjacques}},
  \bibinfo {author} {\bibfnamefont {G.}~\bibnamefont {Franciolini}}, \bibinfo
  {author} {\bibfnamefont {A.}~\bibnamefont {Malhotra}},\ and\ \bibinfo
  {author} {\bibfnamefont {A.}~\bibnamefont {Riotto}},\ }\bibfield  {title}
  {\bibinfo {title} {{The initial spin probability distribution of primordial
  black holes}},\ }\href {https://doi.org/10.1088/1475-7516/2019/05/018}
  {\bibfield  {journal} {\bibinfo  {journal} {JCAP}\ }\textbf {\bibinfo
  {volume} {05}},\ \bibinfo {pages} {018}},\ \Eprint
  {https://arxiv.org/abs/1903.01179} {arXiv:1903.01179 [astro-ph.CO]}
  \BibitemShut {NoStop}%
\bibitem [{\citenamefont {Harada}\ \emph {et~al.}(2021)\citenamefont {Harada},
  \citenamefont {Yoo}, \citenamefont {Kohri}, \citenamefont {Koga},\ and\
  \citenamefont {Monobe}}]{Harada:2020pzb}%
  \BibitemOpen
  \bibfield  {author} {\bibinfo {author} {\bibfnamefont {T.}~\bibnamefont
  {Harada}}, \bibinfo {author} {\bibfnamefont {C.-M.}\ \bibnamefont {Yoo}},
  \bibinfo {author} {\bibfnamefont {K.}~\bibnamefont {Kohri}}, \bibinfo
  {author} {\bibfnamefont {Y.}~\bibnamefont {Koga}},\ and\ \bibinfo {author}
  {\bibfnamefont {T.}~\bibnamefont {Monobe}},\ }\bibfield  {title} {\bibinfo
  {title} {{Spins of primordial black holes formed in the radiation-dominated
  phase of the universe: first-order effect}},\ }\href
  {https://doi.org/10.3847/1538-4357/abd9b9} {\bibfield  {journal} {\bibinfo
  {journal} {Astrophys. J.}\ }\textbf {\bibinfo {volume} {908}},\ \bibinfo
  {pages} {140} (\bibinfo {year} {2021})},\ \Eprint
  {https://arxiv.org/abs/2011.00710} {arXiv:2011.00710 [astro-ph.CO]}
  \BibitemShut {NoStop}%
\bibitem [{\citenamefont {Harada}(2024)}]{Harada:2024jxl}%
  \BibitemOpen
  \bibfield  {author} {\bibinfo {author} {\bibfnamefont {T.}~\bibnamefont
  {Harada}},\ }\bibfield  {title} {\bibinfo {title} {{Primordial Black Holes:
  Formation, Spin and Type II}},\ }\href
  {https://doi.org/10.3390/universe10120444} {\bibfield  {journal} {\bibinfo
  {journal} {Universe}\ }\textbf {\bibinfo {volume} {10}},\ \bibinfo {pages}
  {444} (\bibinfo {year} {2024})},\ \Eprint {https://arxiv.org/abs/2409.01934}
  {arXiv:2409.01934 [gr-qc]} \BibitemShut {NoStop}%
\bibitem [{\citenamefont {Banerjee}\ and\ \citenamefont
  {Harada}(2025)}]{Banerjee:2024nkv}%
  \BibitemOpen
  \bibfield  {author} {\bibinfo {author} {\bibfnamefont {I.~K.}\ \bibnamefont
  {Banerjee}}\ and\ \bibinfo {author} {\bibfnamefont {T.}~\bibnamefont
  {Harada}},\ }\bibfield  {title} {\bibinfo {title} {{Spin of primordial black
  holes from broad power spectrum: radiation dominated universe}},\ }\href
  {https://doi.org/10.1088/1475-7516/2025/05/010} {\bibfield  {journal}
  {\bibinfo  {journal} {JCAP}\ }\textbf {\bibinfo {volume} {05}},\ \bibinfo
  {pages} {010}},\ \Eprint {https://arxiv.org/abs/2409.06494} {arXiv:2409.06494
  [gr-qc]} \BibitemShut {NoStop}%
\bibitem [{\citenamefont {Chongchitnan}\ and\ \citenamefont
  {Silk}(2021)}]{Chongchitnan:2021ehn}%
  \BibitemOpen
  \bibfield  {author} {\bibinfo {author} {\bibfnamefont {S.}~\bibnamefont
  {Chongchitnan}}\ and\ \bibinfo {author} {\bibfnamefont {J.}~\bibnamefont
  {Silk}},\ }\bibfield  {title} {\bibinfo {title} {{Extreme-value statistics of
  the spin of primordial black holes}},\ }\href
  {https://doi.org/10.1103/PhysRevD.104.083018} {\bibfield  {journal} {\bibinfo
   {journal} {Phys. Rev. D}\ }\textbf {\bibinfo {volume} {104}},\ \bibinfo
  {pages} {083018} (\bibinfo {year} {2021})},\ \Eprint
  {https://arxiv.org/abs/2109.12268} {arXiv:2109.12268 [astro-ph.CO]}
  \BibitemShut {NoStop}%
\bibitem [{\citenamefont {Eroshenko}(2021)}]{Eroshenko:2021sez}%
  \BibitemOpen
  \bibfield  {author} {\bibinfo {author} {\bibfnamefont {Y.~N.}\ \bibnamefont
  {Eroshenko}},\ }\bibfield  {title} {\bibinfo {title} {{Spin of primordial
  black holes in the model with collapsing domain walls}},\ }\href
  {https://doi.org/10.1088/1475-7516/2021/12/041} {\bibfield  {journal}
  {\bibinfo  {journal} {JCAP}\ }\textbf {\bibinfo {volume} {12}}\bibfield
  {number} {\bibinfo  {number} { (12)},\ \bibinfo {pages} {041}},\ }\Eprint
  {https://arxiv.org/abs/2111.03403} {arXiv:2111.03403 [astro-ph.CO]}
  \BibitemShut {NoStop}%
\bibitem [{\citenamefont {Mirbabayi}\ \emph {et~al.}(2020)\citenamefont
  {Mirbabayi}, \citenamefont {Gruzinov},\ and\ \citenamefont
  {Nore{\~n}a}}]{Mirbabayi:2019uph}%
  \BibitemOpen
  \bibfield  {author} {\bibinfo {author} {\bibfnamefont {M.}~\bibnamefont
  {Mirbabayi}}, \bibinfo {author} {\bibfnamefont {A.}~\bibnamefont
  {Gruzinov}},\ and\ \bibinfo {author} {\bibfnamefont {J.}~\bibnamefont
  {Nore{\~n}a}},\ }\bibfield  {title} {\bibinfo {title} {{Spin of Primordial
  Black Holes}},\ }\href {https://doi.org/10.1088/1475-7516/2020/03/017}
  {\bibfield  {journal} {\bibinfo  {journal} {JCAP}\ }\textbf {\bibinfo
  {volume} {03}},\ \bibinfo {pages} {017}},\ \Eprint
  {https://arxiv.org/abs/1901.05963} {arXiv:1901.05963 [astro-ph.CO]}
  \BibitemShut {NoStop}%
\bibitem [{\citenamefont {Chiba}\ and\ \citenamefont
  {Yokoyama}(2017)}]{Chiba:2017rvs}%
  \BibitemOpen
  \bibfield  {author} {\bibinfo {author} {\bibfnamefont {T.}~\bibnamefont
  {Chiba}}\ and\ \bibinfo {author} {\bibfnamefont {S.}~\bibnamefont
  {Yokoyama}},\ }\bibfield  {title} {\bibinfo {title} {{Spin Distribution of
  Primordial Black Holes}},\ }\href {https://doi.org/10.1093/ptep/ptx087}
  {\bibfield  {journal} {\bibinfo  {journal} {PTEP}\ }\textbf {\bibinfo
  {volume} {2017}},\ \bibinfo {pages} {083E01} (\bibinfo {year} {2017})},\
  \Eprint {https://arxiv.org/abs/1704.06573} {arXiv:1704.06573 [gr-qc]}
  \BibitemShut {NoStop}%
\bibitem [{\citenamefont {Banerjee}\ and\ \citenamefont
  {Dey}(2024)}]{Banerjee:2023qya}%
  \BibitemOpen
  \bibfield  {author} {\bibinfo {author} {\bibfnamefont {I.~K.}\ \bibnamefont
  {Banerjee}}\ and\ \bibinfo {author} {\bibfnamefont {U.~K.}\ \bibnamefont
  {Dey}},\ }\bibfield  {title} {\bibinfo {title} {{Spinning primordial black
  holes from first order phase transition}},\ }\href
  {https://doi.org/10.1007/JHEP07(2024)006} {\bibfield  {journal} {\bibinfo
  {journal} {JHEP}\ }\textbf {\bibinfo {volume} {07}},\ \bibinfo {pages}
  {006}},\ \bibinfo {note} {[Erratum: JHEP 08, 054 (2024)]},\ \Eprint
  {https://arxiv.org/abs/2311.03406} {arXiv:2311.03406 [gr-qc]} \BibitemShut
  {NoStop}%
\bibitem [{\citenamefont {Ye}\ \emph {et~al.}(2025)\citenamefont {Ye},
  \citenamefont {Gong}, \citenamefont {Harada}, \citenamefont {Kang},
  \citenamefont {Kohri}, \citenamefont {Saito},\ and\ \citenamefont
  {Yoo}}]{Ye:2025wif}%
  \BibitemOpen
  \bibfield  {author} {\bibinfo {author} {\bibfnamefont {W.}~\bibnamefont
  {Ye}}, \bibinfo {author} {\bibfnamefont {Y.}~\bibnamefont {Gong}}, \bibinfo
  {author} {\bibfnamefont {T.}~\bibnamefont {Harada}}, \bibinfo {author}
  {\bibfnamefont {Z.}~\bibnamefont {Kang}}, \bibinfo {author} {\bibfnamefont
  {K.}~\bibnamefont {Kohri}}, \bibinfo {author} {\bibfnamefont
  {D.}~\bibnamefont {Saito}},\ and\ \bibinfo {author} {\bibfnamefont {C.-M.}\
  \bibnamefont {Yoo}},\ }\bibfield  {title} {\bibinfo {title} {{Primordial
  Black Hole Formation and Spin in Matter Domination Revisited}},\ }\href@noop
  {} {\  (\bibinfo {year} {2025})},\ \Eprint {https://arxiv.org/abs/2508.10070}
  {arXiv:2508.10070 [gr-qc]} \BibitemShut {NoStop}%
\bibitem [{\citenamefont {de~Jong}\ \emph {et~al.}(2023)\citenamefont
  {de~Jong}, \citenamefont {Aurrekoetxea}, \citenamefont {Lim},\ and\
  \citenamefont {Fran{\c{c}}a}}]{deJong:2023gsx}%
  \BibitemOpen
  \bibfield  {author} {\bibinfo {author} {\bibfnamefont {E.}~\bibnamefont
  {de~Jong}}, \bibinfo {author} {\bibfnamefont {J.~C.}\ \bibnamefont
  {Aurrekoetxea}}, \bibinfo {author} {\bibfnamefont {E.~A.}\ \bibnamefont
  {Lim}},\ and\ \bibinfo {author} {\bibfnamefont {T.}~\bibnamefont
  {Fran{\c{c}}a}},\ }\bibfield  {title} {\bibinfo {title} {{Spinning primordial
  black holes formed during a matter-dominated era}},\ }\href
  {https://doi.org/10.1088/1475-7516/2023/10/067} {\bibfield  {journal}
  {\bibinfo  {journal} {JCAP}\ }\textbf {\bibinfo {volume} {10}},\ \bibinfo
  {pages} {067}},\ \Eprint {https://arxiv.org/abs/2306.11810} {arXiv:2306.11810
  [astro-ph.CO]} \BibitemShut {NoStop}%
\bibitem [{\citenamefont {Saito}\ \emph {et~al.}(2023)\citenamefont {Saito},
  \citenamefont {Harada}, \citenamefont {Koga},\ and\ \citenamefont
  {Yoo}}]{Saito:2023fpt}%
  \BibitemOpen
  \bibfield  {author} {\bibinfo {author} {\bibfnamefont {D.}~\bibnamefont
  {Saito}}, \bibinfo {author} {\bibfnamefont {T.}~\bibnamefont {Harada}},
  \bibinfo {author} {\bibfnamefont {Y.}~\bibnamefont {Koga}},\ and\ \bibinfo
  {author} {\bibfnamefont {C.-M.}\ \bibnamefont {Yoo}},\ }\bibfield  {title}
  {\bibinfo {title} {{Spins of primordial black holes formed with a soft
  equation of state}},\ }\href {https://doi.org/10.1088/1475-7516/2023/07/030}
  {\bibfield  {journal} {\bibinfo  {journal} {JCAP}\ }\textbf {\bibinfo
  {volume} {07}},\ \bibinfo {pages} {030}},\ \Eprint
  {https://arxiv.org/abs/2305.13830} {arXiv:2305.13830 [gr-qc]} \BibitemShut
  {NoStop}%
\bibitem [{\citenamefont {Saito}\ \emph {et~al.}(2024)\citenamefont {Saito},
  \citenamefont {Harada}, \citenamefont {Koga},\ and\ \citenamefont
  {Yoo}}]{Saito:2024hlj}%
  \BibitemOpen
  \bibfield  {author} {\bibinfo {author} {\bibfnamefont {D.}~\bibnamefont
  {Saito}}, \bibinfo {author} {\bibfnamefont {T.}~\bibnamefont {Harada}},
  \bibinfo {author} {\bibfnamefont {Y.}~\bibnamefont {Koga}},\ and\ \bibinfo
  {author} {\bibfnamefont {C.-M.}\ \bibnamefont {Yoo}},\ }\bibfield  {title}
  {\bibinfo {title} {{Revisiting spins of primordial black holes in a
  matter-dominated era based on peak theory}},\ }\href
  {https://doi.org/10.1088/1475-7516/2024/11/064} {\bibfield  {journal}
  {\bibinfo  {journal} {JCAP}\ }\textbf {\bibinfo {volume} {11}},\ \bibinfo
  {pages} {064}},\ \Eprint {https://arxiv.org/abs/2409.00435} {arXiv:2409.00435
  [gr-qc]} \BibitemShut {NoStop}%
\bibitem [{\citenamefont {{Abbott}}\ \emph {et~al.}(2020)\citenamefont
  {{Abbott}}, \citenamefont {{Abbott}},\ and\ \citenamefont
  {{Abraham}}}]{2020PhRvD.102d3015A}%
  \BibitemOpen
  \bibfield  {author} {\bibinfo {author} {\bibfnamefont {R.}~\bibnamefont
  {{Abbott}}}, \bibinfo {author} {\bibfnamefont {T.~D.}\ \bibnamefont
  {{Abbott}}},\ and\ \bibinfo {author} {\bibfnamefont {e.~a.}\ \bibnamefont
  {{Abraham}}, \bibfnamefont {S.}},\ }\bibfield  {title} {\bibinfo {title}
  {{GW190412: Observation of a binary-black-hole coalescence with asymmetric
  masses}},\ }\href {https://doi.org/10.1103/PhysRevD.102.043015} {\bibfield
  {journal} {\bibinfo  {journal} {\prd}\ }\textbf {\bibinfo {volume} {102}},\
  \bibinfo {eid} {043015} (\bibinfo {year} {2020})},\ \Eprint
  {https://arxiv.org/abs/2004.08342} {arXiv:2004.08342 [astro-ph.HE]}
  \BibitemShut {NoStop}%
\bibitem [{\citenamefont {Abac}\ \emph {et~al.}(2025)\citenamefont {Abac} \emph
  {et~al.}}]{LIGOScientific:2025rsn}%
  \BibitemOpen
  \bibfield  {author} {\bibinfo {author} {\bibfnamefont {A.~G.}\ \bibnamefont
  {Abac}} \emph {et~al.} (\bibinfo {collaboration} {LIGO Scientific, VIRGO,
  KAGRA}),\ }\bibfield  {title} {\bibinfo {title} {{GW231123: A Binary Black
  Hole Merger with Total Mass 190{\textendash}265 M$_{⊙}$}},\ }\href
  {https://doi.org/10.3847/2041-8213/ae0c9c} {\bibfield  {journal} {\bibinfo
  {journal} {Astrophys. J. Lett.}\ }\textbf {\bibinfo {volume} {993}},\
  \bibinfo {pages} {L25} (\bibinfo {year} {2025})},\ \Eprint
  {https://arxiv.org/abs/2507.08219} {arXiv:2507.08219 [astro-ph.HE]}
  \BibitemShut {NoStop}%
\bibitem [{\citenamefont {Abbott}\ \emph {et~al.}(2019)\citenamefont {Abbott}
  \emph {et~al.}}]{2019ApJ...882L..24A}%
  \BibitemOpen
  \bibfield  {author} {\bibinfo {author} {\bibfnamefont {B.~P.}\ \bibnamefont
  {Abbott}} \emph {et~al.} (\bibinfo {collaboration} {LIGO Scientific,
  Virgo}),\ }\bibfield  {title} {\bibinfo {title} {{Binary Black Hole
  Population Properties Inferred from the First and Second Observing Runs of
  Advanced LIGO and Advanced Virgo}},\ }\href
  {https://doi.org/10.3847/2041-8213/ab3800} {\bibfield  {journal} {\bibinfo
  {journal} {Astrophys. J. Lett.}\ }\textbf {\bibinfo {volume} {882}},\
  \bibinfo {pages} {L24} (\bibinfo {year} {2019})},\ \Eprint
  {https://arxiv.org/abs/1811.12940} {arXiv:1811.12940 [astro-ph.HE]}
  \BibitemShut {NoStop}%
\bibitem [{\citenamefont {Caprini}\ and\ \citenamefont
  {Figueroa}(2018)}]{Caprini:2018mtu}%
  \BibitemOpen
  \bibfield  {author} {\bibinfo {author} {\bibfnamefont {C.}~\bibnamefont
  {Caprini}}\ and\ \bibinfo {author} {\bibfnamefont {D.~G.}\ \bibnamefont
  {Figueroa}},\ }\bibfield  {title} {\bibinfo {title} {{Cosmological
  Backgrounds of Gravitational Waves}},\ }\href
  {https://doi.org/10.1088/1361-6382/aac608} {\bibfield  {journal} {\bibinfo
  {journal} {Class. Quant. Grav.}\ }\textbf {\bibinfo {volume} {35}},\ \bibinfo
  {pages} {163001} (\bibinfo {year} {2018})},\ \Eprint
  {https://arxiv.org/abs/1801.04268} {arXiv:1801.04268 [astro-ph.CO]}
  \BibitemShut {NoStop}%
\bibitem [{\citenamefont {Christensen}(2019)}]{Christensen:2018iqi}%
  \BibitemOpen
  \bibfield  {author} {\bibinfo {author} {\bibfnamefont {N.}~\bibnamefont
  {Christensen}},\ }\bibfield  {title} {\bibinfo {title} {{Stochastic
  Gravitational Wave Backgrounds}},\ }\href
  {https://doi.org/10.1088/1361-6633/aae6b5} {\bibfield  {journal} {\bibinfo
  {journal} {Rept. Prog. Phys.}\ }\textbf {\bibinfo {volume} {82}},\ \bibinfo
  {pages} {016903} (\bibinfo {year} {2019})},\ \Eprint
  {https://arxiv.org/abs/1811.08797} {arXiv:1811.08797 [gr-qc]} \BibitemShut
  {NoStop}%
\bibitem [{\citenamefont {Baumann}\ \emph {et~al.}(2007)\citenamefont
  {Baumann}, \citenamefont {Steinhardt}, \citenamefont {Takahashi},\ and\
  \citenamefont {Ichiki}}]{Baumann:2007zm}%
  \BibitemOpen
  \bibfield  {author} {\bibinfo {author} {\bibfnamefont {D.}~\bibnamefont
  {Baumann}}, \bibinfo {author} {\bibfnamefont {P.~J.}\ \bibnamefont
  {Steinhardt}}, \bibinfo {author} {\bibfnamefont {K.}~\bibnamefont
  {Takahashi}},\ and\ \bibinfo {author} {\bibfnamefont {K.}~\bibnamefont
  {Ichiki}},\ }\bibfield  {title} {\bibinfo {title} {{Gravitational Wave
  Spectrum Induced by Primordial Scalar Perturbations}},\ }\href
  {https://doi.org/10.1103/PhysRevD.76.084019} {\bibfield  {journal} {\bibinfo
  {journal} {Phys. Rev. D}\ }\textbf {\bibinfo {volume} {76}},\ \bibinfo
  {pages} {084019} (\bibinfo {year} {2007})},\ \Eprint
  {https://arxiv.org/abs/hep-th/0703290} {arXiv:hep-th/0703290} \BibitemShut
  {NoStop}%
\bibitem [{\citenamefont {Dom\`enech}\ \emph {et~al.}(2024)\citenamefont
  {Dom\`enech}, \citenamefont {Pi}, \citenamefont {Wang},\ and\ \citenamefont
  {Wang}}]{Domenech:2024rks}%
  \BibitemOpen
  \bibfield  {author} {\bibinfo {author} {\bibfnamefont {G.}~\bibnamefont
  {Dom\`enech}}, \bibinfo {author} {\bibfnamefont {S.}~\bibnamefont {Pi}},
  \bibinfo {author} {\bibfnamefont {A.}~\bibnamefont {Wang}},\ and\ \bibinfo
  {author} {\bibfnamefont {J.}~\bibnamefont {Wang}},\ }\bibfield  {title}
  {\bibinfo {title} {{Induced gravitational wave interpretation of PTA data: a
  complete study for general equation of state}},\ }\href
  {https://doi.org/10.1088/1475-7516/2024/08/054} {\bibfield  {journal}
  {\bibinfo  {journal} {JCAP}\ }\textbf {\bibinfo {volume} {08}},\ \bibinfo
  {pages} {054}},\ \Eprint {https://arxiv.org/abs/2402.18965} {arXiv:2402.18965
  [astro-ph.CO]} \BibitemShut {NoStop}%
\bibitem [{\citenamefont {Kohri}\ and\ \citenamefont
  {Terada}(2018)}]{Kohri:2018awv}%
  \BibitemOpen
  \bibfield  {author} {\bibinfo {author} {\bibfnamefont {K.}~\bibnamefont
  {Kohri}}\ and\ \bibinfo {author} {\bibfnamefont {T.}~\bibnamefont {Terada}},\
  }\bibfield  {title} {\bibinfo {title} {{Semianalytic calculation of
  gravitational wave spectrum nonlinearly induced from primordial curvature
  perturbations}},\ }\href {https://doi.org/10.1103/PhysRevD.97.123532}
  {\bibfield  {journal} {\bibinfo  {journal} {Phys. Rev. D}\ }\textbf {\bibinfo
  {volume} {97}},\ \bibinfo {pages} {123532} (\bibinfo {year} {2018})},\
  \Eprint {https://arxiv.org/abs/1804.08577} {arXiv:1804.08577 [gr-qc]}
  \BibitemShut {NoStop}%
\bibitem [{\citenamefont {Dom\`enech}(2021)}]{Domenech:2021ztg}%
  \BibitemOpen
  \bibfield  {author} {\bibinfo {author} {\bibfnamefont {G.}~\bibnamefont
  {Dom\`enech}},\ }\bibfield  {title} {\bibinfo {title} {{Scalar Induced
  Gravitational Waves Review}},\ }\href
  {https://doi.org/10.3390/universe7110398} {\bibfield  {journal} {\bibinfo
  {journal} {Universe}\ }\textbf {\bibinfo {volume} {7}},\ \bibinfo {pages}
  {398} (\bibinfo {year} {2021})},\ \Eprint {https://arxiv.org/abs/2109.01398}
  {arXiv:2109.01398 [gr-qc]} \BibitemShut {NoStop}%
\bibitem [{\citenamefont {Dom\`enech}(2020)}]{Domenech:2019quo}%
  \BibitemOpen
  \bibfield  {author} {\bibinfo {author} {\bibfnamefont {G.}~\bibnamefont
  {Dom\`enech}},\ }\bibfield  {title} {\bibinfo {title} {{Induced gravitational
  waves in a general cosmological background}},\ }\href
  {https://doi.org/10.1142/S0218271820500285} {\bibfield  {journal} {\bibinfo
  {journal} {Int. J. Mod. Phys. D}\ }\textbf {\bibinfo {volume} {29}},\
  \bibinfo {pages} {2050028} (\bibinfo {year} {2020})},\ \Eprint
  {https://arxiv.org/abs/1912.05583} {arXiv:1912.05583 [gr-qc]} \BibitemShut
  {NoStop}%
\bibitem [{\citenamefont {Dom\`enech}\ \emph {et~al.}(2020)\citenamefont
  {Dom\`enech}, \citenamefont {Pi},\ and\ \citenamefont
  {Sasaki}}]{Domenech:2020kqm}%
  \BibitemOpen
  \bibfield  {author} {\bibinfo {author} {\bibfnamefont {G.}~\bibnamefont
  {Dom\`enech}}, \bibinfo {author} {\bibfnamefont {S.}~\bibnamefont {Pi}},\
  and\ \bibinfo {author} {\bibfnamefont {M.}~\bibnamefont {Sasaki}},\
  }\bibfield  {title} {\bibinfo {title} {{Induced gravitational waves as a
  probe of thermal history of the universe}},\ }\href
  {https://doi.org/10.1088/1475-7516/2020/08/017} {\bibfield  {journal}
  {\bibinfo  {journal} {JCAP}\ }\textbf {\bibinfo {volume} {08}},\ \bibinfo
  {pages} {017}},\ \Eprint {https://arxiv.org/abs/2005.12314} {arXiv:2005.12314
  [gr-qc]} \BibitemShut {NoStop}%
\bibitem [{\citenamefont {Dom\`enech}\ \emph {et~al.}(2021)\citenamefont
  {Dom\`enech}, \citenamefont {Takhistov},\ and\ \citenamefont
  {Sasaki}}]{Domenech:2021wkk}%
  \BibitemOpen
  \bibfield  {author} {\bibinfo {author} {\bibfnamefont {G.}~\bibnamefont
  {Dom\`enech}}, \bibinfo {author} {\bibfnamefont {V.}~\bibnamefont
  {Takhistov}},\ and\ \bibinfo {author} {\bibfnamefont {M.}~\bibnamefont
  {Sasaki}},\ }\bibfield  {title} {\bibinfo {title} {{Exploring evaporating
  primordial black holes with gravitational waves}},\ }\href
  {https://doi.org/10.1016/j.physletb.2021.136722} {\bibfield  {journal}
  {\bibinfo  {journal} {Phys. Lett. B}\ }\textbf {\bibinfo {volume} {823}},\
  \bibinfo {pages} {136722} (\bibinfo {year} {2021})},\ \Eprint
  {https://arxiv.org/abs/2105.06816} {arXiv:2105.06816 [astro-ph.CO]}
  \BibitemShut {NoStop}%
\bibitem [{\citenamefont {Pal}\ \emph {et~al.}(2010)\citenamefont {Pal},
  \citenamefont {Pal},\ and\ \citenamefont {Basu}}]{Pal:2009sd}%
  \BibitemOpen
  \bibfield  {author} {\bibinfo {author} {\bibfnamefont {B.~K.}\ \bibnamefont
  {Pal}}, \bibinfo {author} {\bibfnamefont {S.}~\bibnamefont {Pal}},\ and\
  \bibinfo {author} {\bibfnamefont {B.}~\bibnamefont {Basu}},\ }\bibfield
  {title} {\bibinfo {title} {{Mutated Hilltop Inflation : A Natural Choice for
  Early Universe}},\ }\href {https://doi.org/10.1088/1475-7516/2010/01/029}
  {\bibfield  {journal} {\bibinfo  {journal} {JCAP}\ }\textbf {\bibinfo
  {volume} {01}},\ \bibinfo {pages} {029}},\ \Eprint
  {https://arxiv.org/abs/0908.2302} {arXiv:0908.2302 [hep-th]} \BibitemShut
  {NoStop}%
\bibitem [{\citenamefont {Yogesh}\ and\ \citenamefont
  {Mohammadi}(2025)}]{Yogesh:2025hll}%
  \BibitemOpen
  \bibfield  {author} {\bibinfo {author} {\bibnamefont {Yogesh}}\ and\ \bibinfo
  {author} {\bibfnamefont {A.}~\bibnamefont {Mohammadi}},\ }\bibfield  {title}
  {\bibinfo {title} {{Nonstandard Thermal History and Formation of Primordial
  Black Holes and SIGWs in Einstein{\textendash}Gauss{\textendash}Bonnet
  Gravity}},\ }\href {https://doi.org/10.3847/1538-4357/adcee5} {\bibfield
  {journal} {\bibinfo  {journal} {Astrophys. J.}\ }\textbf {\bibinfo {volume}
  {986}},\ \bibinfo {pages} {35} (\bibinfo {year} {2025})},\ \Eprint
  {https://arxiv.org/abs/2501.01867} {arXiv:2501.01867 [gr-qc]} \BibitemShut
  {NoStop}%
\bibitem [{\citenamefont {Hazra}\ \emph {et~al.}(2010)\citenamefont {Hazra},
  \citenamefont {Aich}, \citenamefont {Jain}, \citenamefont {Sriramkumar},\
  and\ \citenamefont {Souradeep}}]{Hazra:2010ve}%
  \BibitemOpen
  \bibfield  {author} {\bibinfo {author} {\bibfnamefont {D.~K.}\ \bibnamefont
  {Hazra}}, \bibinfo {author} {\bibfnamefont {M.}~\bibnamefont {Aich}},
  \bibinfo {author} {\bibfnamefont {R.~K.}\ \bibnamefont {Jain}}, \bibinfo
  {author} {\bibfnamefont {L.}~\bibnamefont {Sriramkumar}},\ and\ \bibinfo
  {author} {\bibfnamefont {T.}~\bibnamefont {Souradeep}},\ }\bibfield  {title}
  {\bibinfo {title} {{Primordial features due to a step in the inflaton
  potential}},\ }\href {https://doi.org/10.1088/1475-7516/2010/10/008}
  {\bibfield  {journal} {\bibinfo  {journal} {JCAP}\ }\textbf {\bibinfo
  {volume} {10}},\ \bibinfo {pages} {008}},\ \Eprint
  {https://arxiv.org/abs/1005.2175} {arXiv:1005.2175 [astro-ph.CO]}
  \BibitemShut {NoStop}%
\bibitem [{\citenamefont {Thomas}\ \emph {et~al.}(2024)\citenamefont {Thomas},
  \citenamefont {Thomas},\ and\ \citenamefont {Joy}}]{Thomas:2024ezg}%
  \BibitemOpen
  \bibfield  {author} {\bibinfo {author} {\bibfnamefont {R.}~\bibnamefont
  {Thomas}}, \bibinfo {author} {\bibfnamefont {J.}~\bibnamefont {Thomas}},\
  and\ \bibinfo {author} {\bibfnamefont {M.}~\bibnamefont {Joy}},\ }\bibfield
  {title} {\bibinfo {title} {{Primordial blackhole formation: exploring chaotic
  potential with a sharp step via the GLMS perspective}},\ }\href
  {https://doi.org/10.1088/1361-6382/ad74d4} {\bibfield  {journal} {\bibinfo
  {journal} {Class. Quant. Grav.}\ }\textbf {\bibinfo {volume} {41}},\ \bibinfo
  {pages} {205001} (\bibinfo {year} {2024})},\ \Eprint
  {https://arxiv.org/abs/2411.10076} {arXiv:2411.10076 [astro-ph.CO]}
  \BibitemShut {NoStop}%
\bibitem [{\citenamefont {Akrami}\ \emph {et~al.}(2020)\citenamefont {Akrami}
  \emph {et~al.}}]{Planck:2018jri}%
  \BibitemOpen
  \bibfield  {author} {\bibinfo {author} {\bibfnamefont {Y.}~\bibnamefont
  {Akrami}} \emph {et~al.} (\bibinfo {collaboration} {Planck}),\ }\bibfield
  {title} {\bibinfo {title} {{Planck 2018 results. X. Constraints on
  inflation}},\ }\href {https://doi.org/10.1051/0004-6361/201833887} {\bibfield
   {journal} {\bibinfo  {journal} {Astron. Astrophys.}\ }\textbf {\bibinfo
  {volume} {641}},\ \bibinfo {pages} {A10} (\bibinfo {year} {2020})},\ \Eprint
  {https://arxiv.org/abs/1807.06211} {arXiv:1807.06211 [astro-ph.CO]}
  \BibitemShut {NoStop}%
\bibitem [{\citenamefont {Fixsen}\ \emph {et~al.}(1996)\citenamefont {Fixsen},
  \citenamefont {Cheng}, \citenamefont {Gales}, \citenamefont {Mather},
  \citenamefont {Shafer},\ and\ \citenamefont {Wright}}]{Fixsen:1996nj}%
  \BibitemOpen
  \bibfield  {author} {\bibinfo {author} {\bibfnamefont {D.~J.}\ \bibnamefont
  {Fixsen}}, \bibinfo {author} {\bibfnamefont {E.~S.}\ \bibnamefont {Cheng}},
  \bibinfo {author} {\bibfnamefont {J.~M.}\ \bibnamefont {Gales}}, \bibinfo
  {author} {\bibfnamefont {J.~C.}\ \bibnamefont {Mather}}, \bibinfo {author}
  {\bibfnamefont {R.~A.}\ \bibnamefont {Shafer}},\ and\ \bibinfo {author}
  {\bibfnamefont {E.~L.}\ \bibnamefont {Wright}},\ }\bibfield  {title}
  {\bibinfo {title} {{The Cosmic Microwave Background spectrum from the full
  COBE FIRAS data set}},\ }\href {https://doi.org/10.1086/178173} {\bibfield
  {journal} {\bibinfo  {journal} {Astrophys. J.}\ }\textbf {\bibinfo {volume}
  {473}},\ \bibinfo {pages} {576} (\bibinfo {year} {1996})},\ \Eprint
  {https://arxiv.org/abs/astro-ph/9605054} {arXiv:astro-ph/9605054}
  \BibitemShut {NoStop}%
\bibitem [{\citenamefont {Chluba}\ \emph {et~al.}(2021)\citenamefont {Chluba}
  \emph {et~al.}}]{Chluba:2019nxa}%
  \BibitemOpen
  \bibfield  {author} {\bibinfo {author} {\bibfnamefont {J.}~\bibnamefont
  {Chluba}} \emph {et~al.},\ }\bibfield  {title} {\bibinfo {title} {{New
  horizons in cosmology with spectral distortions of the cosmic microwave
  background}},\ }\href {https://doi.org/10.1007/s10686-021-09729-5} {\bibfield
   {journal} {\bibinfo  {journal} {Exper. Astron.}\ }\textbf {\bibinfo {volume}
  {51}},\ \bibinfo {pages} {1515} (\bibinfo {year} {2021})},\ \Eprint
  {https://arxiv.org/abs/1909.01593} {arXiv:1909.01593 [astro-ph.CO]}
  \BibitemShut {NoStop}%
\bibitem [{\citenamefont {Inomata}\ and\ \citenamefont
  {Nakama}(2019)}]{Inomata:2018epa}%
  \BibitemOpen
  \bibfield  {author} {\bibinfo {author} {\bibfnamefont {K.}~\bibnamefont
  {Inomata}}\ and\ \bibinfo {author} {\bibfnamefont {T.}~\bibnamefont
  {Nakama}},\ }\bibfield  {title} {\bibinfo {title} {{Gravitational waves
  induced by scalar perturbations as probes of the small-scale primordial
  spectrum}},\ }\href {https://doi.org/10.1103/PhysRevD.99.043511} {\bibfield
  {journal} {\bibinfo  {journal} {Phys. Rev. D}\ }\textbf {\bibinfo {volume}
  {99}},\ \bibinfo {pages} {043511} (\bibinfo {year} {2019})},\ \Eprint
  {https://arxiv.org/abs/1812.00674} {arXiv:1812.00674 [astro-ph.CO]}
  \BibitemShut {NoStop}%
\bibitem [{\citenamefont {Byrnes}\ \emph {et~al.}(2019)\citenamefont {Byrnes},
  \citenamefont {Cole},\ and\ \citenamefont {Patil}}]{Byrnes:2018txb}%
  \BibitemOpen
  \bibfield  {author} {\bibinfo {author} {\bibfnamefont {C.~T.}\ \bibnamefont
  {Byrnes}}, \bibinfo {author} {\bibfnamefont {P.~S.}\ \bibnamefont {Cole}},\
  and\ \bibinfo {author} {\bibfnamefont {S.~P.}\ \bibnamefont {Patil}},\
  }\bibfield  {title} {\bibinfo {title} {{Steepest growth of the power spectrum
  and primordial black holes}},\ }\href
  {https://doi.org/10.1088/1475-7516/2019/06/028} {\bibfield  {journal}
  {\bibinfo  {journal} {JCAP}\ }\textbf {\bibinfo {volume} {06}},\ \bibinfo
  {pages} {028}},\ \Eprint {https://arxiv.org/abs/1811.11158} {arXiv:1811.11158
  [astro-ph.CO]} \BibitemShut {NoStop}%
\bibitem [{\citenamefont {Louis}\ \emph {et~al.}(2025)\citenamefont {Louis}
  \emph {et~al.}}]{ACT:2025fju}%
  \BibitemOpen
  \bibfield  {author} {\bibinfo {author} {\bibfnamefont {T.}~\bibnamefont
  {Louis}} \emph {et~al.} (\bibinfo {collaboration} {ACT}),\ }\bibfield
  {title} {\bibinfo {title} {{The Atacama Cosmology Telescope: DR6 Power
  Spectra, Likelihoods and $\Lambda$CDM Parameters}},\ }\href@noop {}
  {\bibfield  {journal} {\bibinfo  {journal} {{arXiv preprint}}\ } (\bibinfo
  {year} {2025})},\ \Eprint {https://arxiv.org/abs/2503.14452}
  {arXiv:2503.14452 [astro-ph.CO]} \BibitemShut {NoStop}%
\bibitem [{\citenamefont {Calabrese}\ \emph {et~al.}(2025)\citenamefont
  {Calabrese} \emph {et~al.}}]{ACT:2025tim}%
  \BibitemOpen
  \bibfield  {author} {\bibinfo {author} {\bibfnamefont {E.}~\bibnamefont
  {Calabrese}} \emph {et~al.} (\bibinfo {collaboration} {ACT}),\ }\bibfield
  {title} {\bibinfo {title} {{The Atacama Cosmology Telescope: DR6 Constraints
  on Extended Cosmological Models}},\ }\href@noop {} {\bibfield  {journal}
  {\bibinfo  {journal} {{arXiv preprint}}\ } (\bibinfo {year} {2025})},\
  \Eprint {https://arxiv.org/abs/2503.14454} {arXiv:2503.14454 [astro-ph.CO]}
  \BibitemShut {NoStop}%
\bibitem [{\citenamefont {Pi}\ \emph {et~al.}(2025)\citenamefont {Pi},
  \citenamefont {Sasaki}, \citenamefont {Takhistov},\ and\ \citenamefont
  {Wang}}]{Pi:2024ert}%
  \BibitemOpen
  \bibfield  {author} {\bibinfo {author} {\bibfnamefont {S.}~\bibnamefont
  {Pi}}, \bibinfo {author} {\bibfnamefont {M.}~\bibnamefont {Sasaki}}, \bibinfo
  {author} {\bibfnamefont {V.}~\bibnamefont {Takhistov}},\ and\ \bibinfo
  {author} {\bibfnamefont {J.}~\bibnamefont {Wang}},\ }\bibfield  {title}
  {\bibinfo {title} {{Primordial Black Hole formation from power spectrum with
  finite-width}},\ }\href {https://doi.org/10.1088/1475-7516/2025/09/045}
  {\bibfield  {journal} {\bibinfo  {journal} {JCAP}\ }\textbf {\bibinfo
  {volume} {09}},\ \bibinfo {pages} {045}},\ \Eprint
  {https://arxiv.org/abs/2501.00295} {arXiv:2501.00295 [astro-ph.CO]}
  \BibitemShut {NoStop}%
\bibitem [{\citenamefont {Bardeen}\ \emph {et~al.}(1986)\citenamefont
  {Bardeen}, \citenamefont {Bond}, \citenamefont {Kaiser},\ and\ \citenamefont
  {Szalay}}]{Bardeen:1985tr}%
  \BibitemOpen
  \bibfield  {author} {\bibinfo {author} {\bibfnamefont {J.~M.}\ \bibnamefont
  {Bardeen}}, \bibinfo {author} {\bibfnamefont {J.~R.}\ \bibnamefont {Bond}},
  \bibinfo {author} {\bibfnamefont {N.}~\bibnamefont {Kaiser}},\ and\ \bibinfo
  {author} {\bibfnamefont {A.~S.}\ \bibnamefont {Szalay}},\ }\bibfield  {title}
  {\bibinfo {title} {{The Statistics of Peaks of Gaussian Random Fields}},\
  }\href {https://doi.org/10.1086/164143} {\bibfield  {journal} {\bibinfo
  {journal} {Astrophys. J.}\ }\textbf {\bibinfo {volume} {304}},\ \bibinfo
  {pages} {15} (\bibinfo {year} {1986})}\BibitemShut {NoStop}%
\bibitem [{\citenamefont {Peacock}\ and\ \citenamefont
  {Heavens}(1990)}]{Peacock:1990zz}%
  \BibitemOpen
  \bibfield  {author} {\bibinfo {author} {\bibfnamefont {J.~A.}\ \bibnamefont
  {Peacock}}\ and\ \bibinfo {author} {\bibfnamefont {A.~F.}\ \bibnamefont
  {Heavens}},\ }\bibfield  {title} {\bibinfo {title} {{Alternatives to the
  Press-Schechter cosmological mass function}},\ }\href@noop {} {\bibfield
  {journal} {\bibinfo  {journal} {Mon. Not. Roy. Astron. Soc.}\ }\textbf
  {\bibinfo {volume} {243}},\ \bibinfo {pages} {133} (\bibinfo {year}
  {1990})}\BibitemShut {NoStop}%
\bibitem [{\citenamefont {Young}\ and\ \citenamefont
  {Musso}(2020)}]{Young:2020xmk}%
  \BibitemOpen
  \bibfield  {author} {\bibinfo {author} {\bibfnamefont {S.}~\bibnamefont
  {Young}}\ and\ \bibinfo {author} {\bibfnamefont {M.}~\bibnamefont {Musso}},\
  }\bibfield  {title} {\bibinfo {title} {{Application of peaks theory to the
  abundance of primordial black holes}},\ }\href
  {https://doi.org/10.1088/1475-7516/2020/11/022} {\bibfield  {journal}
  {\bibinfo  {journal} {JCAP}\ }\textbf {\bibinfo {volume} {11}},\ \bibinfo
  {pages} {022}},\ \Eprint {https://arxiv.org/abs/2001.06469} {arXiv:2001.06469
  [astro-ph.CO]} \BibitemShut {NoStop}%
\bibitem [{\citenamefont {Ando}\ \emph {et~al.}(2018)\citenamefont {Ando},
  \citenamefont {Inomata},\ and\ \citenamefont {Kawasaki}}]{Ando:2018qdb}%
  \BibitemOpen
  \bibfield  {author} {\bibinfo {author} {\bibfnamefont {K.}~\bibnamefont
  {Ando}}, \bibinfo {author} {\bibfnamefont {K.}~\bibnamefont {Inomata}},\ and\
  \bibinfo {author} {\bibfnamefont {M.}~\bibnamefont {Kawasaki}},\ }\bibfield
  {title} {\bibinfo {title} {{Primordial black holes and uncertainties in the
  choice of the window function}},\ }\href
  {https://doi.org/10.1103/PhysRevD.97.103528} {\bibfield  {journal} {\bibinfo
  {journal} {Phys. Rev. D}\ }\textbf {\bibinfo {volume} {97}},\ \bibinfo
  {pages} {103528} (\bibinfo {year} {2018})},\ \Eprint
  {https://arxiv.org/abs/1802.06393} {arXiv:1802.06393 [astro-ph.CO]}
  \BibitemShut {NoStop}%
\bibitem [{\citenamefont {Young}(2019)}]{Young:2019osy}%
  \BibitemOpen
  \bibfield  {author} {\bibinfo {author} {\bibfnamefont {S.}~\bibnamefont
  {Young}},\ }\bibfield  {title} {\bibinfo {title} {{The primordial black hole
  formation criterion re-examined: Parametrisation, timing and the choice of
  window function}},\ }\href {https://doi.org/10.1142/S0218271820300025}
  {\bibfield  {journal} {\bibinfo  {journal} {Int. J. Mod. Phys. D}\ }\textbf
  {\bibinfo {volume} {29}},\ \bibinfo {pages} {2030002} (\bibinfo {year}
  {2019})},\ \Eprint {https://arxiv.org/abs/1905.01230} {arXiv:1905.01230
  [astro-ph.CO]} \BibitemShut {NoStop}%
\bibitem [{\citenamefont {Escriv\`a}\ \emph {et~al.}(2020)\citenamefont
  {Escriv\`a}, \citenamefont {Germani},\ and\ \citenamefont
  {Sheth}}]{Escriva:2019phb}%
  \BibitemOpen
  \bibfield  {author} {\bibinfo {author} {\bibfnamefont {A.}~\bibnamefont
  {Escriv\`a}}, \bibinfo {author} {\bibfnamefont {C.}~\bibnamefont {Germani}},\
  and\ \bibinfo {author} {\bibfnamefont {R.~K.}\ \bibnamefont {Sheth}},\
  }\bibfield  {title} {\bibinfo {title} {{Universal threshold for primordial
  black hole formation}},\ }\href {https://doi.org/10.1103/PhysRevD.101.044022}
  {\bibfield  {journal} {\bibinfo  {journal} {Phys. Rev. D}\ }\textbf {\bibinfo
  {volume} {101}},\ \bibinfo {pages} {044022} (\bibinfo {year} {2020})},\
  \Eprint {https://arxiv.org/abs/1907.13311} {arXiv:1907.13311 [gr-qc]}
  \BibitemShut {NoStop}%
\bibitem [{\citenamefont {Escriv{\`a}}\ \emph {et~al.}(2022)\citenamefont
  {Escriv{\`a}}, \citenamefont {Tada}, \citenamefont {Yokoyama},\ and\
  \citenamefont {Yoo}}]{Escriva:2022pnz}%
  \BibitemOpen
  \bibfield  {author} {\bibinfo {author} {\bibfnamefont {A.}~\bibnamefont
  {Escriv{\`a}}}, \bibinfo {author} {\bibfnamefont {Y.}~\bibnamefont {Tada}},
  \bibinfo {author} {\bibfnamefont {S.}~\bibnamefont {Yokoyama}},\ and\
  \bibinfo {author} {\bibfnamefont {C.-M.}\ \bibnamefont {Yoo}},\ }\bibfield
  {title} {\bibinfo {title} {{Simulation of primordial black holes with large
  negative non-Gaussianity}},\ }\href
  {https://doi.org/10.1088/1475-7516/2022/05/012} {\bibfield  {journal}
  {\bibinfo  {journal} {JCAP}\ }\textbf {\bibinfo {volume} {05}}\bibfield
  {number} {\bibinfo  {number} { (05)},\ \bibinfo {pages} {012}},\ }\Eprint
  {https://arxiv.org/abs/2202.01028} {arXiv:2202.01028 [astro-ph.CO]}
  \BibitemShut {NoStop}%
\bibitem [{\citenamefont {Uehara}\ \emph
  {et~al.}(2025{\natexlab{a}})\citenamefont {Uehara}, \citenamefont
  {Escriv{\`a}}, \citenamefont {Harada}, \citenamefont {Saito},\ and\
  \citenamefont {Yoo}}]{Uehara:2025idq}%
  \BibitemOpen
  \bibfield  {author} {\bibinfo {author} {\bibfnamefont {K.}~\bibnamefont
  {Uehara}}, \bibinfo {author} {\bibfnamefont {A.}~\bibnamefont {Escriv{\`a}}},
  \bibinfo {author} {\bibfnamefont {T.}~\bibnamefont {Harada}}, \bibinfo
  {author} {\bibfnamefont {D.}~\bibnamefont {Saito}},\ and\ \bibinfo {author}
  {\bibfnamefont {C.-M.}\ \bibnamefont {Yoo}},\ }\bibfield  {title} {\bibinfo
  {title} {{Primordial black hole formation from a type II perturbation in the
  absence and presence of pressure}},\ }\href
  {https://doi.org/10.1088/1475-7516/2025/08/042} {\bibfield  {journal}
  {\bibinfo  {journal} {JCAP}\ }\textbf {\bibinfo {volume} {08}},\ \bibinfo
  {pages} {042}},\ \Eprint {https://arxiv.org/abs/2505.00366} {arXiv:2505.00366
  [gr-qc]} \BibitemShut {NoStop}%
\bibitem [{\citenamefont {Escriv{\`a}}(2025)}]{Escriva:2025rja}%
  \BibitemOpen
  \bibfield  {author} {\bibinfo {author} {\bibfnamefont {A.}~\bibnamefont
  {Escriv{\`a}}},\ }\bibfield  {title} {\bibinfo {title} {{Threshold for PBH
  formation in the type-II region and its analytical estimation}},\ }\href
  {https://doi.org/10.1103/mq67-bbvj} {\bibfield  {journal} {\bibinfo
  {journal} {Phys. Rev. D}\ }\textbf {\bibinfo {volume} {112}},\ \bibinfo
  {pages} {103527} (\bibinfo {year} {2025})},\ \Eprint
  {https://arxiv.org/abs/2504.05814} {arXiv:2504.05814 [astro-ph.CO]}
  \BibitemShut {NoStop}%
\bibitem [{\citenamefont {Shimada}\ \emph {et~al.}(2025)\citenamefont
  {Shimada}, \citenamefont {Escriv{\'a}}, \citenamefont {Saito}, \citenamefont
  {Uehara},\ and\ \citenamefont {Yoo}}]{Shimada:2024eec}%
  \BibitemOpen
  \bibfield  {author} {\bibinfo {author} {\bibfnamefont {M.}~\bibnamefont
  {Shimada}}, \bibinfo {author} {\bibfnamefont {A.}~\bibnamefont
  {Escriv{\'a}}}, \bibinfo {author} {\bibfnamefont {D.}~\bibnamefont {Saito}},
  \bibinfo {author} {\bibfnamefont {K.}~\bibnamefont {Uehara}},\ and\ \bibinfo
  {author} {\bibfnamefont {C.-M.}\ \bibnamefont {Yoo}},\ }\bibfield  {title}
  {\bibinfo {title} {{Primordial black hole formation from type II fluctuations
  with primordial non-Gaussianity}},\ }\href
  {https://doi.org/10.1088/1475-7516/2025/02/018} {\bibfield  {journal}
  {\bibinfo  {journal} {JCAP}\ }\textbf {\bibinfo {volume} {02}},\ \bibinfo
  {pages} {018}},\ \Eprint {https://arxiv.org/abs/2411.07648} {arXiv:2411.07648
  [gr-qc]} \BibitemShut {NoStop}%
\bibitem [{\citenamefont {Uehara}\ \emph
  {et~al.}(2025{\natexlab{b}})\citenamefont {Uehara}, \citenamefont
  {Escriv{\`a}}, \citenamefont {Harada}, \citenamefont {Saito},\ and\
  \citenamefont {Yoo}}]{Uehara:2024yyp}%
  \BibitemOpen
  \bibfield  {author} {\bibinfo {author} {\bibfnamefont {K.}~\bibnamefont
  {Uehara}}, \bibinfo {author} {\bibfnamefont {A.}~\bibnamefont {Escriv{\`a}}},
  \bibinfo {author} {\bibfnamefont {T.}~\bibnamefont {Harada}}, \bibinfo
  {author} {\bibfnamefont {D.}~\bibnamefont {Saito}},\ and\ \bibinfo {author}
  {\bibfnamefont {C.-M.}\ \bibnamefont {Yoo}},\ }\bibfield  {title} {\bibinfo
  {title} {{Numerical simulation of type II primordial black hole formation}},\
  }\href {https://doi.org/10.1088/1475-7516/2025/01/003} {\bibfield  {journal}
  {\bibinfo  {journal} {JCAP}\ }\textbf {\bibinfo {volume} {01}},\ \bibinfo
  {pages} {003}},\ \Eprint {https://arxiv.org/abs/2401.06329} {arXiv:2401.06329
  [gr-qc]} \BibitemShut {NoStop}%
\bibitem [{\citenamefont {Harada}\ \emph {et~al.}(2015)\citenamefont {Harada},
  \citenamefont {Yoo}, \citenamefont {Nakama},\ and\ \citenamefont
  {Koga}}]{Harada:2015yda}%
  \BibitemOpen
  \bibfield  {author} {\bibinfo {author} {\bibfnamefont {T.}~\bibnamefont
  {Harada}}, \bibinfo {author} {\bibfnamefont {C.-M.}\ \bibnamefont {Yoo}},
  \bibinfo {author} {\bibfnamefont {T.}~\bibnamefont {Nakama}},\ and\ \bibinfo
  {author} {\bibfnamefont {Y.}~\bibnamefont {Koga}},\ }\bibfield  {title}
  {\bibinfo {title} {{Cosmological long-wavelength solutions and primordial
  black hole formation}},\ }\href {https://doi.org/10.1103/PhysRevD.91.084057}
  {\bibfield  {journal} {\bibinfo  {journal} {Phys. Rev. D}\ }\textbf {\bibinfo
  {volume} {91}},\ \bibinfo {pages} {084057} (\bibinfo {year} {2015})},\
  \Eprint {https://arxiv.org/abs/1503.03934} {arXiv:1503.03934 [gr-qc]}
  \BibitemShut {NoStop}%
\bibitem [{\citenamefont {Wald}(1984)}]{Wald:1984rg}%
  \BibitemOpen
  \bibfield  {author} {\bibinfo {author} {\bibfnamefont {R.~M.}\ \bibnamefont
  {Wald}},\ }\href {https://doi.org/10.7208/chicago/9780226870373.001.0001}
  {\emph {\bibinfo {title} {{General Relativity}}}}\ (\bibinfo  {publisher}
  {Chicago Univ. Pr.},\ \bibinfo {address} {Chicago, USA},\ \bibinfo {year}
  {1984})\BibitemShut {NoStop}%
\bibitem [{\citenamefont {Choptuik}(1993)}]{Choptuik:1992jv}%
  \BibitemOpen
  \bibfield  {author} {\bibinfo {author} {\bibfnamefont {M.~W.}\ \bibnamefont
  {Choptuik}},\ }\bibfield  {title} {\bibinfo {title} {{Universality and
  scaling in gravitational collapse of a massless scalar field}},\ }\href
  {https://doi.org/10.1103/PhysRevLett.70.9} {\bibfield  {journal} {\bibinfo
  {journal} {Phys. Rev. Lett.}\ }\textbf {\bibinfo {volume} {70}},\ \bibinfo
  {pages} {9} (\bibinfo {year} {1993})}\BibitemShut {NoStop}%
\bibitem [{\citenamefont {Niemeyer}\ and\ \citenamefont
  {Jedamzik}(1998)}]{Niemeyer:1997mt}%
  \BibitemOpen
  \bibfield  {author} {\bibinfo {author} {\bibfnamefont {J.~C.}\ \bibnamefont
  {Niemeyer}}\ and\ \bibinfo {author} {\bibfnamefont {K.}~\bibnamefont
  {Jedamzik}},\ }\bibfield  {title} {\bibinfo {title} {{Near-critical
  gravitational collapse and the initial mass function of primordial black
  holes}},\ }\href {https://doi.org/10.1103/PhysRevLett.80.5481} {\bibfield
  {journal} {\bibinfo  {journal} {Phys. Rev. Lett.}\ }\textbf {\bibinfo
  {volume} {80}},\ \bibinfo {pages} {5481} (\bibinfo {year} {1998})},\ \Eprint
  {https://arxiv.org/abs/astro-ph/9709072} {arXiv:astro-ph/9709072}
  \BibitemShut {NoStop}%
\bibitem [{\citenamefont {Inui}\ \emph {et~al.}(2025)\citenamefont {Inui},
  \citenamefont {Joana}, \citenamefont {Motohashi}, \citenamefont {Pi},
  \citenamefont {Tada},\ and\ \citenamefont {Yokoyama}}]{Inui:2024fgk}%
  \BibitemOpen
  \bibfield  {author} {\bibinfo {author} {\bibfnamefont {R.}~\bibnamefont
  {Inui}}, \bibinfo {author} {\bibfnamefont {C.}~\bibnamefont {Joana}},
  \bibinfo {author} {\bibfnamefont {H.}~\bibnamefont {Motohashi}}, \bibinfo
  {author} {\bibfnamefont {S.}~\bibnamefont {Pi}}, \bibinfo {author}
  {\bibfnamefont {Y.}~\bibnamefont {Tada}},\ and\ \bibinfo {author}
  {\bibfnamefont {S.}~\bibnamefont {Yokoyama}},\ }\bibfield  {title} {\bibinfo
  {title} {{Primordial black holes and induced gravitational waves from
  logarithmic non-Gaussianity}},\ }\href
  {https://doi.org/10.1088/1475-7516/2025/03/021} {\bibfield  {journal}
  {\bibinfo  {journal} {JCAP}\ }\textbf {\bibinfo {volume} {03}},\ \bibinfo
  {pages} {021}},\ \Eprint {https://arxiv.org/abs/2411.07647} {arXiv:2411.07647
  [astro-ph.CO]} \BibitemShut {NoStop}%
\bibitem [{\citenamefont {Hawking}(1974)}]{Hawking:1974rv}%
  \BibitemOpen
  \bibfield  {author} {\bibinfo {author} {\bibfnamefont {S.}~\bibnamefont
  {Hawking}},\ }\bibfield  {title} {\bibinfo {title} {{Black hole
  explosions}},\ }\href {https://doi.org/10.1038/248030a0} {\bibfield
  {journal} {\bibinfo  {journal} {Nature}\ }\textbf {\bibinfo {volume} {248}},\
  \bibinfo {pages} {30} (\bibinfo {year} {1974})}\BibitemShut {NoStop}%
\bibitem [{\citenamefont {Bays}\ \emph {et~al.}(2012)\citenamefont {Bays} \emph
  {et~al.}}]{Super-Kamiokande:2011lwo}%
  \BibitemOpen
  \bibfield  {author} {\bibinfo {author} {\bibfnamefont {K.}~\bibnamefont
  {Bays}} \emph {et~al.} (\bibinfo {collaboration} {Super-Kamiokande}),\
  }\bibfield  {title} {\bibinfo {title} {{Supernova Relic Neutrino Search at
  Super-Kamiokande}},\ }\href {https://doi.org/10.1103/PhysRevD.85.052007}
  {\bibfield  {journal} {\bibinfo  {journal} {Phys. Rev. D}\ }\textbf {\bibinfo
  {volume} {85}},\ \bibinfo {pages} {052007} (\bibinfo {year} {2012})},\
  \Eprint {https://arxiv.org/abs/1111.5031} {arXiv:1111.5031 [hep-ex]}
  \BibitemShut {NoStop}%
\bibitem [{\citenamefont {Carr}\ \emph {et~al.}(2016)\citenamefont {Carr},
  \citenamefont {Kuhnel},\ and\ \citenamefont {Sandstad}}]{Carr:2016drx}%
  \BibitemOpen
  \bibfield  {author} {\bibinfo {author} {\bibfnamefont {B.}~\bibnamefont
  {Carr}}, \bibinfo {author} {\bibfnamefont {F.}~\bibnamefont {Kuhnel}},\ and\
  \bibinfo {author} {\bibfnamefont {M.}~\bibnamefont {Sandstad}},\ }\bibfield
  {title} {\bibinfo {title} {{Primordial Black Holes as Dark Matter}},\ }\href
  {https://doi.org/10.1103/PhysRevD.94.083504} {\bibfield  {journal} {\bibinfo
  {journal} {Phys. Rev. D}\ }\textbf {\bibinfo {volume} {94}},\ \bibinfo
  {pages} {083504} (\bibinfo {year} {2016})},\ \Eprint
  {https://arxiv.org/abs/1607.06077} {arXiv:1607.06077 [astro-ph.CO]}
  \BibitemShut {NoStop}%
\bibitem [{\citenamefont {Carr}\ and\ \citenamefont
  {Kuhnel}(2020)}]{Carr:2020xqk}%
  \BibitemOpen
  \bibfield  {author} {\bibinfo {author} {\bibfnamefont {B.}~\bibnamefont
  {Carr}}\ and\ \bibinfo {author} {\bibfnamefont {F.}~\bibnamefont {Kuhnel}},\
  }\bibfield  {title} {\bibinfo {title} {{Primordial Black Holes as Dark
  Matter: Recent Developments}},\ }\href
  {https://doi.org/10.1146/annurev-nucl-050520-125911} {\bibfield  {journal}
  {\bibinfo  {journal} {Ann. Rev. Nucl. Part. Sci.}\ }\textbf {\bibinfo
  {volume} {70}},\ \bibinfo {pages} {355} (\bibinfo {year} {2020})},\ \Eprint
  {https://arxiv.org/abs/2006.02838} {arXiv:2006.02838 [astro-ph.CO]}
  \BibitemShut {NoStop}%
\bibitem [{\citenamefont {Inomata}\ \emph {et~al.}(2017)\citenamefont
  {Inomata}, \citenamefont {Kawasaki}, \citenamefont {Mukaida}, \citenamefont
  {Tada},\ and\ \citenamefont {Yanagida}}]{Inomata:2017okj}%
  \BibitemOpen
  \bibfield  {author} {\bibinfo {author} {\bibfnamefont {K.}~\bibnamefont
  {Inomata}}, \bibinfo {author} {\bibfnamefont {M.}~\bibnamefont {Kawasaki}},
  \bibinfo {author} {\bibfnamefont {K.}~\bibnamefont {Mukaida}}, \bibinfo
  {author} {\bibfnamefont {Y.}~\bibnamefont {Tada}},\ and\ \bibinfo {author}
  {\bibfnamefont {T.~T.}\ \bibnamefont {Yanagida}},\ }\bibfield  {title}
  {\bibinfo {title} {{Inflationary Primordial Black Holes as All Dark
  Matter}},\ }\href {https://doi.org/10.1103/PhysRevD.96.043504} {\bibfield
  {journal} {\bibinfo  {journal} {Phys. Rev. D}\ }\textbf {\bibinfo {volume}
  {96}},\ \bibinfo {pages} {043504} (\bibinfo {year} {2017})},\ \Eprint
  {https://arxiv.org/abs/1701.02544} {arXiv:1701.02544 [astro-ph.CO]}
  \BibitemShut {NoStop}%
\bibitem [{\citenamefont {Bertone}\ and\ \citenamefont
  {Hooper}(2018)}]{Bertone:2016nfn}%
  \BibitemOpen
  \bibfield  {author} {\bibinfo {author} {\bibfnamefont {G.}~\bibnamefont
  {Bertone}}\ and\ \bibinfo {author} {\bibfnamefont {D.}~\bibnamefont
  {Hooper}},\ }\bibfield  {title} {\bibinfo {title} {{History of dark
  matter}},\ }\href {https://doi.org/10.1103/RevModPhys.90.045002} {\bibfield
  {journal} {\bibinfo  {journal} {Rev. Mod. Phys.}\ }\textbf {\bibinfo {volume}
  {90}},\ \bibinfo {pages} {045002} (\bibinfo {year} {2018})},\ \Eprint
  {https://arxiv.org/abs/1605.04909} {arXiv:1605.04909 [astro-ph.CO]}
  \BibitemShut {NoStop}%
\bibitem [{\citenamefont {Niikura}\ \emph
  {et~al.}(2019{\natexlab{a}})\citenamefont {Niikura} \emph
  {et~al.}}]{Niikura:2017zjd}%
  \BibitemOpen
  \bibfield  {author} {\bibinfo {author} {\bibfnamefont {H.}~\bibnamefont
  {Niikura}} \emph {et~al.},\ }\bibfield  {title} {\bibinfo {title}
  {{Microlensing constraints on primordial black holes with Subaru/HSC
  Andromeda observations}},\ }\href {https://doi.org/10.1038/s41550-019-0723-1}
  {\bibfield  {journal} {\bibinfo  {journal} {Nature Astron.}\ }\textbf
  {\bibinfo {volume} {3}},\ \bibinfo {pages} {524} (\bibinfo {year}
  {2019}{\natexlab{a}})},\ \Eprint {https://arxiv.org/abs/1701.02151}
  {arXiv:1701.02151 [astro-ph.CO]} \BibitemShut {NoStop}%
\bibitem [{\citenamefont {Tisserand}\ \emph {et~al.}(2007)\citenamefont
  {Tisserand} \emph {et~al.}}]{Tisserand:2006zx}%
  \BibitemOpen
  \bibfield  {author} {\bibinfo {author} {\bibfnamefont {P.}~\bibnamefont
  {Tisserand}} \emph {et~al.} (\bibinfo {collaboration} {EROS-2}),\ }\bibfield
  {title} {\bibinfo {title} {{Limits on the Macho Content of the Galactic Halo
  from the EROS-2 Survey of the Magellanic Clouds}},\ }\href
  {https://doi.org/10.1051/0004-6361:20066017} {\bibfield  {journal} {\bibinfo
  {journal} {Astron. Astrophys.}\ }\textbf {\bibinfo {volume} {469}},\ \bibinfo
  {pages} {387} (\bibinfo {year} {2007})},\ \Eprint
  {https://arxiv.org/abs/astro-ph/0607207} {arXiv:astro-ph/0607207}
  \BibitemShut {NoStop}%
\bibitem [{\citenamefont {Niikura}\ \emph
  {et~al.}(2019{\natexlab{b}})\citenamefont {Niikura}, \citenamefont {Takada},
  \citenamefont {Yokoyama}, \citenamefont {Sumi},\ and\ \citenamefont
  {Masaki}}]{Niikura:2019kqi}%
  \BibitemOpen
  \bibfield  {author} {\bibinfo {author} {\bibfnamefont {H.}~\bibnamefont
  {Niikura}}, \bibinfo {author} {\bibfnamefont {M.}~\bibnamefont {Takada}},
  \bibinfo {author} {\bibfnamefont {S.}~\bibnamefont {Yokoyama}}, \bibinfo
  {author} {\bibfnamefont {T.}~\bibnamefont {Sumi}},\ and\ \bibinfo {author}
  {\bibfnamefont {S.}~\bibnamefont {Masaki}},\ }\bibfield  {title} {\bibinfo
  {title} {{Constraints on Earth-mass primordial black holes from OGLE 5-year
  microlensing events}},\ }\href {https://doi.org/10.1103/PhysRevD.99.083503}
  {\bibfield  {journal} {\bibinfo  {journal} {Phys. Rev. D}\ }\textbf {\bibinfo
  {volume} {99}},\ \bibinfo {pages} {083503} (\bibinfo {year}
  {2019}{\natexlab{b}})},\ \Eprint {https://arxiv.org/abs/1901.07120}
  {arXiv:1901.07120 [astro-ph.CO]} \BibitemShut {NoStop}%
\bibitem [{\citenamefont {Mr\'oz}\ \emph
  {et~al.}(2024{\natexlab{a}})\citenamefont {Mr\'oz} \emph
  {et~al.}}]{Mroz:2024wia}%
  \BibitemOpen
  \bibfield  {author} {\bibinfo {author} {\bibfnamefont {P.}~\bibnamefont
  {Mr\'oz}} \emph {et~al.},\ }\bibfield  {title} {\bibinfo {title} {{Limits on
  Planetary-mass Primordial Black Holes from the OGLE High-cadence Survey of
  the Magellanic Clouds}},\ }\href {https://doi.org/10.3847/2041-8213/ad8e68}
  {\bibfield  {journal} {\bibinfo  {journal} {Astrophys. J. Lett.}\ }\textbf
  {\bibinfo {volume} {976}},\ \bibinfo {pages} {L19} (\bibinfo {year}
  {2024}{\natexlab{a}})},\ \Eprint {https://arxiv.org/abs/2410.06251}
  {arXiv:2410.06251 [astro-ph.CO]} \BibitemShut {NoStop}%
\bibitem [{\citenamefont {Mr\'oz}\ \emph
  {et~al.}(2024{\natexlab{b}})\citenamefont {Mr\'oz} \emph
  {et~al.}}]{Mroz:2024mse}%
  \BibitemOpen
  \bibfield  {author} {\bibinfo {author} {\bibfnamefont {P.}~\bibnamefont
  {Mr\'oz}} \emph {et~al.},\ }\bibfield  {title} {\bibinfo {title} {{No massive
  black holes in the Milky Way halo}},\ }\href
  {https://doi.org/10.1038/s41586-024-07704-6} {\bibfield  {journal} {\bibinfo
  {journal} {Nature}\ }\textbf {\bibinfo {volume} {632}},\ \bibinfo {pages}
  {749} (\bibinfo {year} {2024}{\natexlab{b}})},\ \Eprint
  {https://arxiv.org/abs/2403.02386} {arXiv:2403.02386 [astro-ph.GA]}
  \BibitemShut {NoStop}%
\bibitem [{\citenamefont {Smyth}\ \emph {et~al.}(2020)\citenamefont {Smyth},
  \citenamefont {Profumo}, \citenamefont {English}, \citenamefont {Jeltema},
  \citenamefont {McKinnon},\ and\ \citenamefont
  {Guhathakurta}}]{Smyth:2019whb}%
  \BibitemOpen
  \bibfield  {author} {\bibinfo {author} {\bibfnamefont {N.}~\bibnamefont
  {Smyth}}, \bibinfo {author} {\bibfnamefont {S.}~\bibnamefont {Profumo}},
  \bibinfo {author} {\bibfnamefont {S.}~\bibnamefont {English}}, \bibinfo
  {author} {\bibfnamefont {T.}~\bibnamefont {Jeltema}}, \bibinfo {author}
  {\bibfnamefont {K.}~\bibnamefont {McKinnon}},\ and\ \bibinfo {author}
  {\bibfnamefont {P.}~\bibnamefont {Guhathakurta}},\ }\bibfield  {title}
  {\bibinfo {title} {{Updated Constraints on Asteroid-Mass Primordial Black
  Holes as Dark Matter}},\ }\href {https://doi.org/10.1103/PhysRevD.101.063005}
  {\bibfield  {journal} {\bibinfo  {journal} {Phys. Rev. D}\ }\textbf {\bibinfo
  {volume} {101}},\ \bibinfo {pages} {063005} (\bibinfo {year} {2020})},\
  \Eprint {https://arxiv.org/abs/1910.01285} {arXiv:1910.01285 [astro-ph.CO]}
  \BibitemShut {NoStop}%
\bibitem [{\citenamefont {Ali-Haïmoud}\ \emph {et~al.}(2017)\citenamefont
  {Ali-Haïmoud}, \citenamefont {Kovetz},\ and\ \citenamefont
  {Kamionkowski}}]{Ali-Haimoud:2017rtz}%
  \BibitemOpen
  \bibfield  {author} {\bibinfo {author} {\bibfnamefont {Y.}~\bibnamefont
  {Ali-Haïmoud}}, \bibinfo {author} {\bibfnamefont {E.~D.}\ \bibnamefont
  {Kovetz}},\ and\ \bibinfo {author} {\bibfnamefont {M.}~\bibnamefont
  {Kamionkowski}},\ }\bibfield  {title} {\bibinfo {title} {{Merger rate of
  primordial black-hole binaries}},\ }\href
  {https://doi.org/10.1103/PhysRevD.96.123523} {\bibfield  {journal} {\bibinfo
  {journal} {Phys. Rev.}\ }\textbf {\bibinfo {volume} {D96}},\ \bibinfo {pages}
  {123523} (\bibinfo {year} {2017})},\ \Eprint
  {https://arxiv.org/abs/1709.06576} {arXiv:1709.06576 [astro-ph.CO]}
  \BibitemShut {NoStop}%
\bibitem [{\citenamefont {Bird}\ \emph {et~al.}(2016)\citenamefont {Bird},
  \citenamefont {Cholis}, \citenamefont {Mu\~noz}, \citenamefont
  {Ali-Haïmoud}, \citenamefont {Kamionkowski}, \citenamefont {Kovetz},
  \citenamefont {Raccanelli},\ and\ \citenamefont {Riess}}]{Bird:2016dcv}%
  \BibitemOpen
  \bibfield  {author} {\bibinfo {author} {\bibfnamefont {S.}~\bibnamefont
  {Bird}}, \bibinfo {author} {\bibfnamefont {I.}~\bibnamefont {Cholis}},
  \bibinfo {author} {\bibfnamefont {J.~B.}\ \bibnamefont {Mu\~noz}}, \bibinfo
  {author} {\bibfnamefont {Y.}~\bibnamefont {Ali-Haïmoud}}, \bibinfo {author}
  {\bibfnamefont {M.}~\bibnamefont {Kamionkowski}}, \bibinfo {author}
  {\bibfnamefont {E.~D.}\ \bibnamefont {Kovetz}}, \bibinfo {author}
  {\bibfnamefont {A.}~\bibnamefont {Raccanelli}},\ and\ \bibinfo {author}
  {\bibfnamefont {A.~G.}\ \bibnamefont {Riess}},\ }\bibfield  {title} {\bibinfo
  {title} {{Did LIGO detect dark matter?}},\ }\href
  {https://doi.org/10.1103/PhysRevLett.116.201301} {\bibfield  {journal}
  {\bibinfo  {journal} {Phys. Rev. Lett.}\ }\textbf {\bibinfo {volume} {116}},\
  \bibinfo {pages} {201301} (\bibinfo {year} {2016})},\ \Eprint
  {https://arxiv.org/abs/1603.00464} {arXiv:1603.00464 [astro-ph.CO]}
  \BibitemShut {NoStop}%
\bibitem [{\citenamefont {Sasaki}\ \emph {et~al.}(2016)\citenamefont {Sasaki},
  \citenamefont {Suyama}, \citenamefont {Tanaka},\ and\ \citenamefont
  {Yokoyama}}]{Sasaki:2016jop}%
  \BibitemOpen
  \bibfield  {author} {\bibinfo {author} {\bibfnamefont {M.}~\bibnamefont
  {Sasaki}}, \bibinfo {author} {\bibfnamefont {T.}~\bibnamefont {Suyama}},
  \bibinfo {author} {\bibfnamefont {T.}~\bibnamefont {Tanaka}},\ and\ \bibinfo
  {author} {\bibfnamefont {S.}~\bibnamefont {Yokoyama}},\ }\bibfield  {title}
  {\bibinfo {title} {{Primordial Black Hole Scenario for the Gravitational-Wave
  Event GW150914}},\ }\href {https://doi.org/10.1103/PhysRevLett.117.061101}
  {\bibfield  {journal} {\bibinfo  {journal} {Phys. Rev. Lett.}\ }\textbf
  {\bibinfo {volume} {117}},\ \bibinfo {pages} {061101} (\bibinfo {year}
  {2016})},\ \bibinfo {note} {[Erratum: Phys.Rev.Lett. 121, 059901 (2018)]},\
  \Eprint {https://arxiv.org/abs/1603.08338} {arXiv:1603.08338 [astro-ph.CO]}
  \BibitemShut {NoStop}%
\bibitem [{\citenamefont {Cholis}\ \emph {et~al.}(2016)\citenamefont {Cholis},
  \citenamefont {Kovetz}, \citenamefont {Ali-Haïmoud}, \citenamefont {Bird},
  \citenamefont {Kamionkowski}, \citenamefont {Mu\~noz},\ and\ \citenamefont
  {Raccanelli}}]{Cholis:2016kqi}%
  \BibitemOpen
  \bibfield  {author} {\bibinfo {author} {\bibfnamefont {I.}~\bibnamefont
  {Cholis}}, \bibinfo {author} {\bibfnamefont {E.~D.}\ \bibnamefont {Kovetz}},
  \bibinfo {author} {\bibfnamefont {Y.}~\bibnamefont {Ali-Haïmoud}}, \bibinfo
  {author} {\bibfnamefont {S.}~\bibnamefont {Bird}}, \bibinfo {author}
  {\bibfnamefont {M.}~\bibnamefont {Kamionkowski}}, \bibinfo {author}
  {\bibfnamefont {J.~B.}\ \bibnamefont {Mu\~noz}},\ and\ \bibinfo {author}
  {\bibfnamefont {A.}~\bibnamefont {Raccanelli}},\ }\bibfield  {title}
  {\bibinfo {title} {{Orbital eccentricities in primordial black hole
  binaries}},\ }\href {https://doi.org/10.1103/PhysRevD.94.084013} {\bibfield
  {journal} {\bibinfo  {journal} {Phys. Rev. D}\ }\textbf {\bibinfo {volume}
  {94}},\ \bibinfo {pages} {084013} (\bibinfo {year} {2016})},\ \Eprint
  {https://arxiv.org/abs/1606.07437} {arXiv:1606.07437 [astro-ph.HE]}
  \BibitemShut {NoStop}%
\bibitem [{\citenamefont {Clesse}\ and\ \citenamefont
  {Garc\'\i{}a-Bellido}(2017)}]{Clesse:2016vqa}%
  \BibitemOpen
  \bibfield  {author} {\bibinfo {author} {\bibfnamefont {S.}~\bibnamefont
  {Clesse}}\ and\ \bibinfo {author} {\bibfnamefont {J.}~\bibnamefont
  {Garc\'\i{}a-Bellido}},\ }\bibfield  {title} {\bibinfo {title} {{The
  clustering of massive Primordial Black Holes as Dark Matter: measuring their
  mass distribution with Advanced LIGO}},\ }\href
  {https://doi.org/10.1016/j.dark.2016.10.002} {\bibfield  {journal} {\bibinfo
  {journal} {Phys. Dark Univ.}\ }\textbf {\bibinfo {volume} {15}},\ \bibinfo
  {pages} {142} (\bibinfo {year} {2017})},\ \Eprint
  {https://arxiv.org/abs/1603.05234} {arXiv:1603.05234 [astro-ph.CO]}
  \BibitemShut {NoStop}%
\bibitem [{\citenamefont {De~Luca}\ \emph {et~al.}(2020)\citenamefont
  {De~Luca}, \citenamefont {Franciolini}, \citenamefont {Pani},\ and\
  \citenamefont {Riotto}}]{DeLuca:2020qqa}%
  \BibitemOpen
  \bibfield  {author} {\bibinfo {author} {\bibfnamefont {V.}~\bibnamefont
  {De~Luca}}, \bibinfo {author} {\bibfnamefont {G.}~\bibnamefont
  {Franciolini}}, \bibinfo {author} {\bibfnamefont {P.}~\bibnamefont {Pani}},\
  and\ \bibinfo {author} {\bibfnamefont {A.}~\bibnamefont {Riotto}},\
  }\bibfield  {title} {\bibinfo {title} {{Primordial Black Holes Confront
  LIGO/Virgo data: Current situation}},\ }\href
  {https://doi.org/10.1088/1475-7516/2020/06/044} {\bibfield  {journal}
  {\bibinfo  {journal} {JCAP}\ }\textbf {\bibinfo {volume} {06}},\ \bibinfo
  {pages} {044}},\ \Eprint {https://arxiv.org/abs/2005.05641} {arXiv:2005.05641
  [astro-ph.CO]} \BibitemShut {NoStop}%
\bibitem [{\citenamefont {Andr\'es-Carcasona}\ \emph
  {et~al.}(2024)\citenamefont {Andr\'es-Carcasona}, \citenamefont {Iovino},
  \citenamefont {Vaskonen}, \citenamefont {Veerm\"ae}, \citenamefont
  {Mart\'\i{}nez}, \citenamefont {Pujol\`as},\ and\ \citenamefont
  {Mir}}]{Andres-Carcasona:2024wqk}%
  \BibitemOpen
  \bibfield  {author} {\bibinfo {author} {\bibfnamefont {M.}~\bibnamefont
  {Andr\'es-Carcasona}}, \bibinfo {author} {\bibfnamefont {A.~J.}\ \bibnamefont
  {Iovino}}, \bibinfo {author} {\bibfnamefont {V.}~\bibnamefont {Vaskonen}},
  \bibinfo {author} {\bibfnamefont {H.}~\bibnamefont {Veerm\"ae}}, \bibinfo
  {author} {\bibfnamefont {M.}~\bibnamefont {Mart\'\i{}nez}}, \bibinfo {author}
  {\bibfnamefont {O.}~\bibnamefont {Pujol\`as}},\ and\ \bibinfo {author}
  {\bibfnamefont {L.~M.}\ \bibnamefont {Mir}},\ }\bibfield  {title} {\bibinfo
  {title} {{Constraints on primordial black holes from LIGO-Virgo-KAGRA O3
  events}},\ }\href {https://doi.org/10.1103/PhysRevD.110.023040} {\bibfield
  {journal} {\bibinfo  {journal} {Phys. Rev. D}\ }\textbf {\bibinfo {volume}
  {110}},\ \bibinfo {pages} {023040} (\bibinfo {year} {2024})},\ \Eprint
  {https://arxiv.org/abs/2405.05732} {arXiv:2405.05732 [astro-ph.CO]}
  \BibitemShut {NoStop}%
\bibitem [{\citenamefont {Liu}\ \emph {et~al.}(2019{\natexlab{a}})\citenamefont
  {Liu}, \citenamefont {Guo},\ and\ \citenamefont {Cai}}]{Liu:2018ess}%
  \BibitemOpen
  \bibfield  {author} {\bibinfo {author} {\bibfnamefont {L.}~\bibnamefont
  {Liu}}, \bibinfo {author} {\bibfnamefont {Z.-K.}\ \bibnamefont {Guo}},\ and\
  \bibinfo {author} {\bibfnamefont {R.-G.}\ \bibnamefont {Cai}},\ }\bibfield
  {title} {\bibinfo {title} {{Effects of the surrounding primordial black holes
  on the merger rate of primordial black hole binaries}},\ }\href
  {https://doi.org/10.1103/PhysRevD.99.063523} {\bibfield  {journal} {\bibinfo
  {journal} {Phys. Rev. D}\ }\textbf {\bibinfo {volume} {99}},\ \bibinfo
  {pages} {063523} (\bibinfo {year} {2019}{\natexlab{a}})},\ \Eprint
  {https://arxiv.org/abs/1812.05376} {arXiv:1812.05376 [astro-ph.CO]}
  \BibitemShut {NoStop}%
\bibitem [{\citenamefont {Liu}\ \emph {et~al.}(2019{\natexlab{b}})\citenamefont
  {Liu}, \citenamefont {Guo},\ and\ \citenamefont {Cai}}]{Liu:2019rnx}%
  \BibitemOpen
  \bibfield  {author} {\bibinfo {author} {\bibfnamefont {L.}~\bibnamefont
  {Liu}}, \bibinfo {author} {\bibfnamefont {Z.-K.}\ \bibnamefont {Guo}},\ and\
  \bibinfo {author} {\bibfnamefont {R.-G.}\ \bibnamefont {Cai}},\ }\bibfield
  {title} {\bibinfo {title} {{Effects of the merger history on the merger rate
  density of primordial black hole binaries}},\ }\href
  {https://doi.org/10.1140/epjc/s10052-019-7227-0} {\bibfield  {journal}
  {\bibinfo  {journal} {Eur. Phys. J. C}\ }\textbf {\bibinfo {volume} {79}},\
  \bibinfo {pages} {717} (\bibinfo {year} {2019}{\natexlab{b}})},\ \Eprint
  {https://arxiv.org/abs/1901.07672} {arXiv:1901.07672 [astro-ph.CO]}
  \BibitemShut {NoStop}%
\bibitem [{\citenamefont {Carr}(1981)}]{carr1981pregalactic}%
  \BibitemOpen
  \bibfield  {author} {\bibinfo {author} {\bibfnamefont {B.}~\bibnamefont
  {Carr}},\ }\bibfield  {title} {\bibinfo {title} {Pregalactic black hole
  accretion and the thermal history of the universe},\ }\href@noop {}
  {\bibfield  {journal} {\bibinfo  {journal} {Monthly Notices of the Royal
  Astronomical Society}\ }\textbf {\bibinfo {volume} {194}},\ \bibinfo {pages}
  {639} (\bibinfo {year} {1981})}\BibitemShut {NoStop}%
\bibitem [{\citenamefont {Ricotti}\ \emph {et~al.}(2008)\citenamefont
  {Ricotti}, \citenamefont {Ostriker},\ and\ \citenamefont
  {Mack}}]{Ricotti:2007au}%
  \BibitemOpen
  \bibfield  {author} {\bibinfo {author} {\bibfnamefont {M.}~\bibnamefont
  {Ricotti}}, \bibinfo {author} {\bibfnamefont {J.~P.}\ \bibnamefont
  {Ostriker}},\ and\ \bibinfo {author} {\bibfnamefont {K.~J.}\ \bibnamefont
  {Mack}},\ }\bibfield  {title} {\bibinfo {title} {{Effect of Primordial Black
  Holes on the Cosmic Microwave Background and Cosmological Parameter
  Estimates}},\ }\href {https://doi.org/10.1086/587831} {\bibfield  {journal}
  {\bibinfo  {journal} {Astrophys. J.}\ }\textbf {\bibinfo {volume} {680}},\
  \bibinfo {pages} {829} (\bibinfo {year} {2008})},\ \Eprint
  {https://arxiv.org/abs/0709.0524} {arXiv:0709.0524 [astro-ph]} \BibitemShut
  {NoStop}%
\bibitem [{\citenamefont {Serpico}\ \emph {et~al.}(2020)\citenamefont
  {Serpico}, \citenamefont {Poulin}, \citenamefont {Inman},\ and\ \citenamefont
  {Kohri}}]{Serpico:2020ehh}%
  \BibitemOpen
  \bibfield  {author} {\bibinfo {author} {\bibfnamefont {P.~D.}\ \bibnamefont
  {Serpico}}, \bibinfo {author} {\bibfnamefont {V.}~\bibnamefont {Poulin}},
  \bibinfo {author} {\bibfnamefont {D.}~\bibnamefont {Inman}},\ and\ \bibinfo
  {author} {\bibfnamefont {K.}~\bibnamefont {Kohri}},\ }\bibfield  {title}
  {\bibinfo {title} {{Cosmic microwave background bounds on primordial black
  holes including dark matter halo accretion}},\ }\href
  {https://doi.org/10.1103/PhysRevResearch.2.023204} {\bibfield  {journal}
  {\bibinfo  {journal} {Phys. Rev. Res.}\ }\textbf {\bibinfo {volume} {2}},\
  \bibinfo {pages} {023204} (\bibinfo {year} {2020})},\ \Eprint
  {https://arxiv.org/abs/2002.10771} {arXiv:2002.10771 [astro-ph.CO]}
  \BibitemShut {NoStop}%
\bibitem [{\citenamefont {Liu}\ \emph {et~al.}(2024)\citenamefont {Liu},
  \citenamefont {Wu},\ and\ \citenamefont {Chen}}]{Liu:2023hpw}%
  \BibitemOpen
  \bibfield  {author} {\bibinfo {author} {\bibfnamefont {L.}~\bibnamefont
  {Liu}}, \bibinfo {author} {\bibfnamefont {Y.}~\bibnamefont {Wu}},\ and\
  \bibinfo {author} {\bibfnamefont {Z.-C.}\ \bibnamefont {Chen}},\ }\bibfield
  {title} {\bibinfo {title} {{Simultaneously probing the sound speed and
  equation of state of the early Universe with pulsar timing arrays}},\ }\href
  {https://doi.org/10.1088/1475-7516/2024/04/011} {\bibfield  {journal}
  {\bibinfo  {journal} {JCAP}\ }\textbf {\bibinfo {volume} {04}},\ \bibinfo
  {pages} {011}},\ \Eprint {https://arxiv.org/abs/2310.16500} {arXiv:2310.16500
  [astro-ph.CO]} \BibitemShut {NoStop}%
\bibitem [{\citenamefont {Liu}\ \emph {et~al.}(2023)\citenamefont {Liu},
  \citenamefont {Chen},\ and\ \citenamefont {Huang}}]{Liu:2023pau}%
  \BibitemOpen
  \bibfield  {author} {\bibinfo {author} {\bibfnamefont {L.}~\bibnamefont
  {Liu}}, \bibinfo {author} {\bibfnamefont {Z.-C.}\ \bibnamefont {Chen}},\ and\
  \bibinfo {author} {\bibfnamefont {Q.-G.}\ \bibnamefont {Huang}},\ }\bibfield
  {title} {\bibinfo {title} {{Probing the equation of state of the early
  Universe with pulsar timing arrays}},\ }\href
  {https://doi.org/10.1088/1475-7516/2023/11/071} {\bibfield  {journal}
  {\bibinfo  {journal} {JCAP}\ }\textbf {\bibinfo {volume} {11}},\ \bibinfo
  {pages} {071}},\ \Eprint {https://arxiv.org/abs/2307.14911} {arXiv:2307.14911
  [astro-ph.CO]} \BibitemShut {NoStop}%
\bibitem [{\citenamefont {Chen}\ \emph {et~al.}(2024)\citenamefont {Chen},
  \citenamefont {Dimopoulos}, \citenamefont {Er\"oncel},\ and\ \citenamefont
  {Ghoshal}}]{Chen:2024roo}%
  \BibitemOpen
  \bibfield  {author} {\bibinfo {author} {\bibfnamefont {C.}~\bibnamefont
  {Chen}}, \bibinfo {author} {\bibfnamefont {K.}~\bibnamefont {Dimopoulos}},
  \bibinfo {author} {\bibfnamefont {C.}~\bibnamefont {Er\"oncel}},\ and\
  \bibinfo {author} {\bibfnamefont {A.}~\bibnamefont {Ghoshal}},\ }\bibfield
  {title} {\bibinfo {title} {{Enhanced primordial gravitational waves from a
  stiff postinflationary era due to an oscillating inflaton}},\ }\href
  {https://doi.org/10.1103/PhysRevD.110.063554} {\bibfield  {journal} {\bibinfo
   {journal} {Phys. Rev. D}\ }\textbf {\bibinfo {volume} {110}},\ \bibinfo
  {pages} {063554} (\bibinfo {year} {2024})},\ \Eprint
  {https://arxiv.org/abs/2405.01679} {arXiv:2405.01679 [hep-ph]} \BibitemShut
  {NoStop}%
\bibitem [{\citenamefont {Ananda}\ \emph {et~al.}(2007)\citenamefont {Ananda},
  \citenamefont {Clarkson},\ and\ \citenamefont {Wands}}]{Ananda:2006af}%
  \BibitemOpen
  \bibfield  {author} {\bibinfo {author} {\bibfnamefont {K.~N.}\ \bibnamefont
  {Ananda}}, \bibinfo {author} {\bibfnamefont {C.}~\bibnamefont {Clarkson}},\
  and\ \bibinfo {author} {\bibfnamefont {D.}~\bibnamefont {Wands}},\ }\bibfield
   {title} {\bibinfo {title} {{The Cosmological gravitational wave background
  from primordial density perturbations}},\ }\href
  {https://doi.org/10.1103/PhysRevD.75.123518} {\bibfield  {journal} {\bibinfo
  {journal} {Phys. Rev. D}\ }\textbf {\bibinfo {volume} {75}},\ \bibinfo
  {pages} {123518} (\bibinfo {year} {2007})},\ \Eprint
  {https://arxiv.org/abs/gr-qc/0612013} {arXiv:gr-qc/0612013} \BibitemShut
  {NoStop}%
\bibitem [{\citenamefont {Witkowski}(2022)}]{Witkowski:2022mtg}%
  \BibitemOpen
  \bibfield  {author} {\bibinfo {author} {\bibfnamefont {L.~T.}\ \bibnamefont
  {Witkowski}},\ }\bibfield  {title} {\bibinfo {title} {{SIGWfast: a python
  package for the computation of scalar-induced gravitational wave spectra}},\
  }\href@noop {} {\  (\bibinfo {year} {2022})},\ \Eprint
  {https://arxiv.org/abs/2209.05296} {arXiv:2209.05296 [astro-ph.CO]}
  \BibitemShut {NoStop}%
\bibitem [{\citenamefont {Balaji}\ \emph {et~al.}(2023)\citenamefont {Balaji},
  \citenamefont {Dom\`enech},\ and\ \citenamefont
  {Franciolini}}]{Balaji:2023ehk}%
  \BibitemOpen
  \bibfield  {author} {\bibinfo {author} {\bibfnamefont {S.}~\bibnamefont
  {Balaji}}, \bibinfo {author} {\bibfnamefont {G.}~\bibnamefont {Dom\`enech}},\
  and\ \bibinfo {author} {\bibfnamefont {G.}~\bibnamefont {Franciolini}},\
  }\bibfield  {title} {\bibinfo {title} {{Scalar-induced gravitational wave
  interpretation of PTA data: the role of scalar fluctuation propagation
  speed}},\ }\href {https://doi.org/10.1088/1475-7516/2023/10/041} {\bibfield
  {journal} {\bibinfo  {journal} {JCAP}\ }\textbf {\bibinfo {volume} {10}},\
  \bibinfo {pages} {041}},\ \Eprint {https://arxiv.org/abs/2307.08552}
  {arXiv:2307.08552 [gr-qc]} \BibitemShut {NoStop}%
\bibitem [{\citenamefont {Figueroa}\ and\ \citenamefont
  {Tanin}(2019)}]{Figueroa:2019paj}%
  \BibitemOpen
  \bibfield  {author} {\bibinfo {author} {\bibfnamefont {D.~G.}\ \bibnamefont
  {Figueroa}}\ and\ \bibinfo {author} {\bibfnamefont {E.~H.}\ \bibnamefont
  {Tanin}},\ }\bibfield  {title} {\bibinfo {title} {{Ability of LIGO and LISA
  to probe the equation of state of the early Universe}},\ }\href
  {https://doi.org/10.1088/1475-7516/2019/08/011} {\bibfield  {journal}
  {\bibinfo  {journal} {JCAP}\ }\textbf {\bibinfo {volume} {08}},\ \bibinfo
  {pages} {011}},\ \Eprint {https://arxiv.org/abs/1905.11960} {arXiv:1905.11960
  [astro-ph.CO]} \BibitemShut {NoStop}%
\bibitem [{\citenamefont {Bernal}\ and\ \citenamefont
  {Hajkarim}(2019)}]{Bernal:2019lpc}%
  \BibitemOpen
  \bibfield  {author} {\bibinfo {author} {\bibfnamefont {N.}~\bibnamefont
  {Bernal}}\ and\ \bibinfo {author} {\bibfnamefont {F.}~\bibnamefont
  {Hajkarim}},\ }\bibfield  {title} {\bibinfo {title} {{Primordial
  Gravitational Waves in Nonstandard Cosmologies}},\ }\href
  {https://doi.org/10.1103/PhysRevD.100.063502} {\bibfield  {journal} {\bibinfo
   {journal} {Phys. Rev. D}\ }\textbf {\bibinfo {volume} {100}},\ \bibinfo
  {pages} {063502} (\bibinfo {year} {2019})},\ \Eprint
  {https://arxiv.org/abs/1905.10410} {arXiv:1905.10410 [astro-ph.CO]}
  \BibitemShut {NoStop}%
\bibitem [{\citenamefont {Bernal}\ \emph {et~al.}(2020)\citenamefont {Bernal},
  \citenamefont {Ghoshal}, \citenamefont {Hajkarim},\ and\ \citenamefont
  {Lambiase}}]{Bernal:2020ywq}%
  \BibitemOpen
  \bibfield  {author} {\bibinfo {author} {\bibfnamefont {N.}~\bibnamefont
  {Bernal}}, \bibinfo {author} {\bibfnamefont {A.}~\bibnamefont {Ghoshal}},
  \bibinfo {author} {\bibfnamefont {F.}~\bibnamefont {Hajkarim}},\ and\
  \bibinfo {author} {\bibfnamefont {G.}~\bibnamefont {Lambiase}},\ }\bibfield
  {title} {\bibinfo {title} {{Primordial Gravitational Wave Signals in Modified
  Cosmologies}},\ }\href {https://doi.org/10.1088/1475-7516/2020/11/051}
  {\bibfield  {journal} {\bibinfo  {journal} {JCAP}\ }\textbf {\bibinfo
  {volume} {11}},\ \bibinfo {pages} {051}},\ \Eprint
  {https://arxiv.org/abs/2008.04959} {arXiv:2008.04959 [gr-qc]} \BibitemShut
  {NoStop}%
\bibitem [{\citenamefont {Espinosa}\ \emph {et~al.}(2018)\citenamefont
  {Espinosa}, \citenamefont {Racco},\ and\ \citenamefont
  {Riotto}}]{Espinosa:2018eve}%
  \BibitemOpen
  \bibfield  {author} {\bibinfo {author} {\bibfnamefont {J.~R.}\ \bibnamefont
  {Espinosa}}, \bibinfo {author} {\bibfnamefont {D.}~\bibnamefont {Racco}},\
  and\ \bibinfo {author} {\bibfnamefont {A.}~\bibnamefont {Riotto}},\
  }\bibfield  {title} {\bibinfo {title} {{A Cosmological Signature of the SM
  Higgs Instability: Gravitational Waves}},\ }\href
  {https://doi.org/10.1088/1475-7516/2018/09/012} {\bibfield  {journal}
  {\bibinfo  {journal} {JCAP}\ }\textbf {\bibinfo {volume} {09}},\ \bibinfo
  {pages} {012}},\ \Eprint {https://arxiv.org/abs/1804.07732} {arXiv:1804.07732
  [hep-ph]} \BibitemShut {NoStop}%
\bibitem [{\citenamefont {Pi}\ and\ \citenamefont {Sasaki}(2020)}]{Pi:2020otn}%
  \BibitemOpen
  \bibfield  {author} {\bibinfo {author} {\bibfnamefont {S.}~\bibnamefont
  {Pi}}\ and\ \bibinfo {author} {\bibfnamefont {M.}~\bibnamefont {Sasaki}},\
  }\bibfield  {title} {\bibinfo {title} {{Gravitational Waves Induced by Scalar
  Perturbations with a Lognormal Peak}},\ }\href
  {https://doi.org/10.1088/1475-7516/2020/09/037} {\bibfield  {journal}
  {\bibinfo  {journal} {JCAP}\ }\textbf {\bibinfo {volume} {09}},\ \bibinfo
  {pages} {037}},\ \Eprint {https://arxiv.org/abs/2005.12306} {arXiv:2005.12306
  [gr-qc]} \BibitemShut {NoStop}%
\bibitem [{\citenamefont {Danzmann}(1997)}]{Danzmann:1997hm}%
  \BibitemOpen
  \bibfield  {author} {\bibinfo {author} {\bibfnamefont {K.}~\bibnamefont
  {Danzmann}},\ }\bibfield  {title} {\bibinfo {title} {{LISA: An ESA
  cornerstone mission for a gravitational wave observatory}},\ }\href
  {https://doi.org/10.1088/0264-9381/14/6/002} {\bibfield  {journal} {\bibinfo
  {journal} {Class. Quant. Grav.}\ }\textbf {\bibinfo {volume} {14}},\ \bibinfo
  {pages} {1399} (\bibinfo {year} {1997})}\BibitemShut {NoStop}%
\bibitem [{\citenamefont {Harry}(2010)}]{Harry:2010zz}%
  \BibitemOpen
  \bibfield  {author} {\bibinfo {author} {\bibfnamefont {G.~M.}\ \bibnamefont
  {Harry}} (\bibinfo {collaboration} {LIGO Scientific}),\ }\bibfield  {title}
  {\bibinfo {title} {{Advanced LIGO: The next generation of gravitational wave
  detectors}},\ }\href {https://doi.org/10.1088/0264-9381/27/8/084006}
  {\bibfield  {journal} {\bibinfo  {journal} {Class. Quant. Grav.}\ }\textbf
  {\bibinfo {volume} {27}},\ \bibinfo {pages} {084006} (\bibinfo {year}
  {2010})}\BibitemShut {NoStop}%
\bibitem [{\citenamefont {Aasi}\ \emph {et~al.}(2015)\citenamefont {Aasi} \emph
  {et~al.}}]{LIGOScientific:2014qfs}%
  \BibitemOpen
  \bibfield  {author} {\bibinfo {author} {\bibfnamefont {J.}~\bibnamefont
  {Aasi}} \emph {et~al.} (\bibinfo {collaboration} {LIGO Scientific, VIRGO}),\
  }\bibfield  {title} {\bibinfo {title} {{Characterization of the LIGO
  detectors during their sixth science run}},\ }\href
  {https://doi.org/10.1088/0264-9381/32/11/115012} {\bibfield  {journal}
  {\bibinfo  {journal} {Class. Quant. Grav.}\ }\textbf {\bibinfo {volume}
  {32}},\ \bibinfo {pages} {115012} (\bibinfo {year} {2015})},\ \Eprint
  {https://arxiv.org/abs/1410.7764} {arXiv:1410.7764 [gr-qc]} \BibitemShut
  {NoStop}%
\bibitem [{\citenamefont {Amaro-Seoane}\ \emph {et~al.}(2017)\citenamefont
  {Amaro-Seoane} \emph {et~al.}}]{LISA:2017pwj}%
  \BibitemOpen
  \bibfield  {author} {\bibinfo {author} {\bibfnamefont {P.}~\bibnamefont
  {Amaro-Seoane}} \emph {et~al.} (\bibinfo {collaboration} {LISA}),\ }\bibfield
   {title} {\bibinfo {title} {{Laser Interferometer Space Antenna}},\
  }\href@noop {} {\  (\bibinfo {year} {2017})},\ \Eprint
  {https://arxiv.org/abs/1702.00786} {arXiv:1702.00786 [astro-ph.IM]}
  \BibitemShut {NoStop}%
\bibitem [{\citenamefont {Auclair}\ \emph {et~al.}(2023)\citenamefont {Auclair}
  \emph {et~al.}}]{LISACosmologyWorkingGroup:2022jok}%
  \BibitemOpen
  \bibfield  {author} {\bibinfo {author} {\bibfnamefont {P.}~\bibnamefont
  {Auclair}} \emph {et~al.} (\bibinfo {collaboration} {LISA Cosmology Working
  Group}),\ }\bibfield  {title} {\bibinfo {title} {{Cosmology with the Laser
  Interferometer Space Antenna}},\ }\href
  {https://doi.org/10.1007/s41114-023-00045-2} {\bibfield  {journal} {\bibinfo
  {journal} {Living Rev. Rel.}\ }\textbf {\bibinfo {volume} {26}},\ \bibinfo
  {pages} {5} (\bibinfo {year} {2023})},\ \Eprint
  {https://arxiv.org/abs/2204.05434} {arXiv:2204.05434 [astro-ph.CO]}
  \BibitemShut {NoStop}%
\bibitem [{\citenamefont {Seto}\ \emph {et~al.}(2001)\citenamefont {Seto},
  \citenamefont {Kawamura},\ and\ \citenamefont {Nakamura}}]{Seto:2001qf}%
  \BibitemOpen
  \bibfield  {author} {\bibinfo {author} {\bibfnamefont {N.}~\bibnamefont
  {Seto}}, \bibinfo {author} {\bibfnamefont {S.}~\bibnamefont {Kawamura}},\
  and\ \bibinfo {author} {\bibfnamefont {T.}~\bibnamefont {Nakamura}},\
  }\bibfield  {title} {\bibinfo {title} {{Possibility of direct measurement of
  the acceleration of the universe using 0.1-Hz band laser interferometer
  gravitational wave antenna in space}},\ }\href
  {https://doi.org/10.1103/PhysRevLett.87.221103} {\bibfield  {journal}
  {\bibinfo  {journal} {Phys. Rev. Lett.}\ }\textbf {\bibinfo {volume} {87}},\
  \bibinfo {pages} {221103} (\bibinfo {year} {2001})},\ \Eprint
  {https://arxiv.org/abs/astro-ph/0108011} {arXiv:astro-ph/0108011}
  \BibitemShut {NoStop}%
\bibitem [{\citenamefont {Kawamura}\ \emph {et~al.}(2011)\citenamefont
  {Kawamura} \emph {et~al.}}]{Kawamura:2011zz}%
  \BibitemOpen
  \bibfield  {author} {\bibinfo {author} {\bibfnamefont {S.}~\bibnamefont
  {Kawamura}} \emph {et~al.},\ }\bibfield  {title} {\bibinfo {title} {{The
  Japanese space gravitational wave antenna: DECIGO}},\ }\href
  {https://doi.org/10.1088/0264-9381/28/9/094011} {\bibfield  {journal}
  {\bibinfo  {journal} {Class. Quant. Grav.}\ }\textbf {\bibinfo {volume}
  {28}},\ \bibinfo {pages} {094011} (\bibinfo {year} {2011})}\BibitemShut
  {NoStop}%
\bibitem [{\citenamefont {Kawamura}\ \emph {et~al.}(2006)\citenamefont
  {Kawamura} \emph {et~al.}}]{Kawamura:2006up}%
  \BibitemOpen
  \bibfield  {author} {\bibinfo {author} {\bibfnamefont {S.}~\bibnamefont
  {Kawamura}} \emph {et~al.},\ }\bibfield  {title} {\bibinfo {title} {{The
  Japanese space gravitational wave antenna DECIGO}},\ }\href
  {https://doi.org/10.1088/0264-9381/23/8/S17} {\bibfield  {journal} {\bibinfo
  {journal} {Class. Quant. Grav.}\ }\textbf {\bibinfo {volume} {23}},\ \bibinfo
  {pages} {S125} (\bibinfo {year} {2006})}\BibitemShut {NoStop}%
\bibitem [{\citenamefont {Crowder}\ and\ \citenamefont
  {Cornish}(2005)}]{Crowder:2005nr}%
  \BibitemOpen
  \bibfield  {author} {\bibinfo {author} {\bibfnamefont {J.}~\bibnamefont
  {Crowder}}\ and\ \bibinfo {author} {\bibfnamefont {N.~J.}\ \bibnamefont
  {Cornish}},\ }\bibfield  {title} {\bibinfo {title} {{Beyond LISA: Exploring
  future gravitational wave missions}},\ }\href
  {https://doi.org/10.1103/PhysRevD.72.083005} {\bibfield  {journal} {\bibinfo
  {journal} {Phys. Rev. D}\ }\textbf {\bibinfo {volume} {72}},\ \bibinfo
  {pages} {083005} (\bibinfo {year} {2005})},\ \Eprint
  {https://arxiv.org/abs/gr-qc/0506015} {arXiv:gr-qc/0506015} \BibitemShut
  {NoStop}%
\bibitem [{\citenamefont {Corbin}\ and\ \citenamefont
  {Cornish}(2006)}]{Corbin:2005ny}%
  \BibitemOpen
  \bibfield  {author} {\bibinfo {author} {\bibfnamefont {V.}~\bibnamefont
  {Corbin}}\ and\ \bibinfo {author} {\bibfnamefont {N.~J.}\ \bibnamefont
  {Cornish}},\ }\bibfield  {title} {\bibinfo {title} {{Detecting the cosmic
  gravitational wave background with the big bang observer}},\ }\href
  {https://doi.org/10.1088/0264-9381/23/7/014} {\bibfield  {journal} {\bibinfo
  {journal} {Class. Quant. Grav.}\ }\textbf {\bibinfo {volume} {23}},\ \bibinfo
  {pages} {2435} (\bibinfo {year} {2006})},\ \Eprint
  {https://arxiv.org/abs/gr-qc/0512039} {arXiv:gr-qc/0512039} \BibitemShut
  {NoStop}%
\bibitem [{\citenamefont {Harry}\ \emph {et~al.}(2006)\citenamefont {Harry},
  \citenamefont {Fritschel}, \citenamefont {Shaddock}, \citenamefont
  {Folkner},\ and\ \citenamefont {Phinney}}]{Harry:2006fi}%
  \BibitemOpen
  \bibfield  {author} {\bibinfo {author} {\bibfnamefont {G.~M.}\ \bibnamefont
  {Harry}}, \bibinfo {author} {\bibfnamefont {P.}~\bibnamefont {Fritschel}},
  \bibinfo {author} {\bibfnamefont {D.~A.}\ \bibnamefont {Shaddock}}, \bibinfo
  {author} {\bibfnamefont {W.}~\bibnamefont {Folkner}},\ and\ \bibinfo {author}
  {\bibfnamefont {E.~S.}\ \bibnamefont {Phinney}},\ }\bibfield  {title}
  {\bibinfo {title} {{Laser interferometry for the big bang observer}},\ }\href
  {https://doi.org/10.1088/0264-9381/23/15/008} {\bibfield  {journal} {\bibinfo
   {journal} {Class. Quant. Grav.}\ }\textbf {\bibinfo {volume} {23}},\
  \bibinfo {pages} {4887} (\bibinfo {year} {2006})},\ \bibinfo {note}
  {[Erratum: Class.Quant.Grav. 23, 7361 (2006)]}\BibitemShut {NoStop}%
\bibitem [{\citenamefont {Yagi}\ and\ \citenamefont
  {Seto}(2011)}]{Yagi:2011wg}%
  \BibitemOpen
  \bibfield  {author} {\bibinfo {author} {\bibfnamefont {K.}~\bibnamefont
  {Yagi}}\ and\ \bibinfo {author} {\bibfnamefont {N.}~\bibnamefont {Seto}},\
  }\bibfield  {title} {\bibinfo {title} {{Detector configuration of DECIGO/BBO
  and identification of cosmological neutron-star binaries}},\ }\href
  {https://doi.org/10.1103/PhysRevD.83.044011} {\bibfield  {journal} {\bibinfo
  {journal} {Phys. Rev. D}\ }\textbf {\bibinfo {volume} {83}},\ \bibinfo
  {pages} {044011} (\bibinfo {year} {2011})},\ \bibinfo {note} {[Erratum:
  Phys.Rev.D 95, 109901 (2017)]},\ \Eprint {https://arxiv.org/abs/1101.3940}
  {arXiv:1101.3940 [astro-ph.CO]} \BibitemShut {NoStop}%
\bibitem [{\citenamefont {Yagi}\ and\ \citenamefont
  {Seto}(2017)}]{Yagi:2017wg}%
  \BibitemOpen
  \bibfield  {author} {\bibinfo {author} {\bibfnamefont {K.}~\bibnamefont
  {Yagi}}\ and\ \bibinfo {author} {\bibfnamefont {N.}~\bibnamefont {Seto}},\
  }\bibfield  {title} {\bibinfo {title} {Erratum: Detector configuration of
  decigo/bbo and identification of cosmological neutron-star binaries [phys.
  rev. d 83, 044011 (2011)]},\ }\href
  {https://doi.org/10.1103/PhysRevD.95.109901} {\bibfield  {journal} {\bibinfo
  {journal} {Phys. Rev. D}\ }\textbf {\bibinfo {volume} {95}},\ \bibinfo
  {pages} {109901} (\bibinfo {year} {2017})}\BibitemShut {NoStop}%
\bibitem [{\citenamefont {Carilli}\ and\ \citenamefont
  {Rawlings}(2004)}]{Carilli:2004nx}%
  \BibitemOpen
  \bibfield  {author} {\bibinfo {author} {\bibfnamefont {C.~L.}\ \bibnamefont
  {Carilli}}\ and\ \bibinfo {author} {\bibfnamefont {S.}~\bibnamefont
  {Rawlings}},\ }\bibfield  {title} {\bibinfo {title} {{Science with the Square
  Kilometer Array: Motivation, key science projects, standards and
  assumptions}},\ }\href {https://doi.org/10.1016/j.newar.2004.09.001}
  {\bibfield  {journal} {\bibinfo  {journal} {New Astron. Rev.}\ }\textbf
  {\bibinfo {volume} {48}},\ \bibinfo {pages} {979} (\bibinfo {year} {2004})},\
  \Eprint {https://arxiv.org/abs/astro-ph/0409274} {arXiv:astro-ph/0409274}
  \BibitemShut {NoStop}%
\bibitem [{\citenamefont {Moore}\ \emph {et~al.}(2015)\citenamefont {Moore},
  \citenamefont {Cole},\ and\ \citenamefont {Berry}}]{Moore:2014lga}%
  \BibitemOpen
  \bibfield  {author} {\bibinfo {author} {\bibfnamefont {C.~J.}\ \bibnamefont
  {Moore}}, \bibinfo {author} {\bibfnamefont {R.~H.}\ \bibnamefont {Cole}},\
  and\ \bibinfo {author} {\bibfnamefont {C.~P.~L.}\ \bibnamefont {Berry}},\
  }\bibfield  {title} {\bibinfo {title} {{Gravitational-wave sensitivity
  curves}},\ }\href {https://doi.org/10.1088/0264-9381/32/1/015014} {\bibfield
  {journal} {\bibinfo  {journal} {Class. Quant. Grav.}\ }\textbf {\bibinfo
  {volume} {32}},\ \bibinfo {pages} {015014} (\bibinfo {year} {2015})},\
  \Eprint {https://arxiv.org/abs/1408.0740} {arXiv:1408.0740 [gr-qc]}
  \BibitemShut {NoStop}%
\bibitem [{\citenamefont {Weltman}\ \emph {et~al.}(2020)\citenamefont {Weltman}
  \emph {et~al.}}]{Weltman:2018zrl}%
  \BibitemOpen
  \bibfield  {author} {\bibinfo {author} {\bibfnamefont {A.}~\bibnamefont
  {Weltman}} \emph {et~al.},\ }\bibfield  {title} {\bibinfo {title}
  {{Fundamental physics with the Square Kilometre Array}},\ }\href
  {https://doi.org/10.1017/pasa.2019.42} {\bibfield  {journal} {\bibinfo
  {journal} {Publ. Astron. Soc. Austral.}\ }\textbf {\bibinfo {volume} {37}},\
  \bibinfo {pages} {e002} (\bibinfo {year} {2020})},\ \Eprint
  {https://arxiv.org/abs/1810.02680} {arXiv:1810.02680 [astro-ph.CO]}
  \BibitemShut {NoStop}%
\bibitem [{\citenamefont {Correa}\ \emph {et~al.}(2024)\citenamefont {Correa},
  \citenamefont {Gangopadhyay}, \citenamefont {Jaman},\ and\ \citenamefont
  {Mathews}}]{Correa:2023whf}%
  \BibitemOpen
  \bibfield  {author} {\bibinfo {author} {\bibfnamefont {M.}~\bibnamefont
  {Correa}}, \bibinfo {author} {\bibfnamefont {M.~R.}\ \bibnamefont
  {Gangopadhyay}}, \bibinfo {author} {\bibfnamefont {N.}~\bibnamefont
  {Jaman}},\ and\ \bibinfo {author} {\bibfnamefont {G.~J.}\ \bibnamefont
  {Mathews}},\ }\bibfield  {title} {\bibinfo {title} {{Induced gravitational
  waves via warm natural inflation}},\ }\href
  {https://doi.org/10.1103/PhysRevD.109.063539} {\bibfield  {journal} {\bibinfo
   {journal} {Phys. Rev. D}\ }\textbf {\bibinfo {volume} {109}},\ \bibinfo
  {pages} {063539} (\bibinfo {year} {2024})},\ \Eprint
  {https://arxiv.org/abs/2306.09641} {arXiv:2306.09641 [astro-ph.CO]}
  \BibitemShut {NoStop}%
\bibitem [{\citenamefont {Agazie}\ \emph
  {et~al.}(2023{\natexlab{a}})\citenamefont {Agazie} \emph
  {et~al.}}]{NANOGrav:2023gor}%
  \BibitemOpen
  \bibfield  {author} {\bibinfo {author} {\bibfnamefont {G.}~\bibnamefont
  {Agazie}} \emph {et~al.} (\bibinfo {collaboration} {NANOGrav}),\ }\bibfield
  {title} {\bibinfo {title} {{The NANOGrav 15 yr Data Set: Evidence for a
  Gravitational-wave Background}},\ }\href
  {https://doi.org/10.3847/2041-8213/acdac6} {\bibfield  {journal} {\bibinfo
  {journal} {Astrophys. J. Lett.}\ }\textbf {\bibinfo {volume} {951}},\
  \bibinfo {pages} {L8} (\bibinfo {year} {2023}{\natexlab{a}})},\ \Eprint
  {https://arxiv.org/abs/2306.16213} {arXiv:2306.16213 [astro-ph.HE]}
  \BibitemShut {NoStop}%
\bibitem [{\citenamefont {Agazie}\ \emph
  {et~al.}(2023{\natexlab{b}})\citenamefont {Agazie} \emph
  {et~al.}}]{NANOGrav:2023hde}%
  \BibitemOpen
  \bibfield  {author} {\bibinfo {author} {\bibfnamefont {G.}~\bibnamefont
  {Agazie}} \emph {et~al.} (\bibinfo {collaboration} {NANOGrav}),\ }\bibfield
  {title} {\bibinfo {title} {{The NANOGrav 15 yr Data Set: Observations and
  Timing of 68 Millisecond Pulsars}},\ }\href
  {https://doi.org/10.3847/2041-8213/acda9a} {\bibfield  {journal} {\bibinfo
  {journal} {Astrophys. J. Lett.}\ }\textbf {\bibinfo {volume} {951}},\
  \bibinfo {pages} {L9} (\bibinfo {year} {2023}{\natexlab{b}})},\ \Eprint
  {https://arxiv.org/abs/2306.16217} {arXiv:2306.16217 [astro-ph.HE]}
  \BibitemShut {NoStop}%
\bibitem [{\citenamefont {Gangopadhyay}\ \emph {et~al.}(2025)\citenamefont
  {Gangopadhyay}, \citenamefont {Godithi}, \citenamefont {Inui}, \citenamefont
  {Ichiki}, \citenamefont {Kajino}, \citenamefont {Manusankar}, \citenamefont
  {Mathews},\ and\ \citenamefont {Yogesh}}]{Gangopadhyay:2023qjr}%
  \BibitemOpen
  \bibfield  {author} {\bibinfo {author} {\bibfnamefont {M.~R.}\ \bibnamefont
  {Gangopadhyay}}, \bibinfo {author} {\bibfnamefont {V.~V.}\ \bibnamefont
  {Godithi}}, \bibinfo {author} {\bibfnamefont {R.}~\bibnamefont {Inui}},
  \bibinfo {author} {\bibfnamefont {K.}~\bibnamefont {Ichiki}}, \bibinfo
  {author} {\bibfnamefont {T.}~\bibnamefont {Kajino}}, \bibinfo {author}
  {\bibfnamefont {A.}~\bibnamefont {Manusankar}}, \bibinfo {author}
  {\bibfnamefont {G.~J.}\ \bibnamefont {Mathews}},\ and\ \bibinfo {author}
  {\bibnamefont {Yogesh}},\ }\bibfield  {title} {\bibinfo {title} {{Is the
  NANOGrav detection evidence of resonant particle creation during
  inflation?}},\ }\href {https://doi.org/10.1016/j.jheap.2025.100358}
  {\bibfield  {journal} {\bibinfo  {journal} {JHEAp}\ }\textbf {\bibinfo
  {volume} {47}},\ \bibinfo {pages} {100358} (\bibinfo {year} {2025})},\
  \Eprint {https://arxiv.org/abs/2309.03101} {arXiv:2309.03101 [astro-ph.CO]}
  \BibitemShut {NoStop}%
\bibitem [{\citenamefont {Iovino}\ \emph {et~al.}(2024)\citenamefont {Iovino},
  \citenamefont {Perna}, \citenamefont {Riotto},\ and\ \citenamefont
  {Veerm\"ae}}]{Iovino:2024tyg}%
  \BibitemOpen
  \bibfield  {author} {\bibinfo {author} {\bibfnamefont {A.~J.}\ \bibnamefont
  {Iovino}}, \bibinfo {author} {\bibfnamefont {G.}~\bibnamefont {Perna}},
  \bibinfo {author} {\bibfnamefont {A.}~\bibnamefont {Riotto}},\ and\ \bibinfo
  {author} {\bibfnamefont {H.}~\bibnamefont {Veerm\"ae}},\ }\bibfield  {title}
  {\bibinfo {title} {{Curbing PBHs with PTAs}},\ }\href
  {https://doi.org/10.1088/1475-7516/2024/10/050} {\bibfield  {journal}
  {\bibinfo  {journal} {JCAP}\ }\textbf {\bibinfo {volume} {10}},\ \bibinfo
  {pages} {050}},\ \Eprint {https://arxiv.org/abs/2406.20089} {arXiv:2406.20089
  [astro-ph.CO]} \BibitemShut {NoStop}%
\end{thebibliography}%

\end{document}